\documentclass[12pt]{article}   

\usepackage{hyperref}

\newcommand{\be}{\begin{equation}}
\newcommand{\ee}{\end{equation}}
\usepackage{amsmath}
\usepackage{amssymb}
\usepackage{latexsym}
\usepackage{epsfig}

\newcommand{\half}{{\textstyle {1\over 2}}}
\newcommand{\ord}[1]{\mathcal{O}(#1)}

\newcommand{\ket}[1]{|#1\rangle}

\newcommand{\del}{\partial}

\newcommand{\refb}[1]{(\ref{#1})}

\newcommand{\RRR}{{\hbox{\rm R\kern-2.35mm R}}}

\def\ZZZ{{\hbox{ Z\kern-1.6mm Z}}}

\newcommand{\RR}{{\cal R}}

\begin{document}
\noindent

\begin{titlepage}
%\rightline{\today} 
\rightline{\tt arXiv:0805.3701}
\rightline{\tt MIT-CTP-3948}
\begin{center}
\vskip 2.5cm
{\Large \bf {One-Loop Riemann Surfaces in Schnabl Gauge}}\\

\vskip 2.0cm
{\large {Michael Kiermaier and Barton Zwiebach}}
\vskip 1.0cm
{\it {Center for Theoretical Physics}}\\
{\it {Massachusetts Institute of Technology}}\\
{\it {Cambridge, MA 02139, USA}}\\
mkiermai@mit.edu, zwiebach@mit.edu

\vskip 2.0cm
{\bf Abstract}

\vskip 1.0cm
\end{center}

\noindent
\begin{narrower}
Due to a peculiar behavior at the open string
midpoint, loop diagrams in Schnabl gauge
were expected to fail to produce the relevant closed string moduli.
We find that closed string moduli are generated because the
Riemann surfaces are built with {\em slanted} wedges:
semi-infinite strips whose edges have parameterizations
related by scaling.
We examine in detail one-loop string diagrams
and find that the closed string modulus is always produced.
Moreover,  the conformal maps simplify so greatly that both
closed and open moduli become simple calculable functions of the Schwinger parameters, a simplification that occurs neither in Siegel 
gauge nor in light-cone gauge.

\end{narrower}

\medskip

\end{titlepage}

\newpage

\tableofcontents
\baselineskip=17pt
\section{Introduction}

The string field $\Phi$  that represents
the tachyon vacuum in Schnabl's solution~\cite{0511286}
 of open string field theory~\cite{wit1}
satisfies a novel gauge condition.
 The  solution
is not in Siegel gauge~\cite{siegel1}:
$\Phi$  is not annihilated by 
 the zero mode $b_0$ 
 of the antighost field in the canonical
open string frame. 
Rather, $\Phi$ is annihilated by 
the zero mode $B$ of the antighost field
 in the conformal frame of the sliver projector of the
star algebra of open string fields.
The sliver frame is central to the construction
 and analysis
of classical solutions~\cite{0603159}--\cite{Kawano:2008jv}
 but,
as any projector frame, it is singular at the open string midpoint.
One can wonder if the  Schnabl
gauge condition $B\Phi=0$
defines a consistent open string perturbation theory.
In this question, the singular behavior of the open string midpoint
has brought interesting advantages
 but has also introduced some new subtleties.

At tree level, the sliver frame makes all conformal maps from the string
diagrams to the upper-half plane very simple~\cite{Fuji:2006me,0708.2591}.
This is remarkable, if we recall that in Siegel gauge these maps are
extremely complicated and no closed form expressions are
 known
except for four-string  amplitudes~\cite{gid}.
The subtleties arise because there
are delicate contributions whose origin can be traced to the singular behavior at the open string midpoint~\cite{0708.2591}.  These contributions affect the off-shell part of four-string amplitudes and could affect higher-point functions on-shell.
No Feynman rules are known that deal
with these complications
in general tree-level amplitudes.

This state of affairs prompted~\cite{Kiermaier:2007jg}
to introduce a class of {\em regular linear $b$-gauges}
 that produce correct on-shell amplitudes. In this class,
a propagator insertion with  Schwinger parameter approaching
infinity induces an
open string degeneration of the Riemann surface associated with the string diagram-- the desired behavior.
Schnabl gauge does not belong to the class of
regular $b$-gauges,
but there is a simple one-parameter family of regular
 linear
$b$-gauges that
interpolates between Siegel and Schnabl gauge as its parameter $\lambda$ goes
from infinity to zero.
This suggests that
Schnabl gauge amplitudes can be obtained by taking the limit
 $\lambda\to0$ of the well-behaved amplitudes in this $\lambda$-family.

While it is not yet proven that
 moduli space is covered for general tree amplitudes
 in Schnabl gauge, it is no
 mystery how the relevant Riemann surfaces --disks with
 boundary punctures -- carry the moduli and how degenerations
 can be  generated.
Naive arguments, however,
suggest that
Schnabl gauge at loop level only produces surfaces with
degenerate closed string moduli,
thus making it impossible to
reproduce the correct on-shell amplitudes.  In a one-loop amplitude, for
example, the line traced by the open string midpoint is a  nontrivial
closed curve. In the Schnabl propagator the open string midpoint does not move, thus
 naively suggesting
a diagram with a
zero-length closed curve that signals closed string degeneration.

It is the main purpose of this paper to discuss the one-loop string diagrams
in Schnabl gauge.  Our results are quite encouraging.  We find that the
anticipated
problems with closed string moduli are not present.
Our main tool is the regulation provided by the $\lambda$-family
of regular linear
$b$-gauges that yield Schnabl gauge in the limit. Not only
are closed string moduli produced but they are easily calculated,
something that does not happen in Siegel gauge.  Our work focuses only
on the moduli problem; we do not attempt to fully compute any
loop-amplitude.  Such a computation, of course, would be quite interesting.

The analysis shows that closed string moduli arise because
vertical lines
 in the sliver frame that are identified horizontally in
 tree diagrams, require slanted identifications in the case of loops. We recall that wedge surfaces~\cite{Rastelli:2000iu,Schnabl:2002gg}
are semi-infinite strips of fixed width whose vertical edges carry identical
parameterizations.  We are led to introduce {\em slanted} wedges, semi-infinite strips of fixed width whose vertical edges have parameterizations
related by a scale factor.  These slanted
wedges are new, interesting objects in
their own right.  One can glue them
and they are a natural ingredient in
the construction of loop-diagrams.  As opposed to the familiar wedges, however, there are no states associated to them.  With the help of slanted
wedges we develop a formalism that allows us to calculate the moduli
(both open and closed) of arbitrary tree and one-loop amplitudes.
Our analysis also shows that the  BPZ-even gauge condition
$B^+ \Phi = (B + B^\star)\Phi =0$, where $\star$ denotes BPZ
conjugation,  fails to generate the closed string
modulus in one-loop diagrams because
in this gauge
 the identifications in the sliver frame are not slanted.
 Unlike Schnabl gauge, the gauge $B^+ \Phi =0$
 appears to be genuinely inadequate for loop calculations.

\bigskip
This paper is organized as follows.
In Section~\ref{secvac}, we will begin our analysis with the one-loop
vacuum graph
 in general regular linear $b$-gauges,
 focusing on the Riemann surfaces generated by varying
 one of the two Schwinger parameters of the propagator.
We see that the modulus
of the annulus is an exactly calculable function of the Schwinger parameter
and is, in fact, independent of the gauge choice.
 We then study the vacuum graph in Schnabl gauge as a limit in the family
 of regular interpolating gauges.
The role of slanted
identifications in Schnabl gauge
first becomes apparent and the error in the presumption that
diagrams are closed string degenerate is identified.

The situation becomes more nontrivial and challenging for
the one-loop tadpole,
 i.e. the one-loop one-point function.
We study this diagram in Section~\ref{sectp}
 for the family of interpolating gauges parameterized by $\lambda$.
The diagram
only has a closed string modulus; the position of
the open string puncture can be adjusted using rotations.  For any fixed
$\lambda$,
we can use extremal length methods to show that the full moduli space
of annuli is produced as the Schwinger parameter
is varied over its allowed range. In Siegel gauge the modulus
of the annulus is a complicated function of the Schwinger parameter
(defined implicitly by certain elliptic integrals, see, for example~\cite{Zemba:1988rf}).
In the limit that we reach Schnabl gauge the modulus becomes a simple
function of the Schwinger parameter.  In this example one can glean
the main geometrical insight that shows how the two components
of the annulus, each one with its own
 open string boundary, are glued across
 a {\em hidden boundary} at infinity!
 The existence of such a hidden boundary leads us to conclude that
 the operator $L$ (the Virasoro zero mode in the
sliver frame) has an anomalous left/right decomposition, i.e.
 $[L_L,L_R]\neq 0$.

In Section~\ref{secsw}, we will introduce slanted wedges and show how
to glue them together, as suggested by star multiplication, to produce
a closed algebra.
 We discuss how the  operators $L_L$ and $L_R$ and their BPZ conjugates
 act on slanted wedges and derive the action of the full
 Schnabl propagator.
This formalism simplifies tremendously
the construction of string diagrams, as we discuss for the case of trees
in Section~\ref{sectree}.   The moduli for tree diagrams are the positions
of open string punctures and these can be calculated efficiently, as is
demonstrated for the case of
 the 5-point diagram.
 We present the generalization to arbitrary tree diagrams, which
 is surprisingly
 straightforward using the algebra of slanted wedges.

In Section~\ref{secloop} we discuss the Riemann surfaces for
general one-loop string diagrams in Schnabl gauge.  We show how
to construct such a surface by gluing the hidden boundaries of the
surfaces
 associated with each of the
 boundaries of the annulus.
 Both of these surfaces
are naturally built with slanted wedges.
 We determine the closed string modulus and all
 open string moduli as simple functions of the Schwinger parameters.
In particular, we find that the closed string modulus
 only depends on the Schwinger parameters of the propagators running in the loop.
The computations are
illustrated in Section~\ref{secloop2pt}  where we work out
the one-loop diagram  with two external states.
 If both external states are placed on the same boundary component
there are two string diagrams,
and we discuss how they generate together
the relevant open and closed string moduli.

 In Section~\ref{aregvieononelooam}
we  use the family of $\lambda$-regularized gauges
to justify our prescription for the calculation of one-loop moduli.
There are three types of gluing operations that need to be justified
in the Schnabl limit  $\lambda\to0$: (i) the star multiplication of slanted wedges corresponding to external states and propagator surfaces,
(ii) the gluing along hidden boundaries that  forms a
single strip from
the two slanted wedges each of which contains one boundary
component of the
one-loop diagram,
and (iii) the identification of the edges of the resulting strip that
creates the annulus.
We show that all three types of operations can be justified rigorously in the Schnabl limit.
We end in Section~\ref{secconcl}
with some concluding remarks.

\section{The vacuum graph}\label{secvac}
\setcounter{equation}{0}

In this section we discuss the geometry of the vacuum
graph.  Our objectives are to set up notation and
 to calculate the modulus of the vacuum graph as a function of the
 Schwinger parameter for general regular linear $b$-gauges.

\subsection{Gauges, coordinate frames and  the surface $\mathcal{R}(s)$}\label{sec21}
Reference~\cite{Kiermaier:2007jg} studied open string perturbation theory in
  a class of gauges called linear $b$-gauges.
 In these gauges,
 a linear combination of even moded
antighost oscillators  annihilates the classical
string field
$\ket{\psi_{cl}}$ :
\begin{equation}\label{classgaugecond}
   B[v]\,\ket{\psi_{cl}}=0\, .
\end{equation}
Here  $B[v]$ is determined by a vector field $v(\xi)$ via
\begin{equation}\label{vecfield}
      B[v]=\sum_{k\in \mathbb{Z}} v_{2k}b_{2k}=\oint {d\xi\over 2\pi i} v(\xi) b(\xi)\,,
      \qquad \textrm{with} \quad
      v(\xi)=\sum_{k\in\mathbb{Z}}v_{2k}\xi^{2k+1}\,, \quad
      v_{2k}\in\mathbb{R}\,.
\end{equation}
A subset of linear $b$-gauges in which string perturbation is guaranteed to produce the correct
on-shell amplitudes was identified in~\cite{Kiermaier:2007jg}. In this subset the vector field
$v(\xi)$ is analytic in a neighborhood of the unit circle $|\xi|=1$, and
satisfies the
condition
\begin{equation}
     \Re\left(\bar\xi v(\xi)\right)>0\, \qquad  \textrm{for} \quad
     |\xi|=1\,.
\end{equation}
These gauges were called  \emph{regular linear $b$-gauges}.
One also defines
\begin{equation}
   L[v]\equiv\bigl\{Q,B[v]\bigr\}=\oint {d\xi\over 2\pi i} v(\xi) T(\xi)=\sum_{k\in \mathbb{Z}} v_{2k}L_{2k} \,.
\end{equation}
In a certain
frame $w= g(\xi)$
 the operator $L[v]$ generates
translations~\cite{Kiermaier:2007jg,0611200}.
The map $g(\xi)$ is related to the vector field  $v(\xi)$ through
\begin{equation}
\label{vmiahpcss}
{dg\over d\xi} = -{1\over v(\xi)} \,.
\end{equation}
 Normalizing $v(\xi)$ appropriately, we can
  impose on $g(\xi)$ the convenient boundary conditions
 \begin{equation}
 \label{vmiahpcssnorm}
     g(-1)=0  \,, \quad  g(1)=i\pi\,.
 \end{equation}
We also use the frame   $z=f(\xi)$ where the operator
 $L[v]$ is the zero mode Virasoro operator and thus generates
 scaling. This frame is only determined up to an overall factor. We choose the
 normalization
\begin{equation}
\label{zframeaj}
f(\pm 1) = \pm\, \half \,.
\end{equation}
Given such a frame $z=f(\xi)$, one can determine the associated vector field $v(\xi)$ as
\begin{equation}
\label{vfieldaj}
v(\xi) = {f(\xi)\over f'(\xi)}\,.
\end{equation}
The defining property of this vector field is that the operators
$L[v]$ and $B[v]$
are, respectively, the zero modes of the Virasoro and antighost operators
in the $z$ frame.
Use of (\ref{vmiahpcss}) and (\ref{vfieldaj}) immediately shows that the $w$ and $z$ frames
are related by
$g = - \ln f + c$, where $c$ is a constant.
 This constant is determined by our boundary conditions on $g(\xi)$ and $f(\xi)$ in~(\ref{vmiahpcssnorm}) and~(\ref{zframeaj}). We obtain
\begin{equation}
\label{wzrel}
\boxed{\phantom{\Biggl(}
w= g(\xi) =  - \ln (2 f(\xi)) + i\pi  = - \ln (2z) + i \pi \,.~}
\end{equation}
In this map $z$
 is always in the upper-half plane and the
branch of the logarithm is taken using $0 \leq \hbox{Arg}\, z \leq \pi$.
Inverting (\ref{wzrel}) we get
\begin{equation}
\label{zfromwaj}
z = f(\xi) = - \half\, \,e^{-w}\,.
\end{equation}

Picking a gauge condition~(\ref{classgaugecond}) for the
 classical string field $\ket{\psi_{cl}}$ of ghost number one is only the
 first step in the gauge fixing
 procedure~\cite{boch,thorn,preit,Kiermaier:2007jg}.
 Appropriate vector
 fields $v(\xi)$
 must be chosen for each ghost number and the gauge condition
 is that the corresponding $B[v]$
 operator must annihilate the
 string field at the given ghost number.
 We will return to this issue when we address general one-loop amplitudes in
 Section~\ref{secloop}.

\begin{figure}[t]
\centerline{
  \epsfig{figure=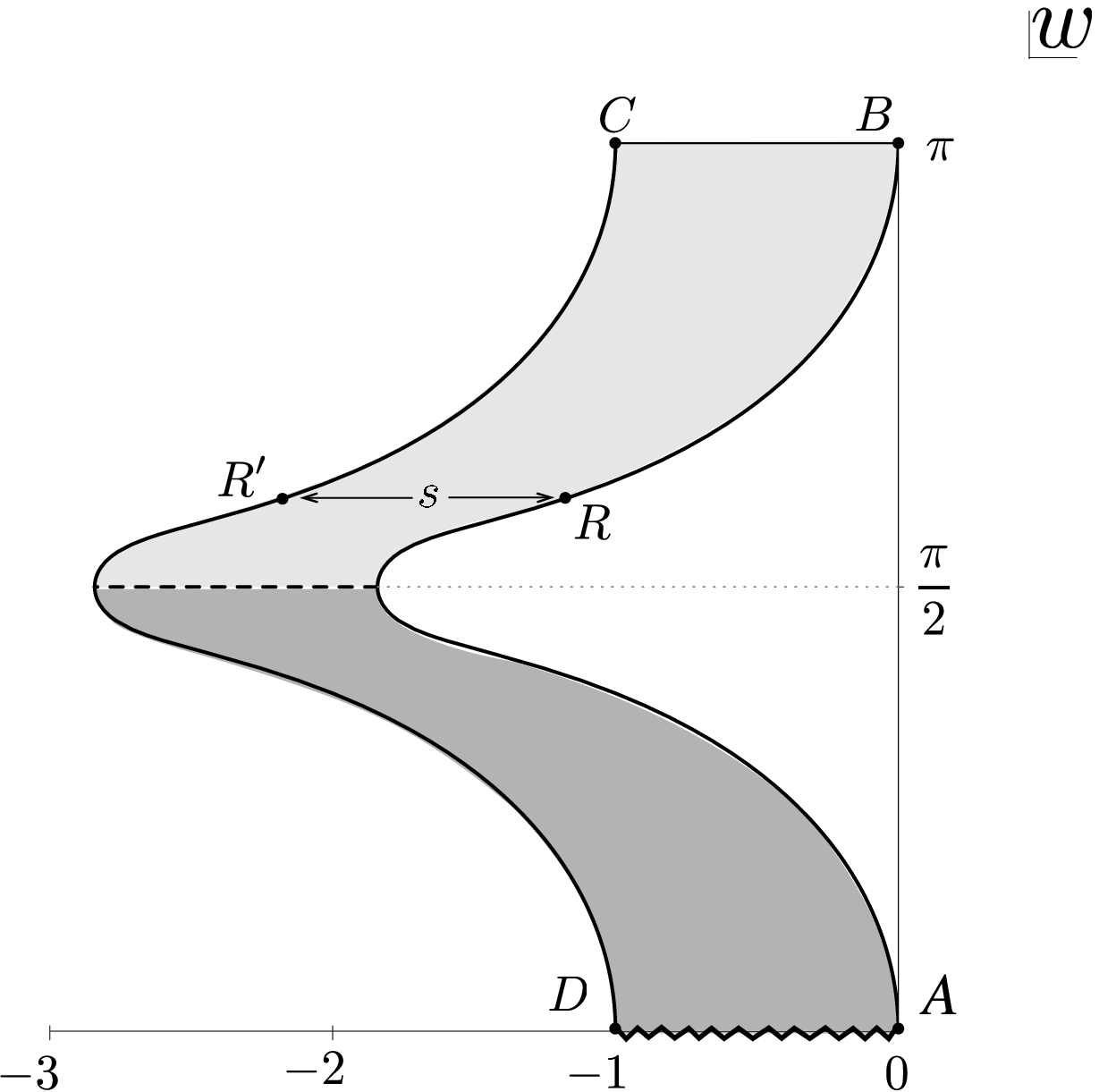, height=8cm}
  \hskip .75cm
  \epsfig{figure=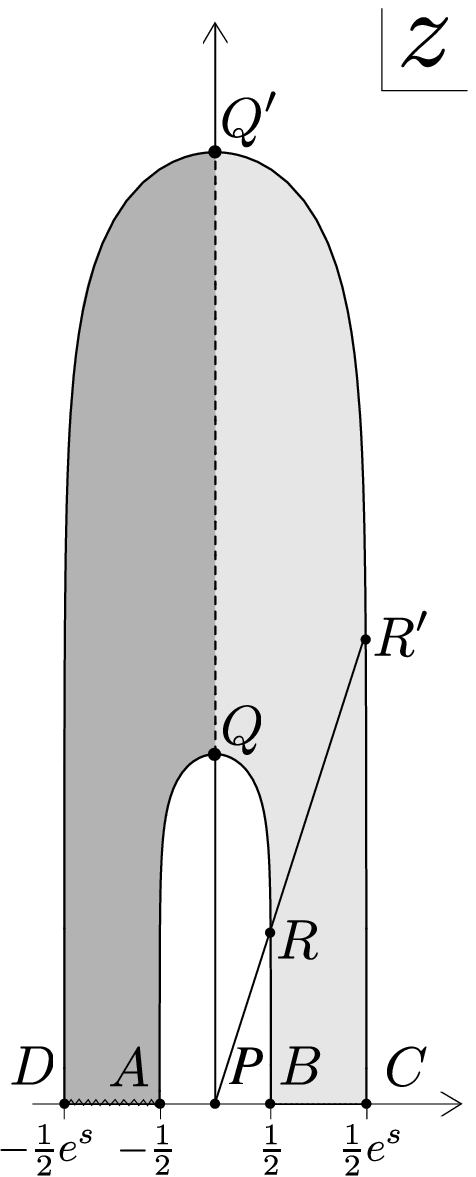, height=8cm}
  \hskip 1.35cm
  \epsfig{figure=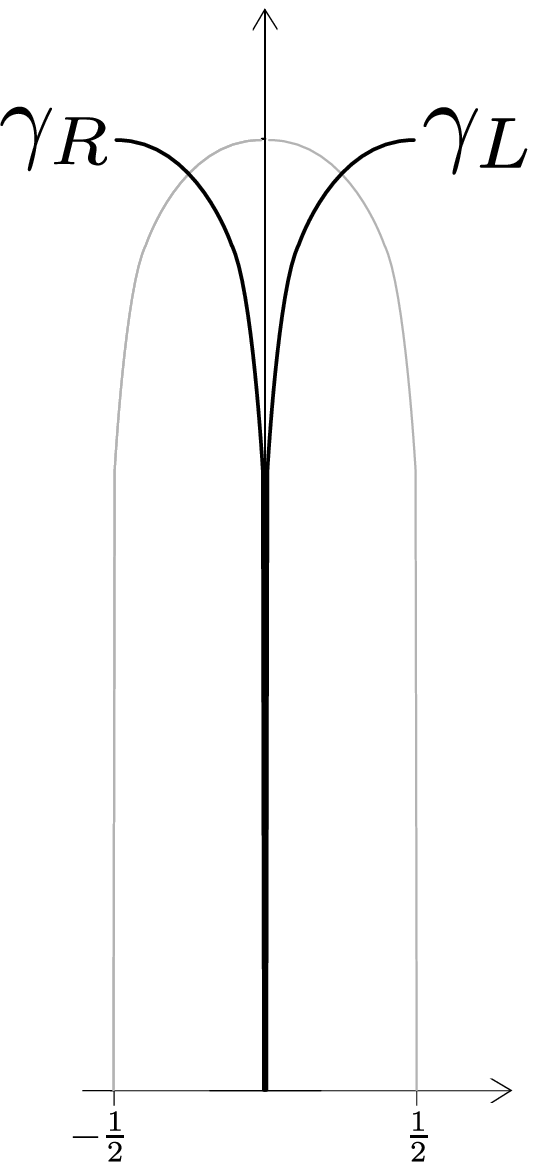, height=8cm}
}
\centerline{ \hskip 4cm  (a) \hskip 6.2cm (b) \hskip 4.3cm (c)\hskip 2.5cm }
\caption{The surface $\RR(s)$ created by
$e^{-sL[v]}$ in the  $w$ frame (a)  and  in the $z$ frame (b).
Points $R$ and $R'$ related
by horizontal translation $w\to w-s$ in the $w$ frame
are related by
scaling $z \to e^s z$ in the $z$ frame.
The surface $\RR(s)$ is displayed for $L[v]=L^\lambda$ with $\lambda=10^{-4}$ and $s=1$.\quad
The curves $\gamma_L$ and $\gamma_R$ arise from the
coordinate curve $f(e^{i\theta})$, as illustrated in (c).
}
\label{sl01fig}
\end{figure}

We noted above that the operator $L[v]$ generates rescalings in the 
$z$ frame and
translations in the $w$ frame.
As a differential operator we
 thus
have
\begin{equation}
\label{diffaction}
L[v] = - z {d\over dz}  = {d\over dw}\,.
\end{equation}
The operator $e^{-sL[v]}$ creates a strip $\mathcal{R}(s)$
of length $s$ in the $w$ frame with two horizontal open string boundaries~\cite{Kiermaier:2007jg}, as depicted in Figure~\ref{sl01fig}(a).
The boundary conditions~(\ref{vmiahpcssnorm}) ensure that the
 width of the strip is normalized to $\pi$.
 Furthermore, the strip domain
 $\RR(s)$
has as right boundary the curve
$w= g(\xi = e^{i\theta})$
 with $0\leq \theta\leq \pi$;
  this is just the $w$-plane image of the coordinate curve.
It is clear from (\ref{diffaction}) that
$e^{-sL[v]}$
translates by a distance $s$ to the left.  It follows that the left boundary of $\mathcal{R}(s)$
is the right boundary copied a distance $s$ to the left.\footnote{
This representation of $\RR(s)$ differs from the representation in~\cite{Kiermaier:2007jg}
by a rescaling of $\frac{1}{2}e^s$ in the $z$ frame and by a translation of $-s$ in the $w$ frame.}

Using the relation~(\ref{zfromwaj}),
we can map the strip $\RR(s)$ to the $z$ frame.
The right boundary of $\RR(s)$ in the $w$ frame becomes the \emph{coordinate curve} $z=f(e^{i\theta})$
with $0\leq\theta\leq\pi$ in the $z$ frame.
As $L[v]$ generates rescalings in this
frame, the surface $\RR(s)$ is swept out by rescalings of the coordinate curve
with scale factors ranging from one to $e^s$.
As we can decompose the operator $L[v]$
into left and right pieces,
 \be
 \label{iwltflckmytngttassfcl}
 L[v] = L[v]_L + L[v]_R\,,
 \ee
we can similarly divide $\RR(s)$ into two components, one associated with the action of $e^{-sL[v]_L}$ and the other associated with the action of
$e^{-sL[v]_R}$.
The component associated with $e^{-sL[v]_R}$ is the part of $\RR(s)$ in the region
$\Re(z)>0$
and is shaded in light grey in Figure~\ref{sl01fig}(b). It is swept out by rescalings of the right
part of the coordinate curve, which we parameterize as
\begin{equation}\label{gammaR}
 ~~\half+\gamma_R(\theta)\equiv f\bigl(e^{i\theta}\bigr)\,, ~\quad\qquad 0\leq\theta\leq\textstyle{\frac{\pi}{2}}\,.
\end{equation}
Similarly, the component of $\RR(s)$ associated with $e^{-sL[v]_L}$,
shaded in dark grey in the figure,
 is located in the region
 $\Re(z)<0$,
and is swept out by rescalings of the left part of the coordinate line, which
we parameterize as
\begin{equation}\label{gammaL}
   -\half+\gamma_L(\theta)\equiv f\bigl(e^{i(\pi-\theta)}\bigr)\,, \qquad 0\leq\theta\leq\textstyle{\frac{\pi}{2}}\,.
\end{equation}
Note that the curves $\gamma_R(\theta)$ and $\gamma_L(\theta)$
introduced above
are, respectively, the right and left parts of the coordinate curve,  displaced horizontally
so that for $\theta=0$ they are
at the origin (Figure~\ref{sl01fig}(c)).
The left component of $\RR(s)$ is
simply a reflection of the
right component around the axis $\Re(z)=0$, because the general form~(\ref{vecfield}) of the vector field
$v(\xi)$ implies
\begin{equation}
   \gamma_L(\theta) =-\overline{\gamma_R(\theta)}\,.
\end{equation}
The left and right components of $\RR(s)$  need to be glued
on the imaginary axis along the line $QQ'$, which stretches
from $f(i)$ to $e^sf(i)$. For regular linear $b$-gauges $f(i)$ is finite, resulting in a finite boundary
$QQ'$ generated by $e^{-sL[v]_L}$ and $e^{-sL[v]_R}$.
Thus,
 $e^{-sL[v]_L}$ and $e^{-sL[v]_R}$ do not give the surface associated with
$e^{-sL[v]}$ until they are glued along $QQ'$.
This can be traced to the non-commutativity of $L[v]_L$ and $L[v]_R$,
  \be
 \label{iwltflckmytngttassfvm}
 \bigl[\,L[v]_L \, , \, L[v]_R\,\bigr] \not= 0 \,,
 \ee
which in turn implies $e^{-sL[v]}\not= e^{-sL[v]_R}e^{-sL[v]_L}$
 for regular linear $b$-gauges.  The operators in (\ref{iwltflckmytngttassfvm})
 fail to commute because the vector field $v$ does not vanish at the open
 string midpoint (see~\cite{0606131}).

\bigskip
In this paper
the family of $\lambda$-regulated
gauges introduced in~\cite{Kiermaier:2007jg}
plays an important role.
 This family is defined through the one-parameter family of
 vector fields
 \begin{equation}
    v^\lambda(\xi) =
    e^\lambda (1+ e^{-2\lambda}\xi^2)
     \tan^{-1} (e^{-\lambda}\xi)
    \,, \quad \text{ with }\quad \lambda>0\,.
 \end{equation}
 The surface $\RR(s)$ in this gauge is then generated by the operator
 \begin{equation}
    L^\lambda\equiv L[v^\lambda]
    = L_0 + 2\sum_{k=1}^\infty {(-1)^{k+1}\over 4 k^2 -1} \, e^{-2k\lambda}
\, L_{2k}\, .
 \end{equation}
This family interpolates
from
Siegel gauge as $\lambda\to\infty$ to Schnabl gauge
 which arises in the limit
$\lambda \to 0$.  In fact, these gauges are regular
 linear $b$-gauges for all values $\lambda>0$.
Schnabl gauge is not regular --
this is why there is no proof yet that amplitudes arise correctly.

For the $\lambda$-regulated $z$ 
frames we have the $\lambda$-regulated
functions
\begin{equation}
\label{flambdadef}
f^\lambda (\xi) = {1\over 2} \, {\tan^{-1} (e^{-\lambda} \xi) \over
\tan^{-1} (e^{-\lambda})}
 \,,\quad \text{with }\quad \lambda>0 \,.
\end{equation}
 While in general regular linear $b$-gauges the functions $f(\xi)$, just like
 $v(\xi)$, need only be analytic in a neighborhood of $|\xi|=1$, the
 functions~(\ref{flambdadef})
 have the nice property that they are analytic on the entire unit disk
 $|\xi|\leq 1$.
 They map the real axis between $\xi=-1$ and $\xi=1$ to the real axis between $z=-\half$ and
 $z=\half$, and map $\xi=0$ to $z=0$. The region in the $z$ frame between the real axis and the
 curve $z=f(e^{i\theta})$ with $0\leq \theta\leq \pi$ can thus be interpreted as a canonical coordinate patch that
 glues nicely to the boundary of $\RR(s)$. The maps $f^\lambda (\xi)$ are thus \emph{coordinate functions}.
 In the Schnabl limit $\lambda\to 0$, we obtain
 \begin{equation}
   \label{f0def}
   f(\xi) \equiv \lim_{\lambda\to0}\, f^\lambda(\xi)= {2\over \pi} \, \tan^{-1} \xi\,.
 \end{equation}
 This is the familiar coordinate function of the sliver frame which is well defined for all $|\xi|\leq1$ except for
 $\xi=i$. The open string midpoint $\xi=i$ is mapped to $i\infty$.

\begin{figure}[t]
\begin{center}
\parbox[b]{6cm}{\epsfig{figure=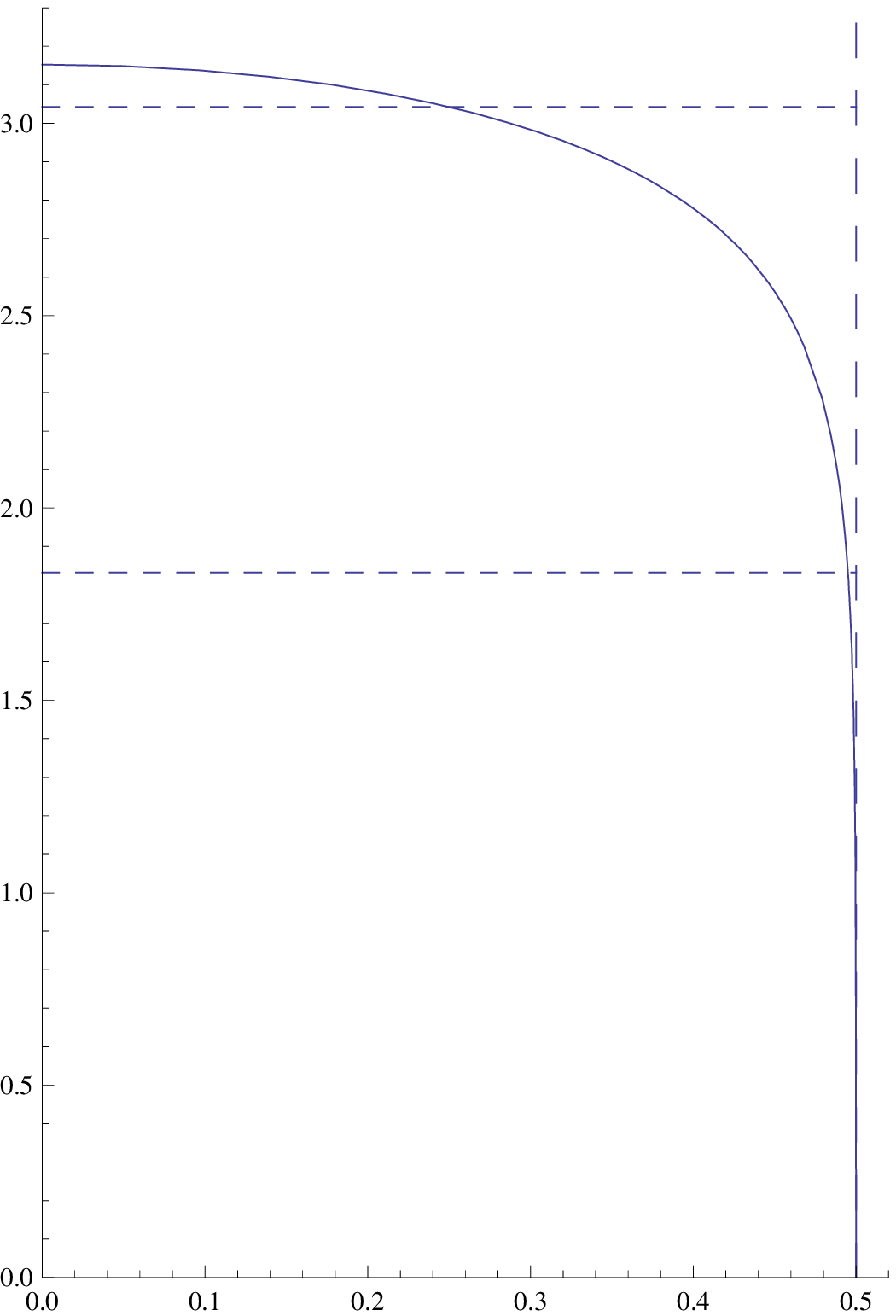, height=9cm} }
\hskip 3cm
\parbox[b]{6cm}{\epsfig{figure=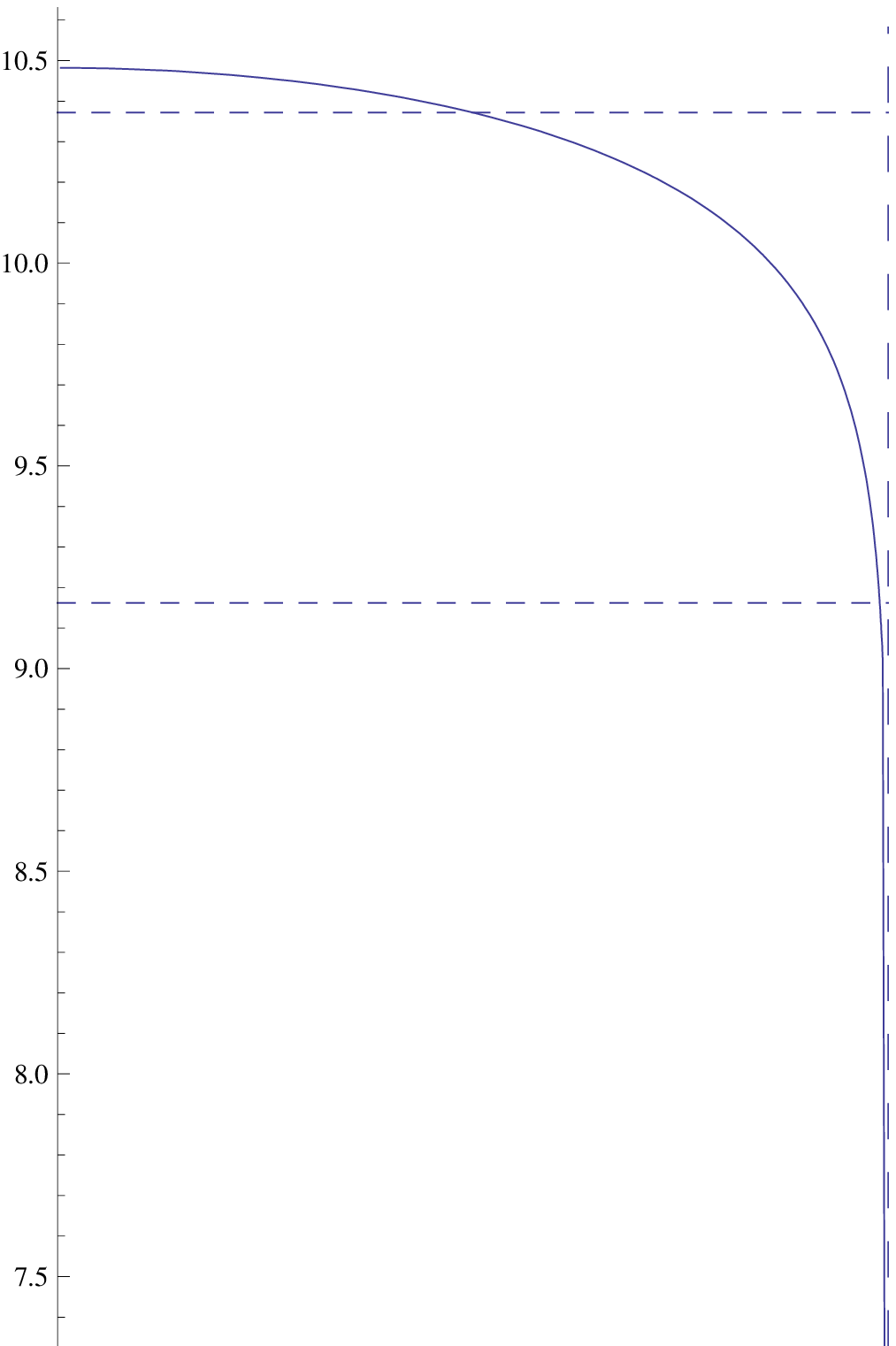, height=8.75cm}\vskip.25cm}
\end{center}
\caption{Left: The coordinate curve $f^\lambda(e^{i\theta})$ for
$\theta \in [0, \pi/2]$ and $\lambda = 10^{-4}$. The intersection
with the bottom dashed line indicates that by the time
the curve has dropped about 1.32 from the top, it is within
1\% of the vertical line $\Re (z) = 1/2$
that defines the sliver frame. Right:  the same portion of the
coordinate curve for $\lambda = 10^{-14}$.  The top part of the
coordinate curve is quite accurately the same as the one shown
to the left, but is displaced upwards.
}
\label{fregfig}
\end{figure}

\medskip
The behavior of the coordinate function $f^\lambda (\xi)$ for
very small $\lambda$
(near Schnabl gauge) will be of
interest. We focus on the coordinate curve
$f^\lambda(\xi = e^{i\theta})$ with $0\leq \theta\leq \pi$.
It is convenient to use the angular variable $\hat \theta$  that
measures angles with respect to the imaginary axis
\begin{equation}
\hat \theta = {\pi\over 2} - \theta \,.
\end{equation}
The coordinate function (\ref{flambdadef}) admits a simple
expansion when both $\lambda$ and $\hat \theta$ are small,
regardless of their ratio.
One then
finds\footnote{We
 follow the convention that terms
of order $\lambda \ln \lambda$ are written as $\mathcal{O}(\lambda)$.}
\begin{equation}
\label{ilas99}
f^\lambda (e^{i\theta}) = - {i\over \pi}  \ln  \Bigl( {\lambda + i \hat \theta\over
2}\, \Bigr)
 \,+\,\ord{\lambda}\,+\,\ord{\hat \theta} \,.
\end{equation}
We define $i \Lambda (\lambda)$ as the value of the
coordinate function at $\xi = i$:
\begin{equation}
\label{iwltfsksss}
i\Lambda \equiv  f^\lambda (i) =  - {i\over \pi} \ln {\lambda\over 2}
 \,+\,\ord{\lambda}   \,.
\end{equation}
Happily,
the regularized curve $f^\lambda (e^{i\theta})$ only differs
appreciably
 from the  sliver
curve $f(e^{i\theta})$
for $\hat\theta=\ord\lambda$. For $\lambda\ll 1$, the part of the curve
$f^\lambda (e^{i\theta})$ which deviates significantly
from $f(e^{i\theta})$ is thus entirely captured by (\ref{ilas99}).
We can write  the leading dependence as
\begin{equation}
\label{ilas999}
f^\lambda (e^{i\theta}) ~\simeq~ i\Lambda (\lambda)  - {i\over 2\pi}  \ln
\Bigl[ 1+ \Bigl( {\hat \theta \over \lambda} \Bigr)^2\,\Bigr]  + {1\over \pi}
\tan^{-1} \Bigl( {\hat \theta \over \lambda} \Bigr)   \,.
\end{equation}
The nature of the curve $f^\lambda (e^{i\theta})$
is quite interesting.  As illustrated in Figure~\ref{fregfig},
for any  $\lambda\ll 1$ the
coordinate curve near the top takes the {\em same} shape.
This is so because,
apart from the $i\Lambda (\lambda)$ term that sets
the height,
the rest of $f^\lambda$ depends only on the
ratio $\hat \theta/\lambda$, which spans the same values as
$\hat \theta$ grows from zero to some multiple of $\lambda$.
For $\hat \theta = \lambda$
the coordinate curve has come down about $0.11$ from the top
and
is 50\% of the way to the maximum real value of $1/2$ (top dashed
lines).
For $\hat \theta = 64\lambda$
the coordinate curve has come down about $1.32$ from the top and
is 99\% of the way to the maximum real value
(lower dashed lines).
Clearly, for sufficiently small $\lambda$, the coordinate curve
deviates from the vertical lines that define the sliver frame only
for $\hat \theta\ll 1$.

The curves $\gamma^\lambda_R$ and $\gamma^\lambda_L$ which parameterize the coordinate curve
$f^\lambda(e^{i\theta})$ for $\lambda$-regulated gauges will play
an important role in our analysis. They are defined by
\begin{equation}\label{gammalLR}
\half+\gamma^\lambda_R(\theta)\equiv f^\lambda\bigl(e^{i\theta}\bigr)\,, \qquad
-\half+\gamma^\lambda_L(\theta)\equiv f^\lambda\bigl(e^{i(\pi-\theta)}\bigr)\,,
\end{equation}
a particular example of the general definitions~(\ref{gammaR}) and~(\ref{gammaL}).
In the Schnabl limit $\lambda\to0$, $\gamma_R^\lambda$ and $\gamma_L^\lambda$ coincide, and we therefore define
\begin{equation}\label{gamma0}
     \gamma(\theta)\equiv \lim_{\lambda\to0}\gamma_R^\lambda(\theta) =\lim_{\lambda\to0}\gamma_L^\lambda(\theta)
     =i\,{2\over \pi} \tanh^{-1}  \Bigl( \tan {\theta\over 2} \Bigr)\,.
\end{equation}
As expected, this is the parameterized vertical line that defines the
left and right parts $\half - \gamma$ and $\half +\gamma$ of the
coordinate curve
of the sliver projector.
Notice, however,  that the limit~(\ref{gamma0}) is not uniform in $\theta$. In fact, for all $\lambda>0$ we have
\begin{equation}
 \lim_{\theta\to\frac{\pi}{2}}\Re\bigl(\gamma_{L/R}^\lambda(\theta)\bigr)=\pm\frac{1}{2}\,,
\end{equation}
while $\Re(\gamma(\theta))=0$, independent of $\theta$.

We now ask how much the coordinate curve of $\lambda$-regulated gauges still
deviates from the vertical line that defines the sliver
by the time its imaginary part has been reduced to
$\Lambda/2$, that is, half the value it has at the top.
To leading order in $\lambda$, the angle corresponding to this point on the curve is given by
\begin{equation}
    \hat \theta_{\frac{1}{2}}=  \sqrt{2\lambda} \quad \to \quad
    \theta_{\frac{1}{2}}=\textstyle{\frac{\pi}{2}}- \sqrt{2\lambda}\,.
\end{equation}
A short calculation then shows
\begin{equation}\label{thatahalf}
    \gamma^\lambda_R\bigl(\theta_{\frac{1}{2}}\bigr)=-{1\over 2\pi}\sqrt{2\lambda} \,+\, i \, {\Lambda \over 2}\,+\, \ord{\lambda} \, .
\end{equation}
As we can see, $\gamma^\lambda_R(\theta)$ only deviates by $\mathcal{O} (\sqrt{\lambda})$ from the imaginary axis by the time its height has dropped by half.

\subsection{The annulus and its modulus}\label{secannmod}
The surfaces associated with the one-loop vacuum graph
are obtained by gluing the two parameterized edges of the
 propagator to itself.
 The propagator associated with regular linear $b$-gauges is in general a  complicated object. Its geometric interpretation depends on the
 ghost number of the state it acts on. In alternating
 gauge~\cite{Kiermaier:2007jg} the surface of the propagator is built by gluing
 the strips associated with $e^{-sL[v]}$ and $e^{-s^\star L^\star[v]}$ in
 some order (that depends on ghost number)
 and  by
 including the action of the BRST operator $Q$, that acts
 as a total derivative on moduli.
 The details of this construction will be important for our general analysis in
 Section~\ref{secloop}. For now, we focus on one term that arises from the
 propagator:  it can be described by setting $s^\star=0$
 and gives
 the strip $\RR(s)$ associated with $e^{-sL[v]}$.
 The generalization to the full propagator will not introduce further conceptual
 problems in our Riemann surface analysis. We restrict ourselves to the simplified propagator in the discussion of
 the vacuum and the tadpole diagrams
 because it suffices
 to demonstrate the main features of loop diagrams in Schnabl gauge.

\begin{figure}
\centerline{
  \epsfig{figure=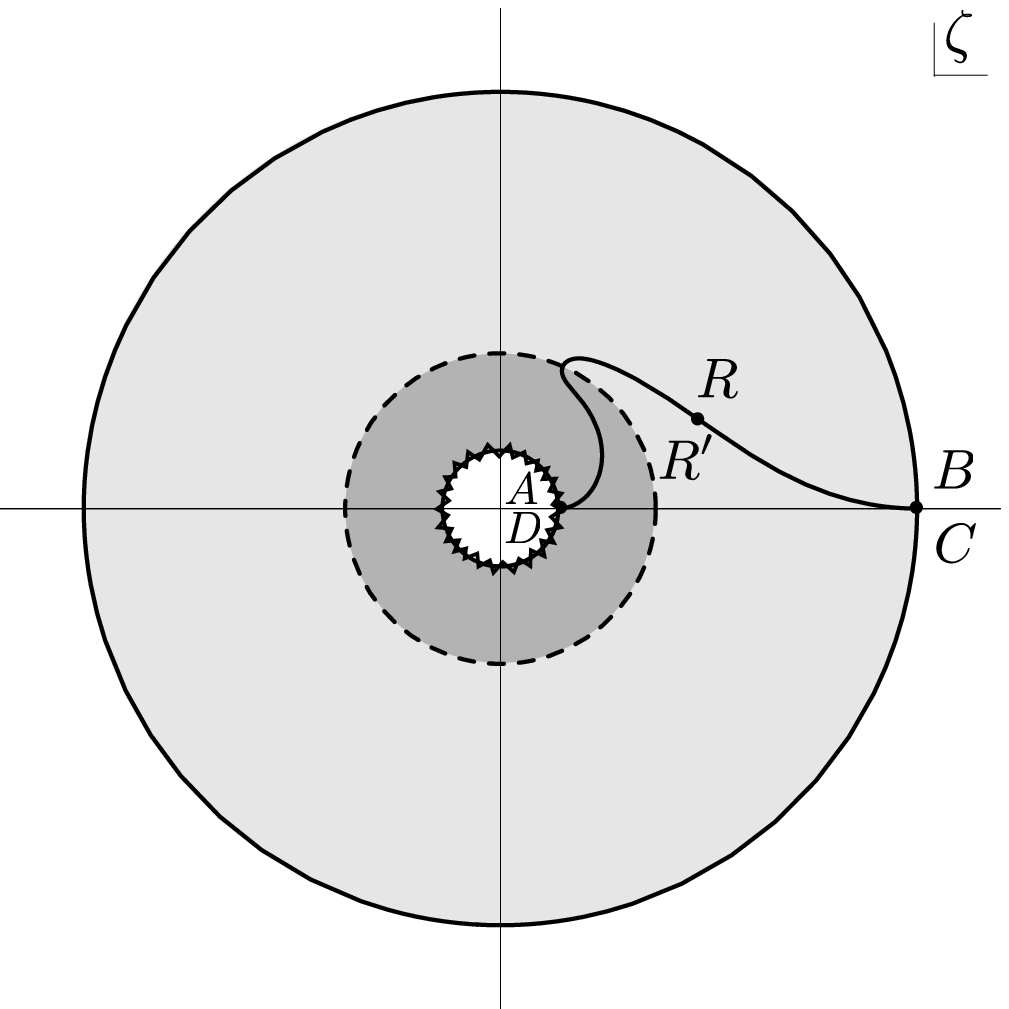, height=8cm}
  \hskip 2.75cm
 \epsfig{figure=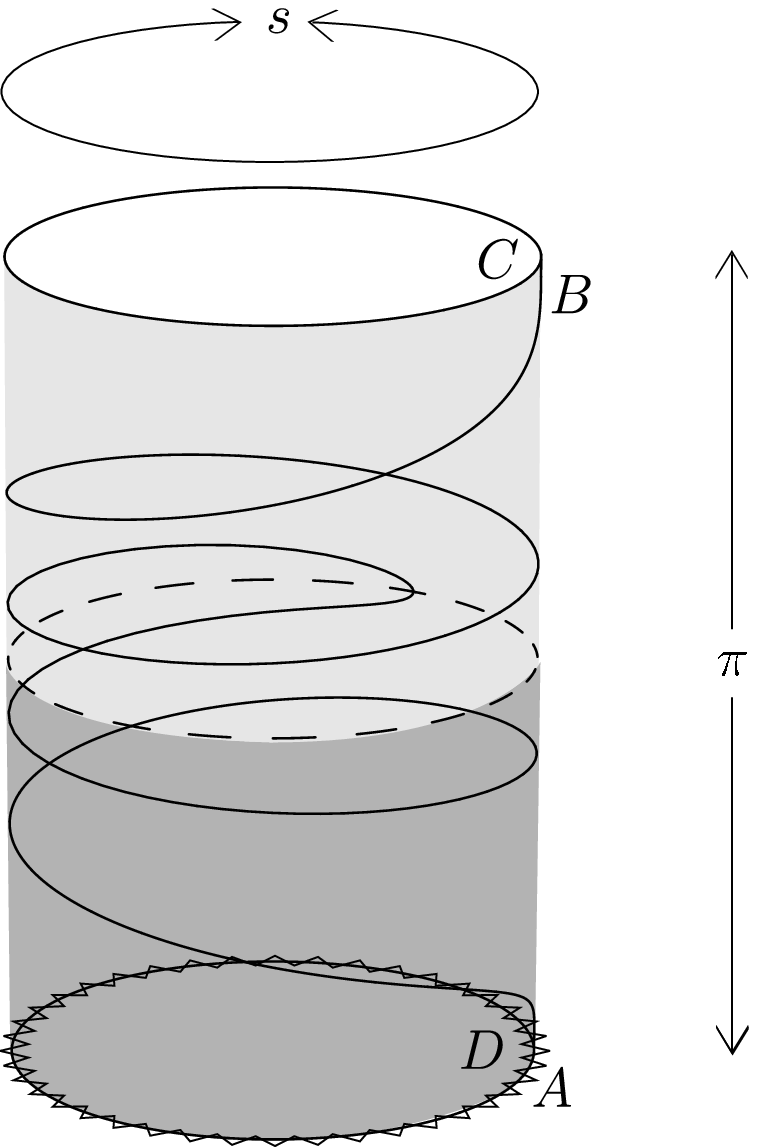, height=8cm}
}
 \centerline{ \hskip 3.25cm  (a) \hskip 8cm (b) \hskip2.75cm}
\caption{
The vacuum graph obtained from gluing the edges of $\RR(s)$, illustrated
for $L[v]=L^\lambda$ with $\lambda=10^{-4}$. In (a) the surface
is displayed as a canonical annulus in the $\zeta$ frame for $s=10$. The cutting curve
is shown explicitly. In (b) the surface is
displayed as a cylinder obtained from the identification
 $w\sim w-s$
in the $w$ frame for $s=1$. This should be compared to Figure~\ref{sl01fig}(a).
}
\label{sl01bfig}
\end{figure}

For any regular linear $b$-gauge,
 the gluing of the simplified propagator  $\RR(s)$
  to itself  is implemented in the $w$ frame
by the identification $w \sim w-s$.
The result, for each value of $s$, is an annulus. In this annulus
the boundaries are the horizontal segments  $BC$ and $AD$,
shown in Figure~\ref{sl01fig}(a)
 for $\lambda$-regularized gauges.
The map from
this annulus to a canonically presented annulus in the $\zeta$
frame is
\begin{equation}
\label{iwltktcntaj}
\zeta = \exp \Bigl( -{2\pi i \over s} (w-i\pi)\Bigr)  =
 \exp \Bigl( -{2\pi^2  \over s} \Bigr) \exp \Bigl( -{2\pi i w\over s} \Bigr) \,.
\end{equation}
See Figure~\ref{sl01bfig}(a).
We can also write, using (\ref{wzrel}),
\begin{equation}
\label{iwltktcntaj99}
\zeta = \exp \Bigl( {2\pi i \over s} \ln 2z \Bigr)   \,.
\end{equation}
The map (\ref{iwltktcntaj}) takes $BC$ into the unit circle $|\zeta| = 1$ and
$AD$ into the inner circle $|\zeta| = \exp(-2\pi^2/s)$.  Since the
strip $\mathcal{R}(s)$ is foliated
 in the $w$ frame
by horizontal lines of length $s$
at heights that go from zero to $\pi$ it is clear that the map
(\ref{iwltktcntaj}) takes the interior of the strip to the region between
the two $\zeta$ circles mentioned above.  The shape of the edges
of $\mathcal{R}(s)$ is irrelevant to the map;
their image under the map is  a cutting curve for the annulus.
Shown to the right in
Figure~\ref{sl01bfig}(b) is the $w$-frame picture
of $\mathcal{R}(s)$ rolled up into a cylinder of height $\pi$ and
circumference $s$. The cutting curve is shown in both presentations.

\begin{figure}[t]
\centerline{
  \epsfig{figure=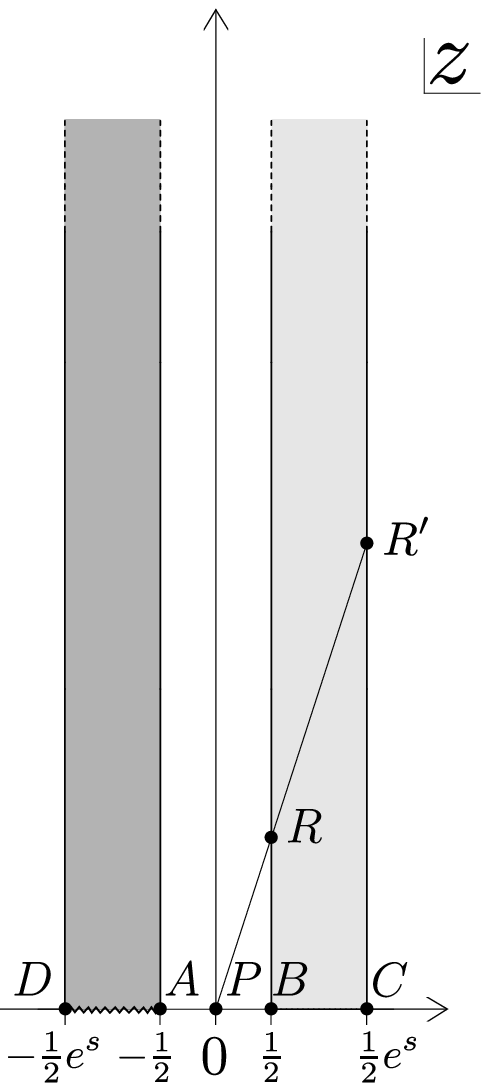, height=9cm}
  \hskip 2cm
 \epsfig{figure=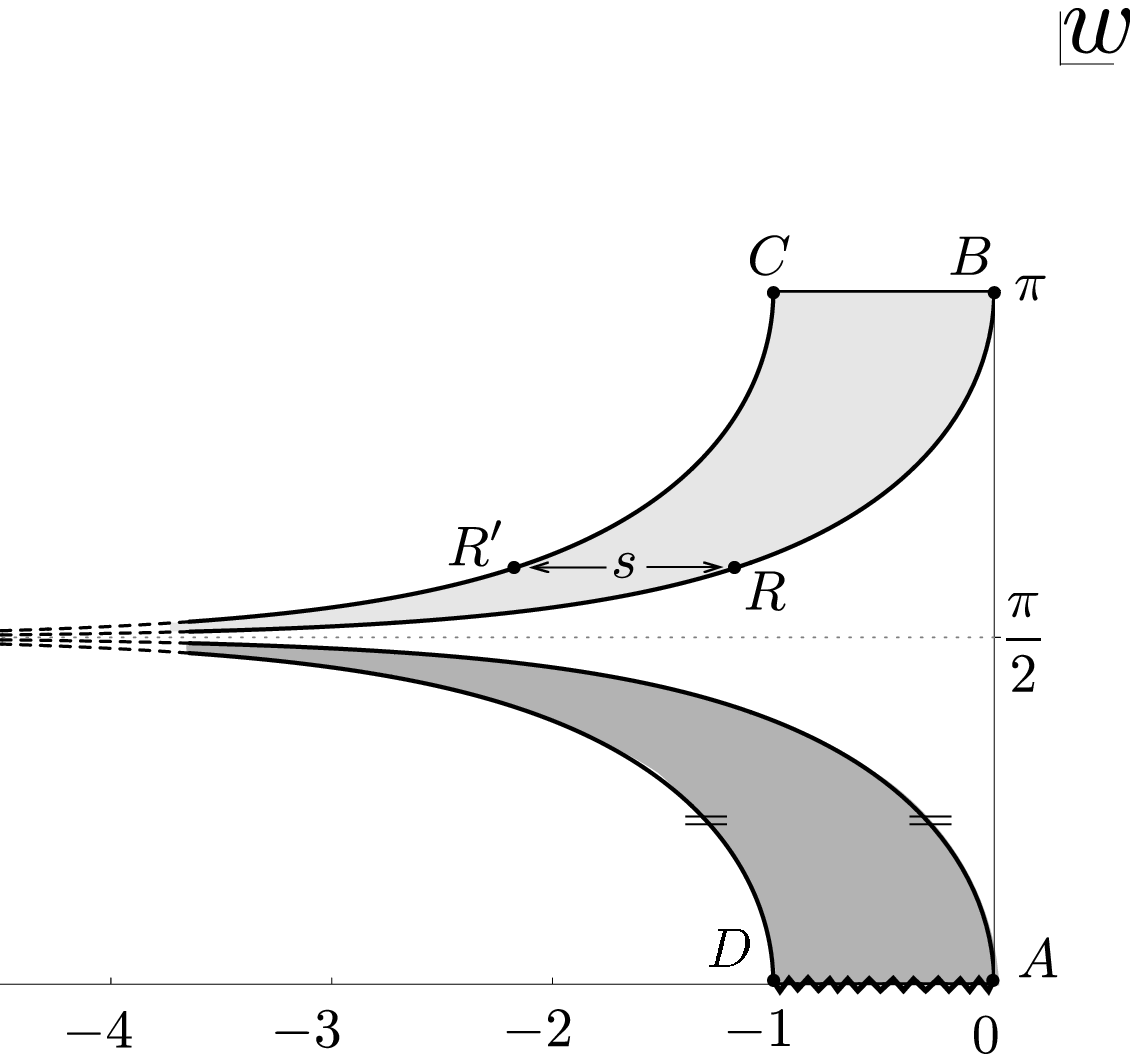, height=8.8cm}}
 \centerline{ \hskip 2.3cm  (a) \hskip 8.7cm (b) \hskip5cm}
\caption{The $\mathcal{R}(s)$ strip in the Schnabl limit
 $\lambda =0$ both in the $w$ and in the $z$ frames, displayed for $s=1$.
 The gluing of the free edges of the strip gives rise to an
 annulus of finite modulus (see Figure~\ref{sl02bfig}).  The gluing identification in the
 $z$ frame is that induced by radial lines emerging from
 the origin.}
\label{sl02fig}
\end{figure}

\begin{figure}[t]
\centerline{
  \epsfig{figure=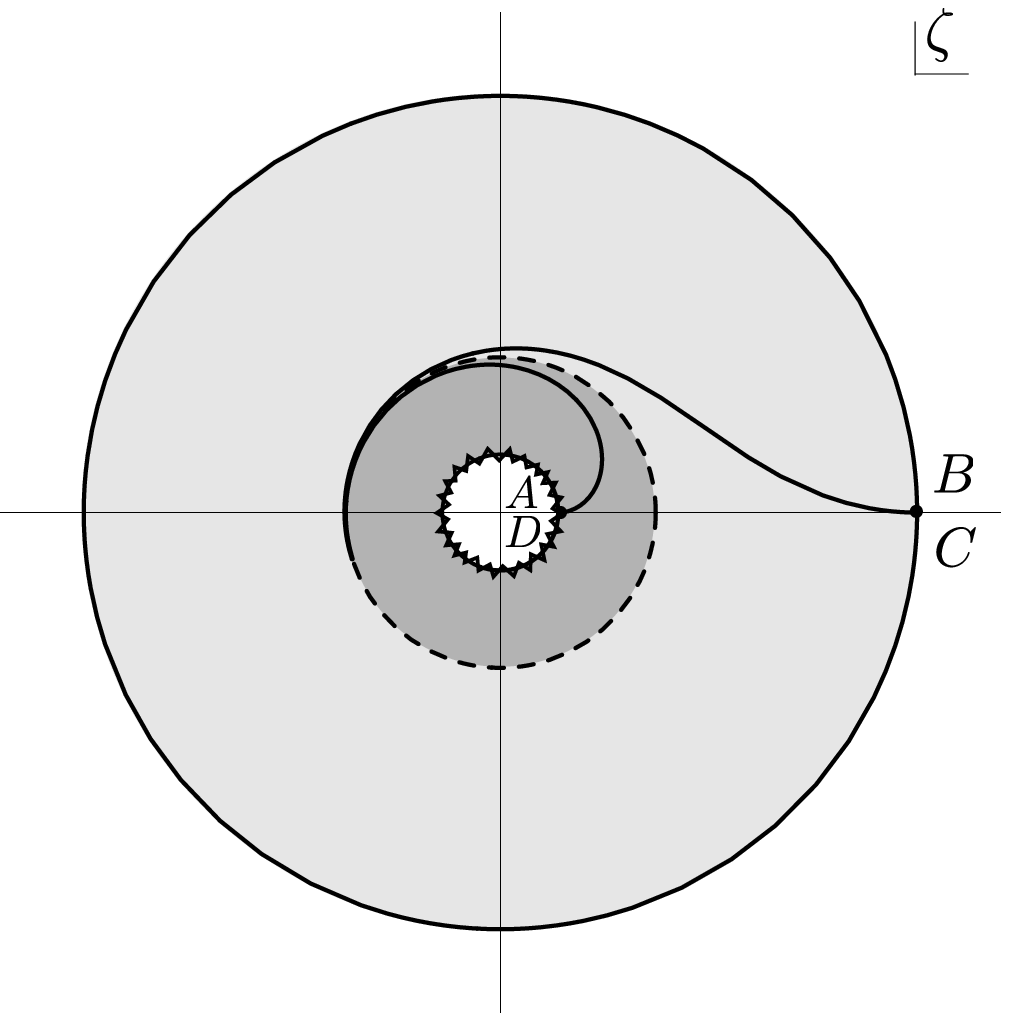, height=8cm}
  \hskip 2.75cm
 \epsfig{figure=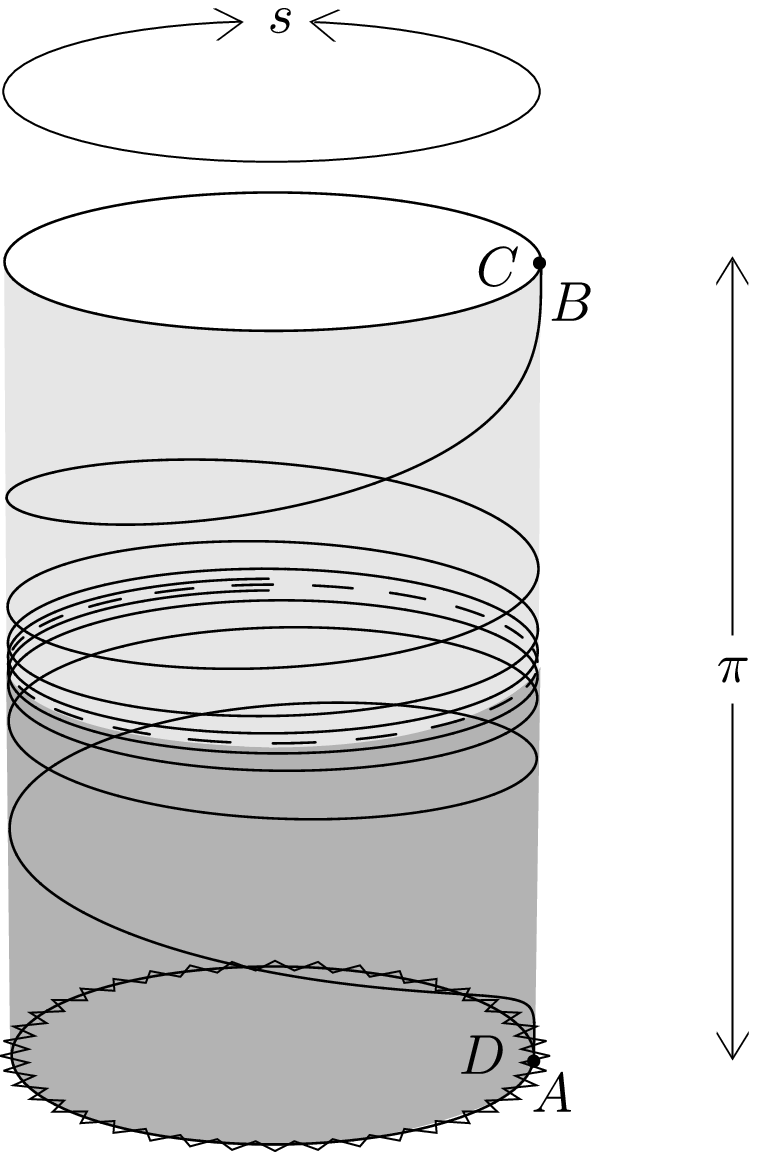, height=8cm}}
 \centerline{ \hskip 3.25cm  (a) \hskip 8cm (b) \hskip2.75cm}
\caption{
The vacuum graph obtained from gluing the edges of $\RR(s)$ in the Schnabl limit
$\lambda\to 0$. In (a) the surface
is displayed as a canonical annulus in the $\zeta$ frame for $s=10$. In (b) the surface is displayed as a cylinder obtained from the identification $w\sim w-s$ in the
$w$ frame for $s=1$. These surfaces differ from the corresponding
finite-$\lambda$
 surfaces in Figure~\ref{sl01bfig} only through the
shape of the cutting curve.  }
\label{sl02bfig}
\end{figure}

The modulus $M$ of an annulus with radii
$r_{\rm{in}}$ and $r_{\rm{out}}$  with $r_{\rm{in}}<r_{\rm{out}}$
is usually defined by
\begin{equation}\label{defM}
M \equiv {1\over 2\pi } \ln {r_{\rm{out}}\over r_{\rm{in}}} \,.
\end{equation}
The moduli space of annuli is the set
\begin{equation}
\label{modspaann}
0\leq M \leq \infty\,.
\end{equation}
For our annulus the modulus is
\begin{equation}
\label{modourann}
M =  {\pi\over s} \,.
\end{equation}
 This result for the annulus modulus is valid for any regular linear $b$-gauge.
 In particular,
the modulus $M$ of the annulus produced by
the gluing of the edges of $\mathcal{R}(s)$ is the same
for all values of $\lambda$ in the $\lambda$-regularized gauges
and depends
only on $s$.
As $s\to 0$, $M\to \infty$, the inner circle goes to zero size, and we
approach closed string degeneration.  As $s\to \infty$ the inner circle
approaches the outer circle, $M$ goes to zero, and we approach
open string degeneration.   The full moduli space (\ref{modspaann}) is therefore covered.
It thus follows that in the Schnabl limit $\lambda \to 0$ the gluing of
$\mathcal{R}(s)$ also gives an annulus
of $M = \pi/s$  and that moduli space is covered in this case as well.
The limit $\lambda \to 0$
of  Figure~\ref{sl01fig} is shown in Figure~\ref{sl02fig}.
Moreover, Figure~\ref{sl02bfig} shows the map to the $\zeta$ plane
and the cylinder view of the $w$-presentation.
Note that we could have calculated the annulus modulus in Schnabl gauge
 using any other family of regular linear $b$-gauges which approaches
 Schnabl gauge when the regulator is removed. The result for  $M$ would
 have been the same.
 \medskip

A few remarks
 about this construction in the Schnabl limit
are in order.
\begin{itemize}
\item
The map (\ref{iwltktcntaj})
still takes the shaded domain in the $w$ plane to the circular annulus
because the identification $w\sim w-s$ still holds.  The cutting curve is infinitely long (Figure~\ref{sl02bfig}).

\item  In the $z$ plane the vertical strip
to the right produces the
upper half of the annulus
(the upper half of
the vertical cylinder of height $\pi$ and circumference $s$).
The vertical strip to the left produces the lower half of the annulus.
The two halves are glued.

\item  The identifications $w\sim w-s$ become {\em slanted}
identifications
$z \sim e^s z$ of the vertical lines through $B$  and $C$,  and
of the vertical lines through $A$ and $D$. If the identifications
had been horizontal
 ($z \sim z - \half+\half e^s$) 
both the right and left
strips would have each given rise to a (closed string) degenerate
 annulus.
 In fact, such a problematic horizontal identification happens for the
 gauge condition $B^+ \Phi=0$ in the sliver frame.
It is the  slanted identification that makes the $z$ frame picture
 in Schnabl gauge
consistent with a finite modulus annulus.

\end{itemize}

 In the previous section we remarked that the propagator strip $\RR(s)$ for
 regular linear $b$-gauges can be decomposed into two components associated with
 $e^{-sL[v]_L}$ and $e^{-sL[v]_R}$, respectively. These components are glued along the
 boundary $QQ'$ in Figure~\ref{sl01fig}(b).
 $L[v]_L$ and $L[v]_R$  generate this unmatched boundary that needs to be glued by hand
 because they do not commute.
 The operators $L$ and $L^\star$ in Schnabl gauge  can also be decomposed
 into  left and right parts.  We write $L = L_L + L_R$,  $L^\star = L^\star_L
 +L^\star_R$.
 In the Schnabl limit, the unmatched boundary is hidden at $i\infty$
 in the $z$- frame, but arises in the annulus frame $\zeta$ as the circle
 $|\zeta|=\exp(-\pi^2/s)$,  shown dashed
 in Figure~\ref{sl02bfig}(a).
 We are led to conclude that
  while both $L$ and $L^\star$ arise from vector fields
 that vanish at the open string midpoint, they do not vanish fast enough
 to ensure that
 $L_L$ and $L_R$ commute and that  $L^\star_L$ and $L^\star_R$
 commute:\footnote{In fact
   the linear combination
  $L^+ = L+L^*$ arises from a vector that,
  as we approach the midpoint, vanishes sufficiently fast to ensure
  that
  $L^+_L$
  and $L^+_R$ commute.}
\begin{equation}
\label{o9rhd}
  \boxed{\phantom{\Bigl(}[L_L,L_R]\neq0\,,  \qquad [L^\star_L,L^\star_R]\neq0\,. ~~ }
\end{equation}

We conclude this subsection by recalling the relation
 of the modulus  $M$ with the
conformal invariant known as the extremal length~\cite{ahlfors}.
The extremal length is an invariant associated to a
 given set  of curves $\Gamma$ on a Riemann surface.
Let $\rho$ denote a conformal metric
(a metric for which  $ds=\rho(z,\bar z)|dz|$) on the
Riemann surface.
The length $\ell(\gamma, \rho)$ of a curve
$\gamma\in \Gamma$ and the area $A(\Omega, \rho)$ of the
Riemann surface $\Omega$
are given by:
\be
\ell(\gamma, \rho) = \int_\gamma \rho |dz|\,, \quad
A(\Omega, \rho) = \int\hskip-7pt\int_\Omega \rho^2 \, dx dy\,.
\ee
We define $\ell(\Gamma, \rho)$ as the length of the shortest
curve in $\Gamma$
 with respect to the metric $\rho$:
\be
\ell(\Gamma, \rho) = \inf_{\gamma\in \Gamma}  \ell(\gamma, \rho)\,.
\ee
The extremal length
$\lambda_\Gamma$  is defined
as~\cite{ahlfors}
\be
\label{def_extremal_length}
 \lambda_\Gamma
=  \sup_\rho  \Biggl( {\ell^2(\Gamma, \rho)
\over A(\Omega, \rho)} \Biggr) \,.
\ee
To evaluate
$\lambda_\Gamma$
one must  search over metrics
until the quantity inside parenthesis on the right-hand side
is maximized.  The extremal metric $\rho$ for which the maximum
is attained is a minimal area metric:  it is the metric with least area
consistent with all curves in the set having a length greater than
or equal to a certain prescribed value.
{}From the definition \refb{def_extremal_length} it is clear that
the extremal length
 $\lambda_\Gamma$
is a conformal invariant.

\smallskip
Let us now return to
the  vacuum graph of regular linear $b$-gauges.
Imagine the domain $\mathcal{R}(s)$,
 glued to itself to form the vacuum graph,
as a cylinder of circumference $s$ and height $\pi$. This is, in fact,
the $w$ frame picture in Figure~\ref{sl01bfig}(b).
There are two types of curves on this cylinder (or annulus):
open curves that stretch from one boundary to the other and
closed curves that go around the cylinder. We thus have an extremal
length $\lambda_{\rm{open}}$ associated with the set of open curves
and an extremal length $\lambda_{\rm{closed}}$ associated with
the set of closed curves. It is a familiar result
that in the $w$ frame the {\em same} metric
$\rho=1$ is extremal for {\em both} open and closed curves~\cite{gardiner}.
It is clear that in this flat metric the shortest open curves have length $\pi$ and the
shortest closed
curves have length $s$.  The area, moreover, is $\pi s$.  It follows that
the extremal lengths are
\begin{equation}
\lambda_{\rm{open}} =  {\pi^2 \over \pi s} =  {\pi\over s}\,,
 \qquad
\lambda_{\rm{closed}} =  {s^2\over \pi s} =  {s\over \pi}\,.
\end{equation}
It is interesting to note that
\begin{equation}\label{openclosed}
\lambda_{\rm{open}} \lambda_{\rm{closed}} = 1 \,, \quad
\hbox{and} \quad  M = \lambda_{\rm{open}}
= {1\over \lambda_{\rm{closed}} }\,.
\end{equation}
 The relations~(\ref{openclosed}) are general and
valid for any annulus.
Note that degeneration of a given type means
vanishing extremal length for the curves of associated type.
Thus closed string degeneration ($s\to 0$) happens for $\lambda_{\rm{closed}} \to 0$ and open string degeneration ($s\to \infty$)
happens for $\lambda_{\rm{open}} \to 0$.

\section{One-loop tadpole graph}\label{sectp}
\setcounter{equation}{0}

In this section we discuss the one-loop tadpole graph.
The underlying Riemann surface is an annulus with
an open string puncture, that is, a puncture on one
of the boundary components of the annulus.
The puncture, which represents the external state,
 introduces significant complications in the geometry.
 Indeed, it is well known that in Siegel gauge the map
of the string diagram to the round annulus is nontrivial
and the modulus of the annulus cannot be calculated
in simple closed form.

As in the previous section
 we restrict ourselves to the contribution from the propagator
 surface $\RR(s)$ generated by $e^{-sL[v]}$.
We discuss the graph
for the family of interpolating gauges.  We first show that
for any value of the regulator $\lambda$
the moduli space of annuli is generated when the
Schwinger parameter
 $s$ covers the range from zero to infinity.
We then study
the geometry as the regulator parameter
 $\lambda$
goes to zero and we approach Schnabl gauge.
 We present a construction which allows us
 to  exactly map the  tadpole string diagram
 to the round annulus in the limit $\lambda\to0$.
The modulus of the annulus becomes exactly calculable
in Schnabl gauge.

\subsection{Covering moduli space in the $\lambda$-regulated gauges}
\label{comospitherega}

\begin{figure}[t]
\begin{center}
  \parbox{3.5cm}{\epsfig{figure=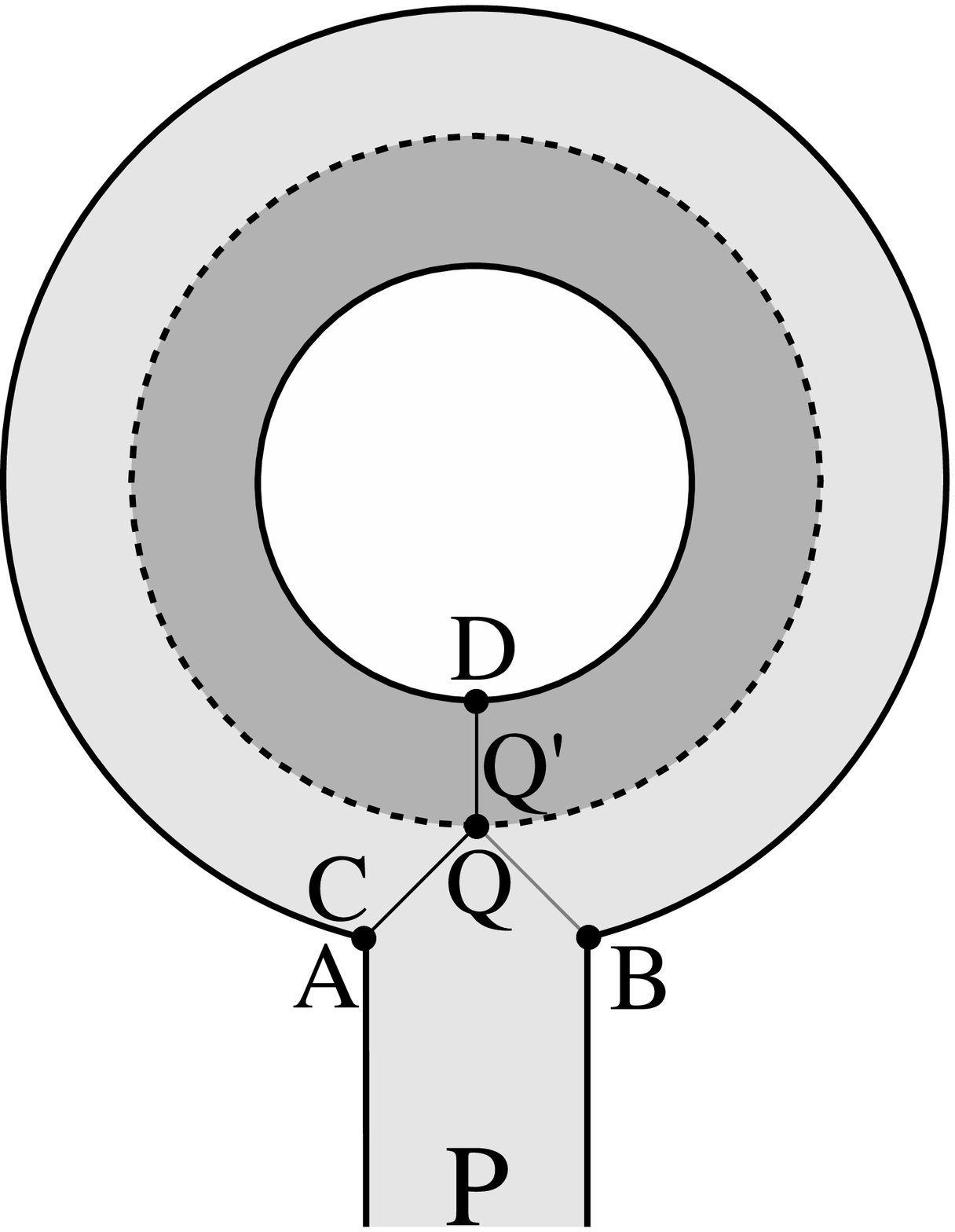, height=5.5cm}}
  \hskip 1.2cm
  \parbox{3.5cm}{\epsfig{figure=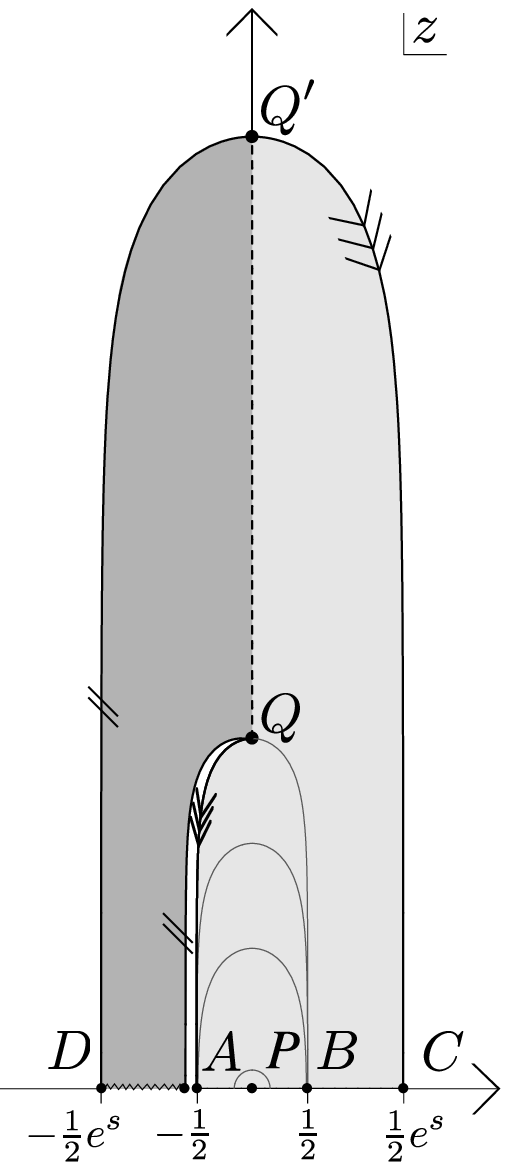, height=8cm}}
  \hskip .8cm
  \parbox{7cm}{\epsfig{figure=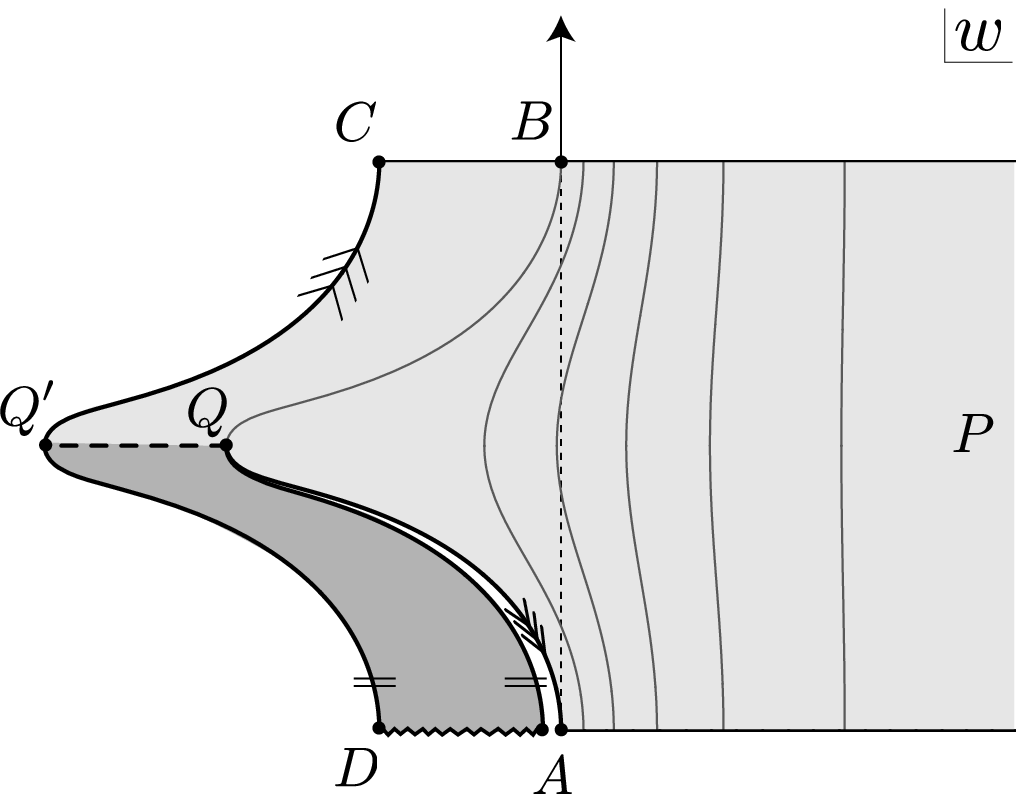, height=5cm}}
\end{center}
 \centerline{ \hskip 2.5cm  (a) \hskip 3.8cm (b) \hskip6.2cm (c)\hskip 3.5cm}
\caption{ (a) The topology of the one-loop tadpole diagram obtained by gluing the external
strip for the Fock space state to the propagator strip.  (b)  The tadpole diagram for a $\lambda$-regulated gauge in
the $z$ frame, displayed for $\lambda=10^{-4}$ and $s=1$.  Note
the cut from $Q$ to $A$ that separates the two boundary components.
 (c) The tadpole diagram in the $w$ frame.}
\label{sl03fig}
\end{figure}

Let us consider the one-loop tadpole graph
 with  propagator $e^{-sL[v]}$.
 It is useful to first examine the surface obtained
 in the $\lambda$-regulated gauges.
The way to assemble the surface
 is illustrated using Figure~\ref{sl03fig}.  We need the
part of the surface associated with the external state and
the propagator strip $\mathcal{R}(s)$.

As we can see in Figure~\ref{sl03fig}(b),
the placement of $\RR(s)$ in the $z$ frame is the same one
used for the vacuum graph in the
last section~(Figure~\ref{sl01fig}(b)).
 As discussed above equation (\ref{o9rhd}),
it is convenient to view
the surface $\RR(s)$ as built by gluing together
 two pieces -- one associated with
 $e^{-sL^\lambda_R}$ and one associated with $e^{-sL^\lambda_L}$.
 These two pieces are glued along the dashed line $QQ'$ to form the complete
 surface $\RR(s)$.

 The two curved boundaries of $e^{-sL^\lambda_L}$
 are identified, just as for the vacuum graph. This time, however,
 the two curved boundaries of
 $e^{-sL^\lambda_R}$
 are not glued to each other.
 To form the tadpole, we need to glue
 these two boundaries to the left and right boundaries of the external state.
 As the functions $f^\lambda(\xi)$ are coordinate functions and
 thus well defined for all $|\xi|\leq1$, we can conveniently place
 this external state in the region between the real axis and the
 coordinate curve $f^\lambda(e^{i\theta})$. The operator insertion is
 then located at $z=f(0)=0$ (see Figure~\ref{sl03fig}(b)).

The gluing patterns
both in the $z$ and $w$ frames are readily obtained from
the graph in Figure~\ref{sl03fig}(a).
The only slightly nontrivial gluing operation
 is that identifying the  curves $AQ$ and $CQ'$
in the $z$ plane (the lines with triple arrows).
 We can express these two curves using  $\gamma_{L/R}^\lambda$
 defined in~(\ref{gammalLR}):
 \begin{equation}\label{AQCQ}
       AQ=-\half +\gamma^\lambda_L\,, \qquad
       CQ'=e^s(\half +\gamma^\lambda_R)\,.
 \end{equation}
 It then follows that the identification between $AQ$ and $CQ'$ is given
 by the map
 \begin{equation}
      z=-\half +\gamma^\lambda_L(\theta) ~\to~ z'=e^s\bigl(\half +\gamma^\lambda_R(\theta)\bigr)\,.
 \end{equation}
 Recalling $\gamma_L(\theta) =-\overline{\gamma_R(\theta)}$, we find that
a point $z\in AQ$
is identified with the point $z'\in CQ'$, where $z'$ is obtained
by first reflecting $z$ across the vertical axis $z\to - {\bar z}$, and then applying
 the expansion factor $e^s$:
\begin{equation}
\label{gindents}
z \to z ' = - e^s \,\bar{z} \,.
\end{equation}
There should be no concern that $z'$ appears to be a non-analytic
function of $z$.  The above relation is not a {\em sewing} relation, but just
a relation valid on the curve (for example, the analytic relation
$\xi \xi' = -1$ becomes $\xi' = - \bar {\xi}$ on the unit circle).
The analytic gluing relation is determined by the sequence of conformal
maps $z\to f^{-1}(z) $ back to the coordinate circle,  $\xi \to -1/\xi$,
followed by the action of $f$ and, finally, multiplication by $e^s$.
 The analytic gluing relation corresponding to the identification~(\ref{gindents}) is thus
 \begin{equation}
     z\sim e^sf\Bigl(-{1\over f^{-1}(z)}\Bigr)\,.
 \end{equation}

 \medskip
Since there is no simple closed
form expression for the modulus $M(s)$ of the annulus
in Siegel gauge,  we cannot hope to calculate
explicitly $M(s)$ for arbitrary finite $\lambda$.  Extremal length,
however, gives a very simple
proof that moduli space will be covered.   Consider
the $w$-frame picture in Figure~\ref{sl03fig}(c).
The extremal metric cannot be found, but let us use
the
metric
$\rho=1$ on the lower half of the
strip $\mathcal{R}(s)$ (below $Q'Q$)  and
 $\rho=0$
elsewhere.
In other words, we are setting $\rho=1$ only on the part
of the surface corresponding to $e^{-sL_L^\lambda}$
 (shaded in dark grey in the figure).
The area
of the surface in this metric is
 $A = {1\over 2} \pi s$.
In this metric the shortest open curves
have length $\frac{\pi}{2}$.  This gives the following
inequality
 for the open string extremal length
\begin{equation}
\label{vmitgfitw}
\lambda_{\rm{open}}
\geq {\bigl({\pi\over 2}\bigr)^2 \over {\pi s\over 2}}
= {\pi\over 2s} \,. \end{equation}
For closed curves we take $\rho=1$
all over the propagator strip $\mathcal{R}(s)$ {\em and}
over the portion of the external state strip that lies to the left of the
vertical line $AB$
 in  Figure~\ref{sl03fig}(c).
 In other words, we set $\rho=1$ in the region $\Re(w)<0$.
We  set $\rho=0$ elsewhere.
A little thought
shows that in this metric the shortest closed curve has length $s$.  The area is $\pi s + A(\lambda)$, where $A(\lambda)$ is the area of
the external state strip in the chosen metric. We thus get
\begin{equation}
\label{vmitgf}
\lambda_{\rm{closed}} \geq {s^2 \over \pi s + A(\lambda)}
 \quad \to \quad \lambda_{\rm{open}} \leq {\pi \over s} + {A(\lambda)\over s^2} \,.
\end{equation}
In the Siegel limit
$\lambda \to \infty$,
the vertical line $AB$
 in the $w$ frame
coincides with the
right boundary of $\mathcal{R}(s)$ so that the area $A(\infty) =0$.
It is easy to see that
the area $A(\lambda)$ grows as $\lambda$ decreases, but it stays finite even in the limit $\lambda\to0$. In fact, the relevant integral can be exactly calculated
and one finds that
\begin{equation}
A_0 \equiv \lim_{\lambda\to0}A(\lambda)= \pi  \ln 2\,.
\end{equation}
Back in (\ref{vmitgf}),
 we use $A(\lambda)\leq A_0$ and find
\begin{equation}
\label{vmitgf99}
 \lambda_{\rm{open}} \leq {\pi \over s} + {A_0\over s^2} =
 {\pi \over s} + {\pi \ln 2\over s^2} \,.
\end{equation}
Combining (\ref{vmitgfitw}) and (\ref{vmitgf99})  and recalling that
$M = \lambda_{\rm{open}}$ we get
\begin{equation}\label{inequ}
 {\pi\over 2s}\, \leq M(s) \,\leq {\pi\over s} \Bigl( 1 +  {\ln 2\over s}\Bigr) \,.
\end{equation}
The above inequalities imply that
$M(s)\to 0$ as $s\to \infty$ and $M(s) \to \infty$ as
$s\to 0$, so the full moduli space will be covered for $s\in [0, \infty)$.
 This is consistent with the results of~\cite{Kiermaier:2007jg} which showed that regular
 linear $b$-gauges, such as the $\lambda$-regulated gauges,
 give correct on-shell string amplitudes.
The
inequalities~(\ref{inequ}) hold for all
 $\lambda>0$.  We thus
 conclude that moduli space is covered in the Schnabl limit $\lambda\to0$.

\begin{figure}[t]
\begin{center}
  \parbox[b]{3cm}{
    \epsfig{figure=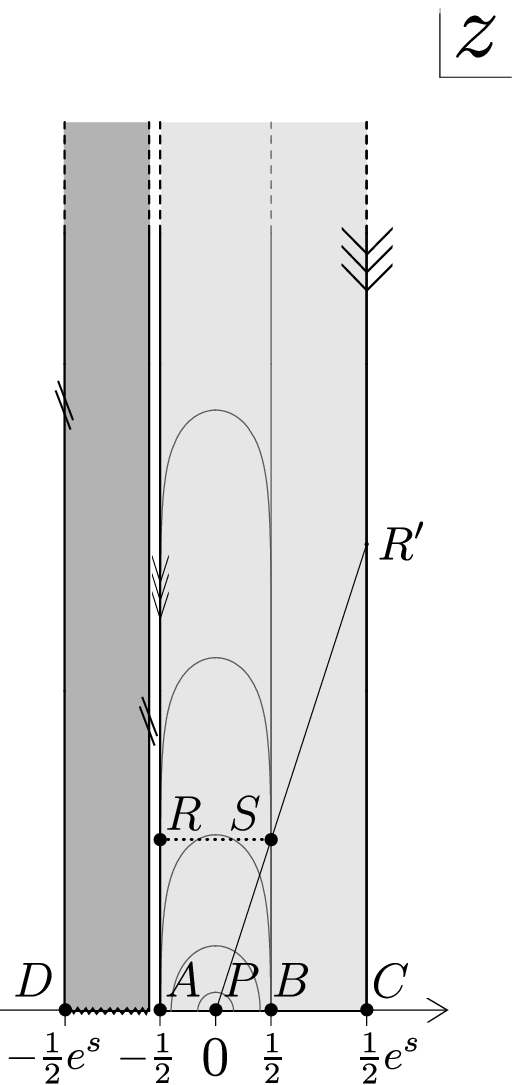, height=8.0cm}}
  \hskip 0.7cm
  \parbox[b]{5cm}{\epsfig{figure=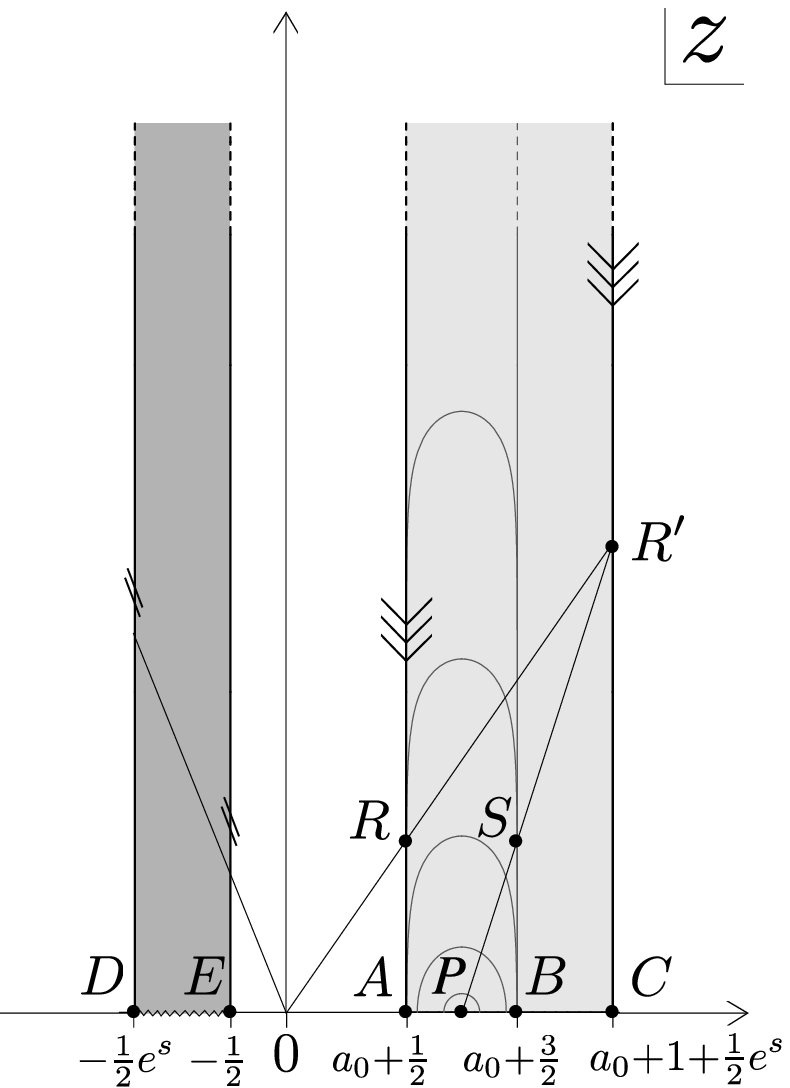, height=7.5cm}}
  \hskip 0.7cm
  \parbox[b]{6.5cm}{  \epsfig{figure=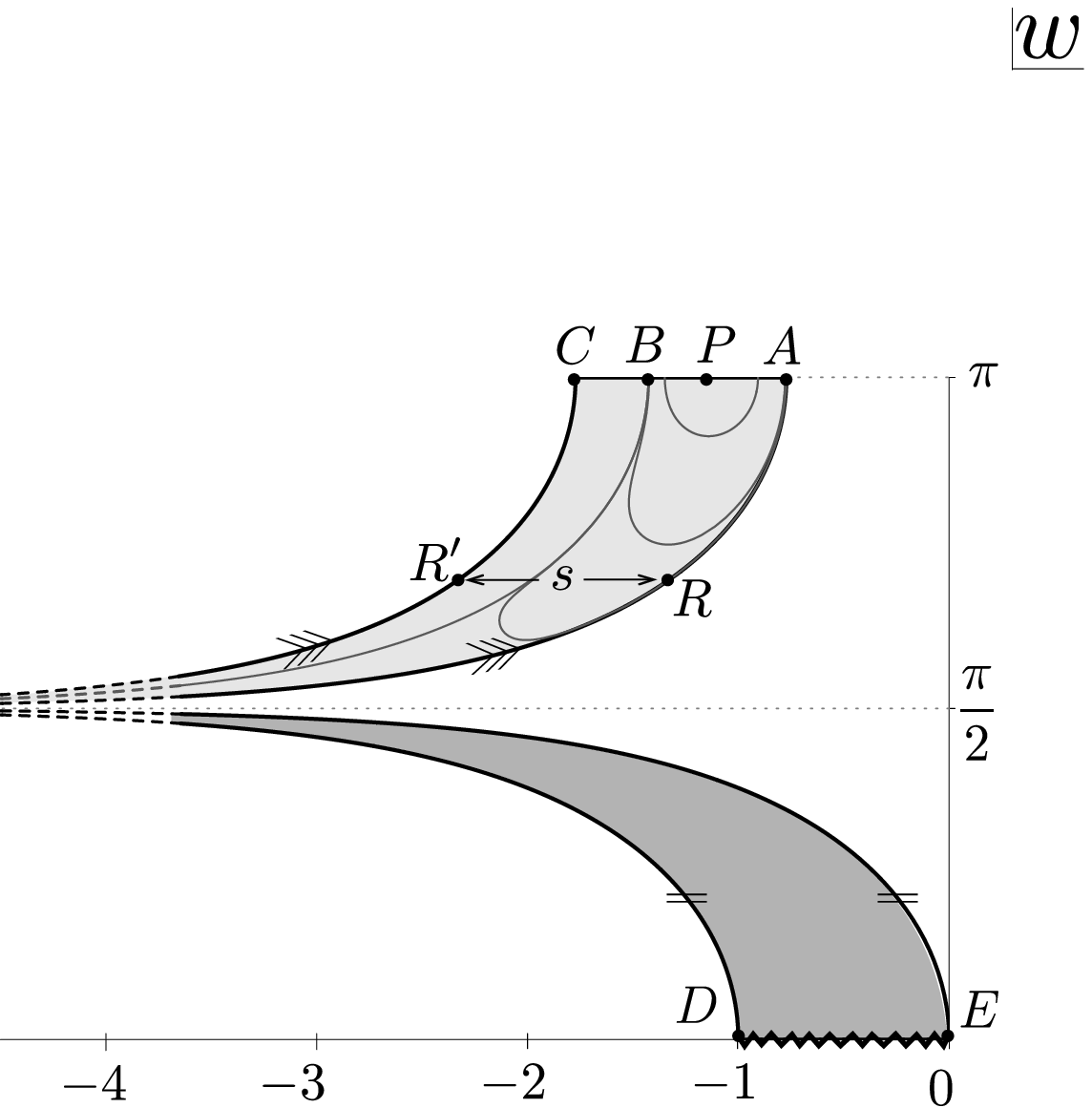, height=7.5cm}}
\end{center}
 \centerline{
 \hskip 2cm  (a) \hskip 4.30cm
 (b) \hskip6.2cm (c)\hskip 3.5cm}
\caption{(a) The one-loop tadpole in the
 $z$ frame.
The surface is composed of two separate strips: one above
$z\in [-{1\over 2}e^s, -{1\over 2} ]$ and the other above $z\in [-{1\over 2},
{1\over 2} e^s]$.  These two strips are joined at $i\infty$. The figure is
displayed for $s=1$.
  (b)  The same surface with the
right strip translated to the right by a distance $a_0+1$, which
depends on $s$ and makes the identifications of the left and
right boundaries work with rays through the origin.
(c) The middle figure mapped to the $w$ frame with $w= -\ln 2z + i\pi$.}
\label{sl04fig}
\end{figure}

\subsection{Modulus in Schnabl gauge}

The estimates done in the previous subsection bound $M(s)$
and allow us to confirm that moduli space is covered for any
value of the deformation parameter $\lambda$.  We now claim
that
the value of the modulus $M(s)$  becomes calculable in simple
closed form in the Schnabl limit $\lambda \to 0$.  The
 derivation
requires careful analysis of a conformal map in the limit $\lambda \to 0$.
Since the final result is simple, we will present it here, without proof.
In the following subsection we justify our claim.

We begin with Figure~\ref{sl04fig}(a), where we see that the
surface of the tadpole diagram
appears as two  {\em disconnected} vertical strips
 in the $z$ frame.
The strip above the real segment $[-{1\over 2} e^s , -{1\over 2}]$
 represents $e^{-sL_L}$
 and the strip
 above the real segment $[-{1\over 2}, {1\over 2} e^s]$
 represents the external state and $e^{-sL_R}$.
These real segments are the boundaries of the annulus.
On the left strip the identification of the edges is
$z \sim e^s z$.  On the right strip the identification is
more nontrivial.
 Its left boundary carries the ordinary sliver parameterization
 and is given by  $-\frac{1}{2}+\gamma(\theta)$, with $\gamma(\theta)$ defined in~(\ref{gamma0}).
 The right boundary of the right strip is given by $e^s(\half+\gamma(\theta))$ and thus carries a parameterization which is rescaled
 by $e^s$. It follows that
a point $R$ on the line above $z=-{1\over 2}$  and
a point $R'$ on the line above $z={1\over 2} e^s$ are identified
if the copy $S$ of
$R$ on the line above $z={1\over 2} $
 is related to $R'$ via the scaling
$z \sim e^s z$.
This is, in fact, the gluing prescription discussed around
equation~(\ref{gindents}).
The two separate strips are supposed to be glued together
at $i\infty$ but it is not obvious how to glue these hidden
boundaries.

 We could proceed as we did in the previous section and map this configuration of surfaces directly to the $w$ frame via~(\ref{wzrel}).
 Just like in
  Figure~\ref{sl03fig}(c)), the external state would  be represented
   in the $w$ frame by an infinite strip
 of height $\pi$. In Schnabl gauge, however,
 we can construct a different map of the tadpole diagram to
 the $w$ frame, one
 in which the whole surface is foliated by
  horizontal lines of length $s$.
  It is then possible to use the map $\zeta(w)$ in (\ref{iwltktcntaj})
 to get a round annulus.
 We will now show how this is done.

 In the $z$ frame we
translate the right strip towards the right by a
distance that makes the line through
 the identified points
$R$ and $R'$ go through the origin. Since the heights of $R$ and $R'$ are related by $e^s$ it follows,
 by similar triangles, that $R\sim R'$ are related by $z \sim e^s z$
 (see Figure~\ref{sl04fig}(b)).  The requisite displacement, called
 $a_0 +1$ for later convenience, is determined from the similar
 triangles:
\begin{equation}
e^s
=   {CR'\over AR}
= { a_0+1+{1\over 2} e^s  \over a_0+{1\over 2}} \,.
\end{equation}
One readily finds that
\begin{equation}
\label{aval99}
a_0 = {1\over e^s -1} \,, \quad   a_0 + 1
= {1\over 1-e^{-s}}\,.
\end{equation}
With this result one can check that the two vertical lines for the right strip
 are located at
\begin{equation}
\label{iwltfkchlw}
 \Re(z)=
a_0 +\frac{1}{2}= \frac{1}{2}
\coth \left(\frac{s}{2}\right)
\,, \quad
\hbox{and} \quad
 \Re(z)=
a_0 + 1+ {1\over 2} e^s   = e^s \cdot{1\over 2}
\coth \left(\frac{s}{2}\right) \,.
\end{equation}
The map
$w = - \ln (2z)+ i\pi$  in (\ref{wzrel})
takes the full left and
right strips to the
 $w$-frame picture in Figure~\ref{sl04fig}(c).
This picture is similar to that in Figure~\ref{sl02fig}(b), which
refers to the vacuum graph.
There is only one minor difference:
 the image of the
right strip in Figure~\ref{sl04fig}(c)
is displaced some distance to the left.
 This happens
because the coordinate $z(A)$ of the point $A$ satisfies
\begin{equation}
    z(A) = a_0+{1\over 2} > {1\over 2}
    \quad \to\quad \Re\bigl(w(A)\bigr)< 0\,.
\end{equation}
Since both strips in
Figure~\ref{sl04fig}(b)
work with identification
$z \sim e^s z$, the $w$ plane Figure~\ref{sl04fig}(c)
has the identification $w\sim w-s$.
This $w$ presentation is different from the earlier $w$ presentation in which
the coordinate half-disk for the external state appears as a semi-infinite
strip (Figure~\ref{sl03fig}(c));  the coordinate half-disk has been pushed
up!
The identification $w\sim w-s$ ensures that the map (\ref{iwltktcntaj}) takes the $w$-plane region to the annulus with modulus
\begin{equation}
\label{Mtp}
     \boxed{ \phantom{\Bigl(} M = \frac{\pi}{s}\,. ~}
\end{equation}
 Inserting an external state to form the tadpole graph therefore did not affect the modulus of the annulus -- the modulus~(\ref{Mtp}) coincides with our
 result~(\ref{modourann}) for the modulus of the vacuum graph.
The only evidence of the
 external state
is that
the top boundary of the
annulus is split between the boundary  $AB$ of the
coordinate half-disk
 with the puncture
and the
 boundary $BC$ generated by $e^{-sL_R}$.
 The surprisingly simple form of
 the
 modulus will turn out to be
 generic for one-loop diagrams in Schnabl gauge.
 In fact, we will find that the annulus modulus of a general one-loop diagram is a simple function that depends
 only on the Schwinger parameters of the propagators running in the loop;
 the Schwinger parameters of trees attached to the loop do not affect the
 modulus of the annulus.
\begin{figure}
\centerline{\epsfig{figure=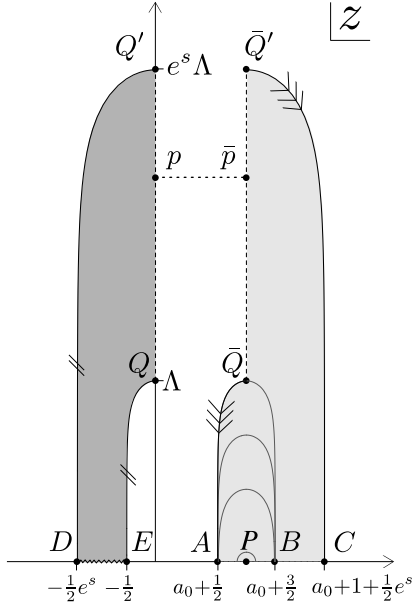, height=8cm}}
\caption{The $z$-frame $\lambda$-regulated one-loop tadpole of
Figure~\ref{sl03fig} cut along $QQ'$ and with the right piece
translated to the right a distance $a_0+1$ so that the identification
of $A$ and $C$ is through scaling by
 $e^s$.
 The figure is displayed for $\lambda=10^{-4}$ and $s=1$.}
\label{sl05fig}
\end{figure}

\subsection{Taking the $\lambda \to 0$ limit}\label{dofeoruhk}

 We will now justify our construction of the map of the Schnabl tadpole diagram to the round annulus.
Let us
consider the $\lambda$-regulated version of the one-loop tadpole
graph, first shown in Figure~\ref{sl03fig}(b).  We cut the diagram along the
$QQ'$ line to produce two disconnected pieces.  Just like we did for
the Schnabl tadpole, we displace the right part of the figure to the right a distance $a_0+1$.  The identifications on the left part of the surface still
work with $z \sim e^s z$, but on the right they do not anymore.  Choosing
$a_0$ as before (see (\ref{aval99})) we ensure that the points
$A$ and $C$ are still  identified with $z\sim e^s z$, but this identification is only approximate for the other points on the curves $A\bar Q $ and
$C{\bar Q}' $.

\begin{figure}
\rightline{\epsfig{figure=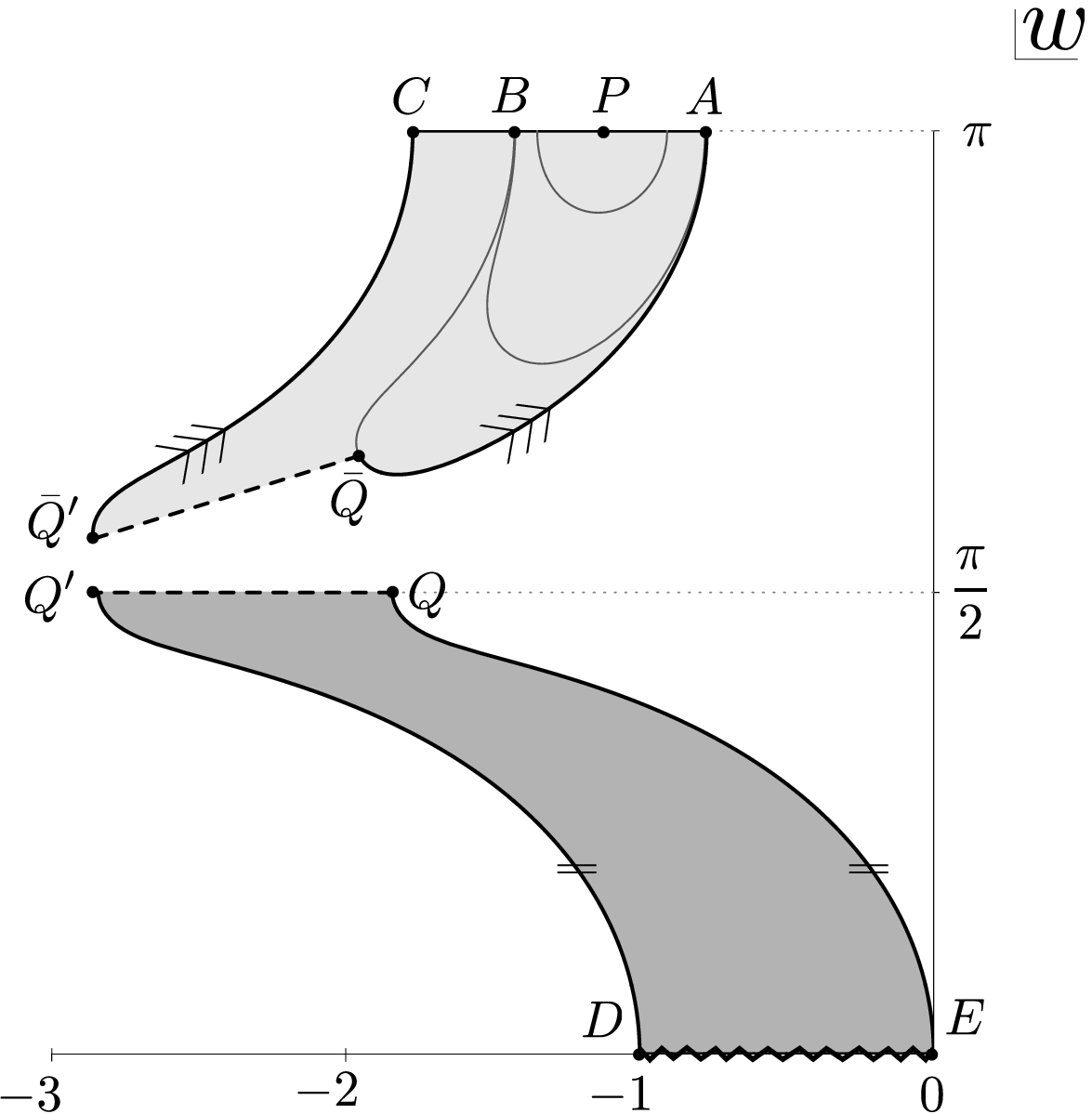, height=9.5cm}\hskip 42pt}
\vskip 2.0cm
\rightline{\epsfig{figure=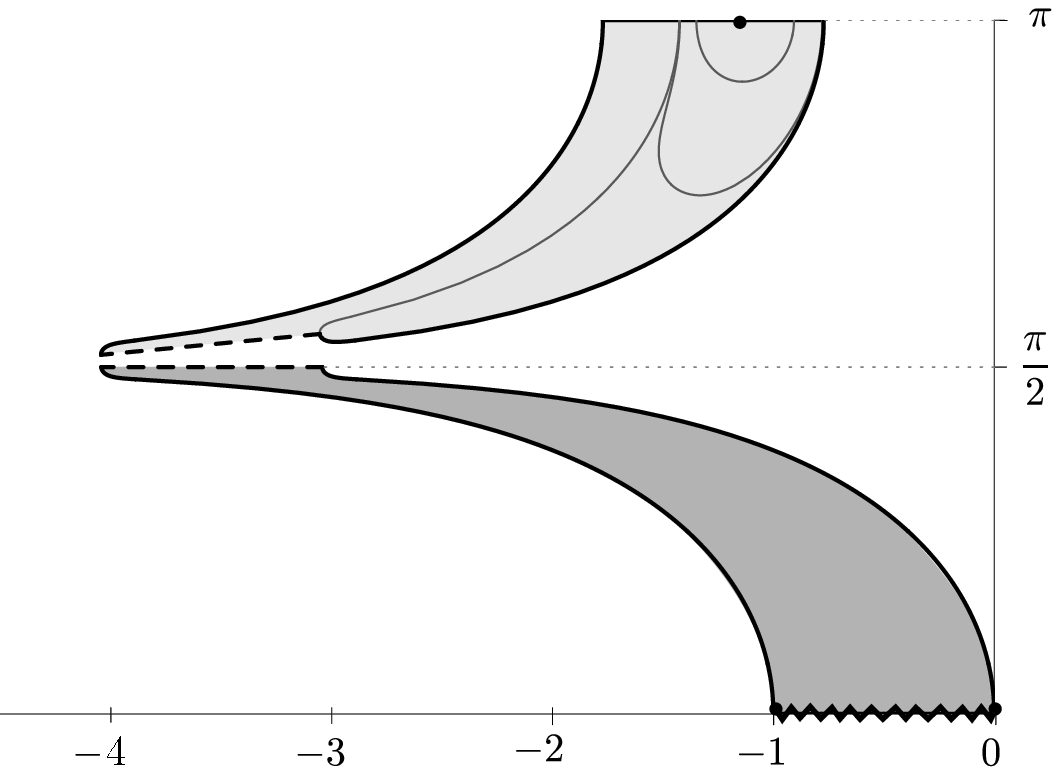, height=9.5cm}\hskip 60pt}
\caption{The $\lambda$-regulated one-loop
tadpole in the plane $w = - \ln (2z) + i\pi$.
The top figure arises for  $\lambda=10^{-4}$ and the bottom figure arises for $\lambda=10^{-14}$. Both figures use $s=1$. }
\label{sl06fig}
\end{figure}

As before, the map $w = - \ln (2z) + i\pi$ takes the left part of
Figure~\ref{sl05fig} (the surface associated with $e^{-sL_L^\lambda}$)
to the familiar annular domain with identifications
exactly given by $w\sim w-s$ (see Figure~\ref{sl06fig}).
Since
 $z(Q) = i\Lambda$
(see (\ref{iwltfsksss}))
the image of $QQ'$
 in the $w$ frame
is shifted $\ln 2\Lambda$ to the left with respect
to the image of the inner boundary $DE$.

For the map of the right part of Figure~\ref{sl05fig} we have
to be a bit more careful.  We will use the same map
$w= - \ln (2z) + i\pi$, which
 results in a surface whose identification is
  not quite $w\sim w-s$ and
 thus cannot be interpreted as an annular region for general $\lambda$.
 Furthermore, the image of $\bar Q\bar Q'$ in the $w$ frame does not quite coincide with the image of $QQ'$.
What we are
going to show is that in the limit as $\lambda \to 0$ (and
consequently $\Lambda \to \infty$) the identifications needed to
form the full annulus become exact.  More precisely, as
$\lambda \to 0$ two things should happen:

\begin{enumerate}

\item
All points
$p\in QQ'$ and $\bar p\in {\bar Q}{\bar Q}'$
that are at the same height (and should therefore be identified),
are mapped to points on the $w$ frame
that approach each other as $\lambda \to 0$.
This convergence is uniform on $QQ'$
ensuring that the top and bottom parts of the annulus glue well.

\item  Points $q\in A{\bar Q}$ and $q'\in A{\bar Q}'$
that must be identified will map to coordinates $w$ that satisfy
$w(q) - w(q') = s$ in the limit $\lambda \to 0$.
  This convergence is uniform on $A{\bar Q}$,
 ensuring that the top part of the annulus works with the same
 identification $w\sim w-s$
as the bottom part.

\end{enumerate}
If these two claims hold, it justifies the prescription
given in the previous subsection for the Schnabl limit.  In the remainder
of this subsection we will prove (1) and (2).

Consider first claim (1) regarding the
gluing of $QQ'$ to ${\bar Q} {\bar Q}'$.
Let $i x \Lambda$, with $x$ a real number, denote the imaginary part of a point $p\in QQ'$ that must
 be identified with a point $\bar p \in {\bar Q} {\bar Q}'$
with the same imaginary part.  Since the imaginary part of any
point $p$ (or $\bar p$) ranges between $\Lambda$ and $e^s \Lambda$ we have
\begin{equation}
1 \leq  x  \leq e^s \quad \to \quad
\Lambda \leq x \Lambda \leq e^s \Lambda \,.
\end{equation}
We then have
\begin{equation}
z(p) = i\, x \Lambda \,, \quad   z (\bar p) =  a_0 + 1 + i \,x\Lambda
= {1\over 1 -e^{-s} } +  i \,x\Lambda\,,
\end{equation}
where we made use of (\ref{aval99}).
Using (\ref{wzrel})  we get
\begin{equation}
\label{iwltsskstts}
 w({\bar p})  -  w(p) = -  \ln \Bigl[\, {z({\bar p})\over z(p)}\Bigr]
 = -\ln \Bigl[ 1 +  {1 \over ix\Lambda (1-e^{-s})} \Bigr]
\end{equation}
As $\lambda \to 0$ we have $\Lambda \to \infty$. It is then clear that for any fixed value of
 $s>0$
and any $x\in [1, e^s]$ the above gives
$w(\bar p) - w (p) \to 0$.
 Furthermore, it follows from~(\ref{iwltsskstts}) and $x\geq1$
 that the convergence of $\bar Q\bar Q'$ to $QQ'$ is uniform.
This proves claim (1).

It is interesting to discuss the above result
in more detail.  We show in Figure~\ref{sl06fig} two examples of
the $w$ plane surface, both for $s=1$.  The top figure uses
$\lambda = 10^{-4}$ and the bottom one uses $\lambda = 10^{-14}$.
One can see the image of ${\bar Q} {\bar Q}'$ as the sloping edge
that approaches
 (as we go from the top figure to the bottom figure)
the horizontal image
of $Q Q'$.
  Expanding the logarithm in (\ref{iwltsskstts}) we get
\begin{equation}\label{pbarpconv}
 w(\bar p)  -  w(p) = i \,\,{1 \over x\Lambda (1-e^{-s})} - {1\over 2}
 \, { 1\over x^2\Lambda^2 (1-e^{-s})^2} + \mathcal{O}(\Lambda^{-3})\,.
\end{equation}
  The vertical distance
between the images of $p$ and $\bar p$ vanishes as $\Lambda^{-1}$.
The horizontal distance
vanishes faster,  as fast as $\Lambda^{-2}$.
These features are clearly seen in the figure for the pair
$\bar Q$, $Q$, and for the pair ${\bar Q}'$, $Q'$.
Furthermore, 
the vertical convergence of $Q'$ to $\bar Q'$ in the $w$ frame
is faster by a factor of $e^s$ than the vertical convergence of
$Q$ to $Q'$. This is due to the suppression factor $\frac{1}{x}$ in
the imaginary part of~(\ref{pbarpconv}), and is clearly visible
in the figure.

\medskip
Let us now address claim (2).
Before the translation is performed (see
Figure~\ref{sl03fig}(b)), the identified curves $AQ$ and $CQ'$ are parameterized as shown in~(\ref{AQCQ}).
After the translation by $a_0+1$, we obtain Figure~\ref{sl05fig} with the curves $A\bar Q$ and $C\bar {Q'}$ given by
\begin{equation}\label{AbarQCbarQ}
       A\bar Q=a_0+\half +\gamma^\lambda_L\,, \qquad
       C\bar{Q'}=a_0+1+e^s\bigl(\half +\gamma^\lambda_R\bigr)\,.
\end{equation}
These parameterized curves are identified.
The (complex)
ratio $r(\theta)$ between identified points on the curves is given by
\begin{equation}
\label{iwltskthcltofsk}
r(\theta) = {a_0+1+e^s\bigl(\half +\gamma^\lambda_R(\theta)\bigr)\over  a_0+\half +\gamma^\lambda_L(\theta) }
=e^s\cdot\frac{\gamma_R(\theta)+\half \coth\tfrac{s}{2}}{\gamma_L(\theta)+\half \coth\tfrac{s}{2}}\,,
\end{equation}
where we used the definition~(\ref{aval99})  of the shift $a_0+1$
as well as~(\ref{iwltfkchlw}).
We must show that this ratio has
the limit
\begin{equation}\label{limittoshow}
    r(\theta)\to e^s\,\qquad\text{for }\quad\lambda \to 0\,,\quad0\leq\theta\leq\tfrac{\pi}{2}\,.
\end{equation}
If this is so, the map to the
$w$ plane (via the logarithm) will imply that the points corresponding to $\theta$ are separated
by a horizontal
translation by $s$.
 To make the map to the annulus well defined in the limit $\lambda\to0$, we need this horizontal
 separation by $s$ to hold to arbitrary precision for all points on the identified curves, i.e. we need the
 limit~(\ref{limittoshow}) to hold \emph{uniformly} on $0\leq\theta\leq\tfrac{\pi}{2}$.

One finds $r(\theta=0) = e^s$, exactly, as expected for the ratio of the base points $A$ and $C$ of the two curves.
Indeed, the translation was designed to make the identification $z\sim e^s z $ work on the
real axis.  For general $\theta$, a short calculation gives
\begin{equation}
\label{iwltskthansofskl}
r(\theta) =e^s \cdot \, {1 +\delta(\theta)\over  1-\delta(\theta)} \,, \quad\hbox{with}\quad
  \delta =  \frac{\gamma_R-\gamma_L}{\gamma_R+\gamma_L+\coth {s\over 2}}=
  {\Re\bigl(\gamma_R\bigr) \over  i\Im\bigl(\gamma_R\bigr) + {1\over 2} \coth {s\over 2}}\,,
\end{equation}
where we used $\gamma_L=-\overline{\gamma_R}$ in the last step.
As we map the two points in question to the $w$ plane, their separation is
 given by $\ln r$.
We obtain
\begin{equation}
\ln r =  s +  \ln \Bigl(   {1 -\delta \over  1+\delta}    \Bigr)\,.
\end{equation}
We want to show that $\delta$
 goes to zero
 uniformly on $0\leq\theta\leq\tfrac{\pi}{2}$
when
$\lambda \to 0$.
We are going to
break the
curve
 $\gamma_R$
 into two parts: (i) the top part for which $\Im(\gamma_R) \in [\Lambda/2, \Lambda]$
and (ii) the bottom part for which $\Im(\gamma_R) \in [0, \Lambda/2]$.
 We recall from~(\ref{thatahalf}), that this corresponds to splitting the range of $\theta$ at
 $\theta=\theta_{\frac{1}{2}}$.
Consider the top part (i).
 In this region we estimate
\begin{equation}
\bigl|\delta \bigr| =
\biggl|
{\Re\bigl(\gamma_R\bigr) \over  i\Im\bigl(\gamma_R\bigr) + {1\over 2} \coth {s\over 2}}
\biggr|
\leq\, {
\hbox{Max}\,\Re\bigl(\gamma_R\bigr) \over \hbox{Min}\,\Im\bigl(\gamma_R\bigr) }
={ {1\over 2} \over \,{\Lambda\over 2}\,} = {1\over \Lambda}
\end{equation}
so that
\begin{equation}\label{regioni}
\bigl|\delta(\theta)\bigr| \leq {1\over \Lambda} \, \qquad \hbox{for }
 \quad\theta\in\bigl[\theta_\frac{1}{2},\tfrac{\pi}{2}\bigr]\,.
\end{equation}
Now consider region (ii), i.e. $0\leq\theta\leq \theta_\frac{1}{2}$.  Recall our earlier estimate
 (\ref{thatahalf})
that at $ \theta=\theta_\frac{1}{2}$
the coordinate curve has indeed risen to a height of $\Lambda/2$ and
 that
 \begin{equation}\label{maxRegammaR}
 \Re \Bigl(\gamma_R\bigl(\theta_\frac{1}{2}\bigr)\Bigr)=-{1\over 2\pi}\sqrt{2\lambda}\,.
 \end{equation}
In  this region
 $\Im(\gamma_R)$
can be arbitrarily small, and
$|\Re(\gamma_R)|$ reaches its maximal value at $\theta=\theta_\frac{1}{2}$.
We thus estimate
\begin{equation}
|\delta| = \biggl| {\Re(\gamma_R) \over  i\Im(\gamma_R) + {1\over 2} \coth {s\over 2}}\biggr|
\leq  \biggl| {\Re(\gamma_R) \over  {1\over 2} \coth {s\over 2}} \biggr|
  \leq\, {  {1\over 2\pi}\sqrt{2\lambda} \over {1\over 2}  }
={1\over \pi}
\sqrt{2\lambda}\,
\end{equation}
 for  region (ii),
so that
\begin{equation}\label{regionii}
\bigl|\delta(\theta)\bigr| \leq {1\over \pi}
\sqrt{2\lambda}\,\qquad \hbox{for }
 \quad\theta\in\bigl[0,\theta_\frac{1}{2}\bigr]\,.
\end{equation}
We now have upper bounds
 on $\delta$
valid for
 the regions
(i) and (ii).
 For any $\lambda < 1$
the upper bound in
 (\ref{regioni}) for region
(i)
is larger than that in
 (\ref{regionii}) for region
(ii).
Therefore we obtain
the uniform upper bound
\begin{equation}
\bigl|\delta(\theta)\bigr| \leq  {1\over \Lambda}
 \quad \hbox{for all } \quad\theta\in\bigl[0,\tfrac{\pi}{2}\bigr]\,,\quad \lambda<1\,.
\end{equation}
This means that $\delta(\theta)$ will vanish
 uniformly on $0\leq\theta\leq\tfrac{\pi}{2}$
as $\lambda\to 0$, as we wanted
to prove.  This establishes the second claim, and thus completes
the argument that shows that regulation leads to the claimed
simple map in Schnabl gauge.

\medskip
We conclude with a comment concerning the Schnabl gauge limit.
In the unregulated case, shown in Figure~\ref{sl04fig}(b), we see that the left and right cylinders are supposed to be glued at $i\infty$.
It may seem as if the gluing involves both the coordinate patch strip of the
external state and the strip to the right of it.  The regulation shows that this is not quite the way things work. The
 coordinate frame for the external state tapers out and does not
glue to the bottom part of the diagram, which arises from the
left cylinder.   The tip $\bar Q$ of the local coordinate
frame (the string midpoint) lies at the end of the gluing line.
As can be seen in Figure~\ref{sl06fig},
at $\bar Q$ the coordinate curve goes both up
towards $B$ and down to
eventually reach $A$.
The behavior at $\bar Q$
 follows from conformality
to the $z$ frame, as shown in Figure~\ref{sl05fig}.

\section{Slanted wedges: A family of surfaces}\label{secsw}
\setcounter{equation}{0}

Loop amplitudes in Schnabl gauge
use surfaces that do not feature in tree amplitudes.
As we have seen in the previous sections, we sometimes
deal with semi-infinite strips that look like the familiar
wedge surfaces, except that the vertical edges are subject
to identifications that are slanted. For wedge
surfaces, presented as vertical semi-infinite strips, the natural
identification of the vertical edges is a horizontal translation
by  the width $a$ of the wedge.

It turns out to be convenient to introduce a set of surfaces
that generalize the wedge surfaces.  They will be called
{\em slanted} wedges and are characterized by two parameters:
the width $a$ of  the underlying wedge and the slant $b$,
to be defined below.  There is one important difference between
wedges and slanted wedges.  Associated with wedges there
are wedge states but there are
no surface states associated
with slanted wedges.

For wedge surfaces, the surface states are based on  once-punctured
disks.  The disk is formed by attaching the left edge of the wedge surface
to the right edge of a unit-width wedge coordinate frame (with a marked
point, or puncture) and gluing the two remaining vertical edges with a horizontal identification.
The resulting surface is a semi-infinite cylinder with a puncture
on the boundary at the real axis.
This surface can be conformally mapped to a disk.    More precisely,
the disk has an inner puncture because it misses one point,
the image of $i\infty$ on the wedges. This missing point can be
ignored.  The situation is far more serious for slanted wedges.
As we have seen in the construction of the one-loop tadpole, a wedge
with a slanted identification has a
hidden boundary at $i\infty$, a boundary that must be glued to another surface.  Instead of having a vanishingly small additional
boundary associated with a missing point, as in the case
for wedges,
slanted wedges have 
an additional boundary that cannot be ignored.  
As a result
 there are no canonical
  surface  states associated
with slanted wedges.  The hidden boundaries of slanted wedges
can be brought into the open by $\lambda$-regularization.

Even without associated states, we can define  a kind
of star algebra of slanted wedges.
While not strictly needed for tree diagrams,
slanted wedges
simplify significantly the construction of the
associated Riemann surfaces.
For loop diagrams
slanted wedges are key to the construction of the relevant Riemann surfaces.

\subsection{Definition and examples}\label{ssecsw}

\begin{figure}
\centerline{
\epsfig{figure=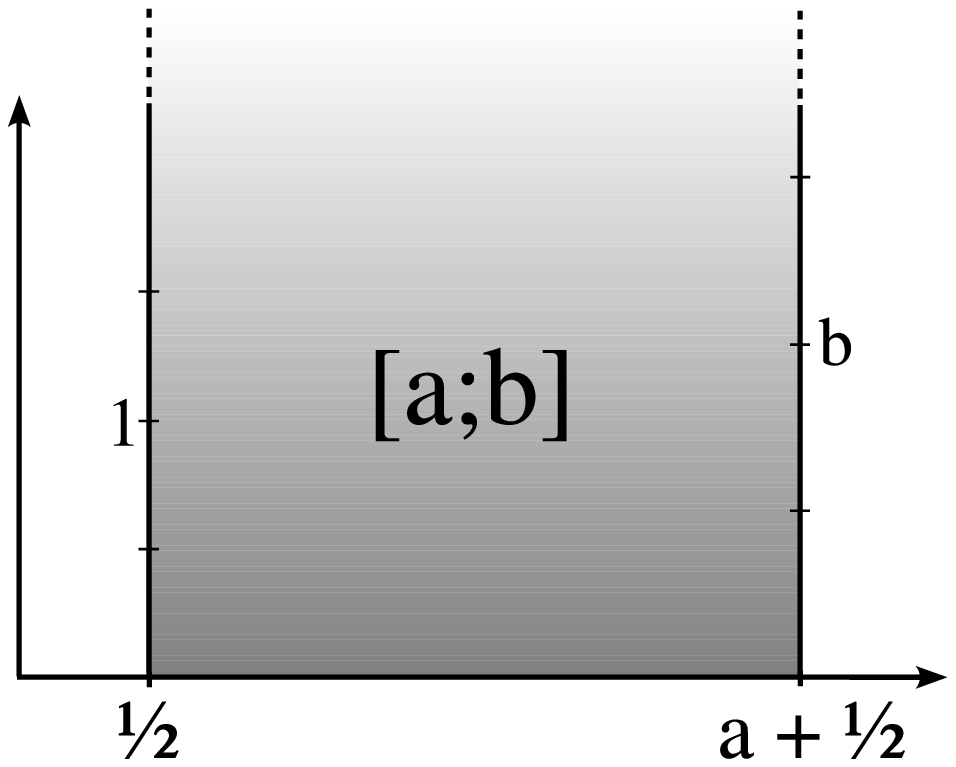, height=5cm}
\hskip 1cm
\epsfig{figure=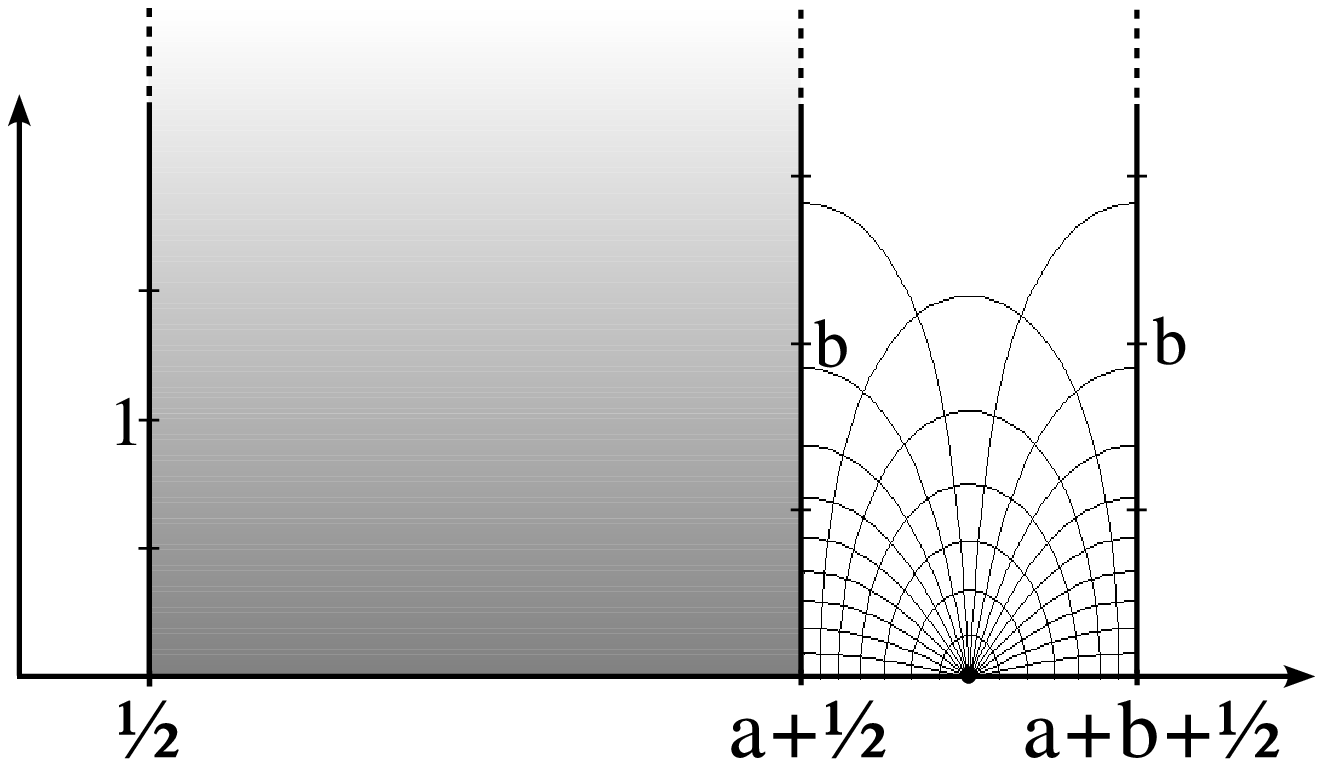, height=5cm}}
 \centerline{(a)\hskip 7.5cm (b) \hskip 1.1cm}
\caption{(a) The slanted wedge $[a;b]$ in the sliver frame $z$.\quad (b) Illustration
of the star multiplication of a slanted wedge $[a;b]$ with a Fock space state
$[1;1]$. The result is a slanted wedge $[a+b;b]$.}
\label{figSigma}
\end{figure}

The slanted wedge $[a;b]$, with $a, b \geq 0$, is defined
on the upper-half plane $z$ as the semi-infinite strip between $\Re(z)=\half$
and $\Re(z)=\half +a$:
\begin{equation}
[a;b]  \equiv \Bigl\{ z \, \Bigl|   \, {1\over 2} \leq \Re (z) \leq {1\over 2} + a\,,
~  \Im (z) \geq 0 \, \Bigr\}\,.
\end{equation}
The above states that, as a region, $[a;b]$ is the wedge
of width $a$, positioned so that the left boundary is
$\Re(z) = {1\over 2}$.  By definition, the left boundary
$\Re(z) = {1\over 2}$ carries  the parameterization induced by the sliver
map
$z=\frac{2}{\pi}\tan^{-1}\xi$.
 More explicitly, the point $\xi = e^{i\theta}$ is mapped to
\begin{equation}
\label{parleftboundary}
~\hbox{Left Boundary:}~~~e^{i\theta} \to
 \half+ \gamma(\theta)
\qquad\qquad \hbox{with} \quad 0\leq \theta \leq {\pi\over 2}\,,
\end{equation}
where the curve $\gamma(\theta)$ was defined in~(\ref{gamma0}).
It follows that the left boundary of $[a;b]$ glues naturally to a coordinate patch
$-\half \leq \Re(z) \leq \half $ of the sliver frame.
The slant parameter $b >0$ is a scaling factor for the parameterization
of the right boundary $\Re (z) = {1\over 2} + a$ of $[a;b]$.
We have
\begin{equation}
\label{parrightboundary}
\hbox{Right Boundary:}~~~e^{i\theta} \to
  \half+ a +b\cdot\gamma(\theta)
\quad \hbox{with} \quad 0\leq \theta \leq {\pi\over 2}
\,.
\end{equation}
This implies that
the parameterization of the right boundary
is obtained by stretching that of the left boundary by the factor $b$.
See Figure~\ref{figSigma}(a) for a representation of the slanted
wedge $[a;b]$.
For $b=1$
 both boundaries of the slanted wedge carry the same parameterization
 and are thus horizontal translations of each other.
Thus
$[a;1]$ is just the familiar
 ordinary
wedge surface of width $a$:
\begin{equation}
[a; 1]  = W_a \,.
\end{equation}
Fock space states are described as $[1;1]$ with a
local operator insertion at $z=1$ between the two
boundaries. The Fock space state insertion is mapped from
$\xi = 0$ to $[1;1]$ via
$z=1+\frac{2}{\pi}\tan^{-1} \xi $.
In general, slanted wedges can carry operator insertions
or line integrals.

Since slanted wedges are Riemann surfaces we have some
equivalence relations that must be noted.  First, the position of
the slanted wedge can be altered.  While $[a;b]$ is always assumed
to have a left boundary $\Re(z) = \half$, a translation by a real
constant can be used
to position the slanted wedge elsewhere.  This is useful to form star
products, for example. Sometimes we have to deal with wedge
regions where both edges carry scaled parameterizations.
We could call such surfaces $[b_L; a; b_R]$ with $b_L$
and $b_R$ denoting the scaling factors
for the left and the right edges, respectively. Explicitly,
this means that
 the parameterizations of the left and right boundaries  in
(\ref{parleftboundary}) and (\ref{parrightboundary}) are replaced
by $\half+b_L\gamma(\theta)$ and $\half+a +b_R\gamma(\theta)$, respectively.
This surface, under the map $z \to z/b_L$ and a possible
translation, gives us the conformal
identification
\begin{equation}
[b_L; a; b_R]~\sim ~ [ a/b_L \,;  b_R/b_L] \,.
\end{equation}
We obtain a wedge of width $a/b_L$ with unit scaling on the
left boundary and scaling $b_R/b_L$ on the right boundary.
The above shows that we do not have to define slanted
wedges with scaled parameterizations on both edges.

\subsection{Operations on slanted wedges}

In order to create the surfaces relevant to the Feynman
rules we need to introduce the ``star-multiplication" of
slanted wedges.  For plain wedges the star multiplication
is homomorphic to the star multiplication
of the corresponding wedge states.  Since we have no states
associated with slanted wedges, their star multiplication is
only a device to
construct
interesting surfaces.

As for surface states, we define star multiplication as the gluing of the right boundary of the
first surface to the left boundary of the second surface. This gluing,
however, requires identical parameterizations.
For two slanted wedges $[a_1;b_1]$ and $[a_2;b_2]$, we define
\begin{equation}\label{alg}
 \boxed{ \phantom{\Bigl(}  [a_1;b_1]\ast [a_2;b_2]\equiv  [\,a_1+b_1a_2\,;\,b_1b_2\,]\,.}
\end{equation}
The logic behind this is clear:  since the right boundary of the first
slanted wedge carries a scaling $b_1$, the second slanted wedge
must be fully scaled by $b_1$ so that its left boundary carries the
same scaling.  In this process its width becomes $b_1 a_2$ and
the scaling of its right boundary $b_1b_2$.
Once the surfaces are glued, we get a total width of $a_1 + b_1a_2$
and a scaling factor $b_1b_2$, which applies to the right boundary.

Clearly, slanted wedges form a closed algebra under the star multiplication and plain wedges form a commutative subalgebra.
The algebra~(\ref{alg}) of slanted wedges $[a;b]$ can also be represented as the algebra of matrices of the form
\begin{equation}\label{matrep}
[a;b] \leftrightarrow     \begin{pmatrix} b & a \\ 0 & 1 \end{pmatrix} \,.
\end{equation}
Indeed, in agreement with (\ref{alg}) we then have
\begin{equation}
\begin{pmatrix} b_1 & a_1 \\ 0 & 1 \end{pmatrix}\begin{pmatrix} b_2 & a_2 \\ 0 & 1 \end{pmatrix}=\begin{pmatrix} b_1b_2 & a_1+ b_1 a_2 \\ 0 & 1 \end{pmatrix}\,.
\end{equation}
A simple and useful particular case of (\ref{alg}) involves a
Fock space state and a slanted wedge:
\begin{equation}
    [a;b]\ast [1;1]=[a+b\,;b\,]\,.
\end{equation}
This example is
illustrated in Figure~\ref{figSigma}(b).
Note that in the final surface the puncture lies at
$z= \half + a + \half b$, the first $\half$ for the conventional
offset, the $a$ due to the first surface and $\half b$ because
the slanting required scaling the unit width of the Fock state
surface by $b$.

\bigskip
We now consider the Schnabl gauge propagator.  As we will
 see,
its various ingredients act naturally on slanted wedges
and can be themselves represented by slanted wedges.
The classical propagator is given by
\begin{equation}
\label{classprop1}
    {\cal P}=\int
    ds ds^\star\, e^{-sL}\,B\,Q\,B^\star e^{-s^\star L^\star}
    = \int
    ds ds^\star\, \,B\,Q\,e^{-sL} e^{-s^\star L^\star} B^\star\,.
\end{equation}
We will focus solely on the Riemann surface interpretation of this propagator, namely the action of $e^{-sL} e^{-s^\star L^\star}$ on
surfaces. The presence
of line integral insertions from $B,Q$, and $B^\star$
 will not play a role in the following analysis.

\medskip
We will construct the action of the propagator step by step, treating the operators  $e^{-sL_R}$, $e^{-sL_L}$, $e^{-s^\star L_R^\star}$, and
$e^{-s^\star L_L^\star}$ separately.
 As discussed in Section~\ref{secannmod} these operators
 generate hidden boundaries, which
 will now be associated with slanted wedges.
For loop diagrams these boundaries
require special attention.

\begin{figure}
\centerline{
\epsfig{figure=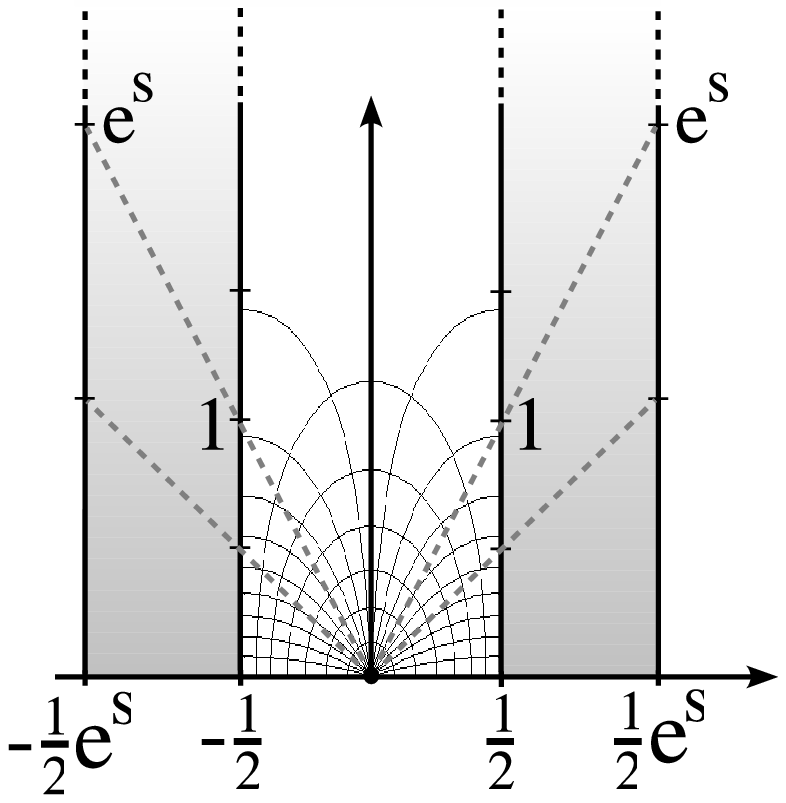, height=5cm}
\hskip .8cm
\epsfig{figure=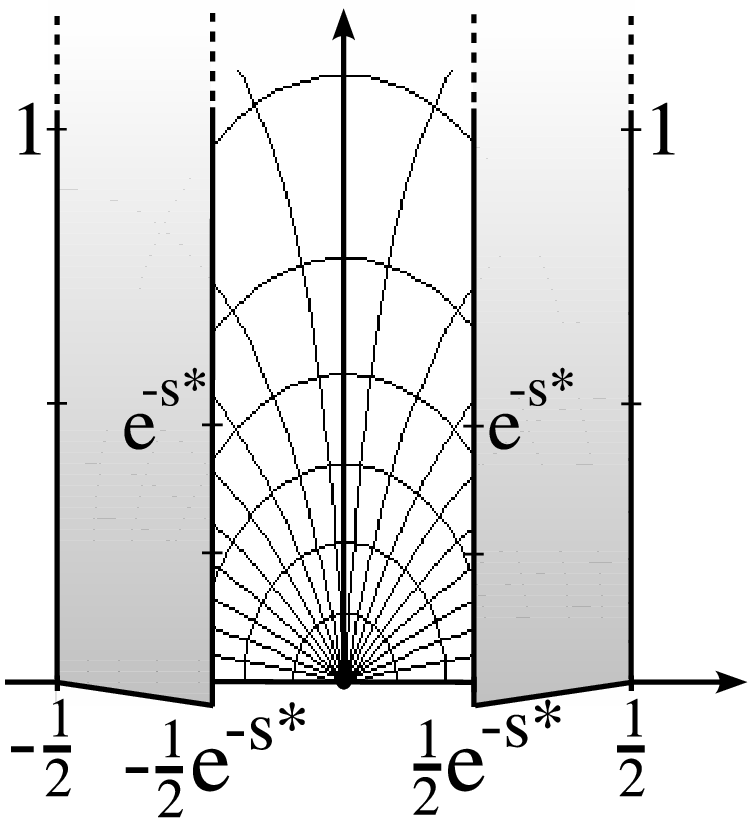, height=5cm}
\hskip .8cm
\epsfig{figure=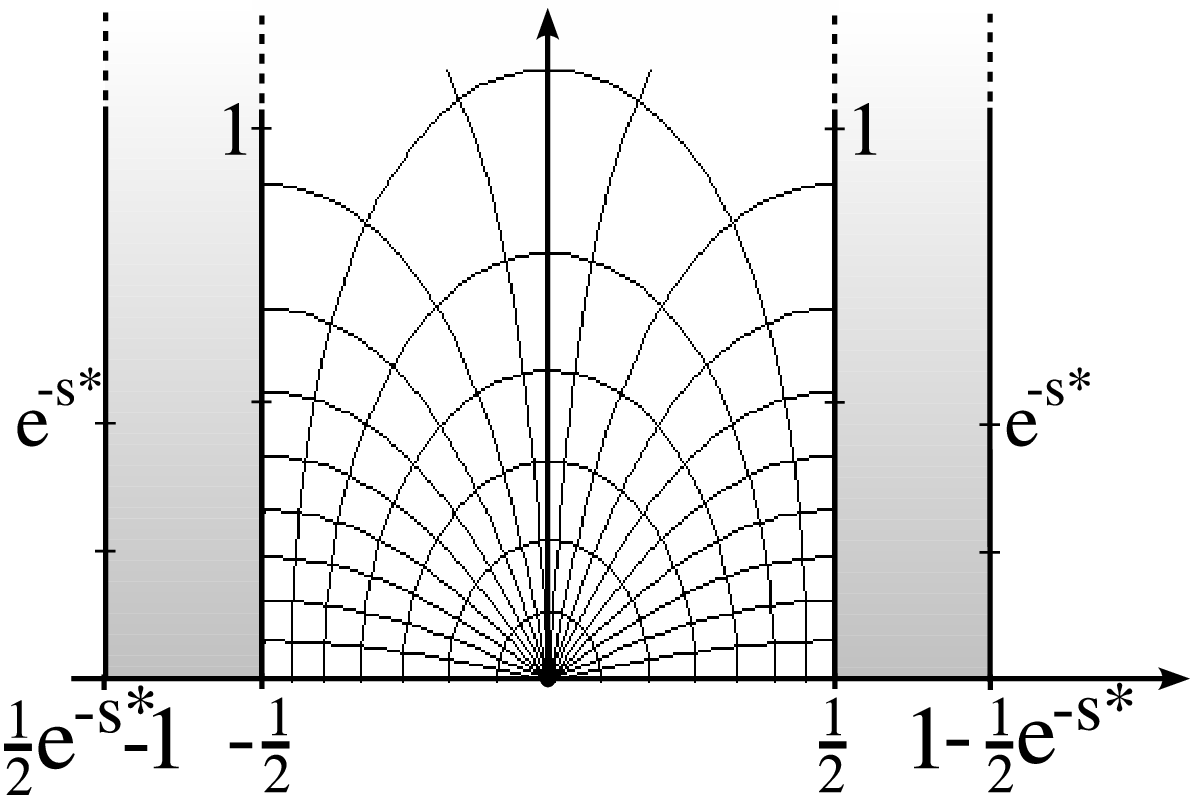, height=5cm}}
 \centerline{(a)\hskip 4.9cm (b) \hskip 5.65cm (c)\hskip 1.15cm}
\caption{(a) The action of $e^{-sL}$ on a Fock space state adds the shaded
strips on both sides of the Fock space state. The shaded strip on the right
represents $e^{-sL_R}$, the
 shaded strip on the left represents $e^{-sL_L}$. The dashed grey lines
illustrate the rescaling of the parameterizations from the inner to the outer boundaries.
(b) The action of $e^{-s^\star L^\star}$ on a Fock space state creates the
shaded strips which
 lie on top of
the Fock space state surface.  (c)
The action of $e^{-s^\star L^\star}$ on a Fock space state after flipping the
strips of the propagator.
}
\label{figemsL}
\end{figure}

Let us  first consider the action of $e^{-sL_R}$ on a general Fock space state $\ket{F}$. We represent $\ket{F}$ in the sliver frame $z$ as
the semi-infinite strip between $\Re(z)=-\half $ and $\Re(z)=\half $.
The operator insertion of the Fock space state is mapped to $z=0$ in the
sliver frame via the map
$z=\frac{2}{\pi}\tan^{-1} \xi$.
Recalling the discussion of
$\mathcal{R}(s)$ in Section~\ref{sectp},
we see that $e^{-sL_R}\ket{F}$ is
represented in the $z$ frame by gluing a strip of width
$\half (e^s-1)$  to the right boundary
of $\ket{F}$ (Figure~\ref{sl02fig}). The parametrization of the right boundary on the resulting
surface, however, has a scaling factor~$e^s$.
We conclude that $e^{-sL_R}$
attaches to the right of  $\ket{F}$ the slanted wedge
$[\half (e^s-1);e^s]$ (see Figure~\ref{figemsL}(a)).
Having determined that $e^{-sL_R}$ is represented by the right attachment
of the
slanted wedge $[\half (e^s-1);e^s]$, it follows that, more generally,
\begin{equation}\label{eslr}
 e^{-sL_R}\, [a;b]=[a;b]\ast [\half (e^s-1);e^s]\,.
\end{equation}
The slanted wedge $[a;b]$ has hidden boundaries and the action
of $e^{-sL_R}$ introduces an additional one that stems from
cutting the propagator surface $\RR(s)$. We have seen this
hidden boundary emerge through
$\lambda$-regularization as the line $\bar{Q} \bar{Q}'$ displayed
in Figure~\ref{sl05fig}.

Similarly,
 $e^{-sL_L}$ glues a strip of width
$\half (e^s-1)$ to the left boundary of $\ket{F}$. Now
the \emph{left} boundary of the resulting surface has a parametrization which
is scaled by $e^s$ (see Figure~\ref{figemsL}(a)). To interpret
this added piece of strip as a slanted wedge, we need to rescale
it by a factor $e^{-s}$ so that its left boundary has canonical
parameterization. We conclude that the action $e^{-sL_L}$
on $\ket{F}$ glues the slanted wedge
$[\half (1-e^{-s});e^{-s}]$ to the left boundary of $\ket{F}$. This generalizes to
\begin{equation}\label{esll}
 e^{-sL_L} \,[a;b]= [\half (1-e^{-s});e^{-s}]\ast[a;b]\,.
\end{equation}
It follows from (\ref{eslr}) and (\ref{esll}) that
\be
 e^{-s_RL_R}   e^{-s_LL_L} [a;b]  =   e^{-s_LL_L}  e^{-s_RL_R}[a;b]\,,
 \ee
for all values of $s_L$ and $s_R$.  We thus conclude that acting on
slanted wedges the operators $L_L$ and $L_R$ commute.
The action of $e^{-sL}$ on a given slanted wedge can
be calculated as follows
\begin{equation}
\label{eroivkjb}
\begin{split}
e^{-sL} [a; b] &=   e^{-s(L_L+L_R)} [a; b] =
e^{-sL_L}e^{-sL_R} [a;b]
\\[1.0ex] &
=  [\half (1-e^{-s});e^{-s}]\ast[a;b] \ast  [\half (e^s-1);e^s]\,,
\end{split}
\end{equation}
and therefore
\begin{equation}
\label{eroivkjbee}
e^{-sL} [a; b] = [\half (1+b) (1-e^{-s}) + a e^{-s} \,; b \,]\,.
\end{equation}
For the particularly important case of $b=1$, the above
reduces to
\begin{equation}
\boxed{\phantom{\biggl(}     e^{-sL} \,[a;1]=[1+(a-1)e^{-s}\,;\,1]\,.~}
\end{equation}
Since $[a;1]$ is a
 wedge state, this identity can be readily confirmed
by familiar methods (see eqn.~(A.27) of~\cite{0708.2591}).

\medskip
The geometric interpretation of
$e^{-s^\star L^\star}\ket{F}$ is somewhat more intricate.
In the construction of
$e^{-s L}\ket{F}$ we glue the {\em right}
boundary of
 the $w$-frame strip $\mathcal{R}(s)$
to the coordinate curve of~$\ket{F}$.
 This
 boundary of $\RR(s)$ is mapped by (\ref{zfromwaj})
 to the coordinate curve $\Re(z)=\pm\half$
 and the gluing to the Fock space state works out naturally,
 as shown in Figure~\ref{figemsL}(a).
It follows from the discussion in \S4.3 of~\cite{Kiermaier:2007jg}
that the surface corresponding to
$e^{-s^\star L^\star}$
can be obtained by gluing the {\em left} boundary of the $w$-frame
$\mathcal{R}(s^\star)$
to the coordinate curve of $\ket{F}$.
 This left boundary of $\RR(s^\star)$ is mapped
by (\ref{zfromwaj}) to $\Re(z)=\pm\half e^{s^\star}$
and the strip develops inwards up to $\Re(z)=\pm\half$.
 To glue the chosen boundary of $\RR(s^\star)$
 to the coordinate curve
 at $\Re(z)=\pm\half$, we rescale  $\RR(s^\star)$
 by $z\to e^{-s^\star}z$.
 The result,  illustrated in Figure~\ref{figemsL}(b),
is that  $e^{-s^\star L_R^\star}$
acting on the sliver-frame $\ket{F}$
 introduces
a strip between $\half e^{-s^\star}$ and $\half$,
 which is glued to the Fock space state at $\Re(z)=\half$.
In this presentation the surface added  by $e^{-s^\star L_R^\star}$
lies on top of the Fock state surface.
The  parameterization of the left
boundary of this added strip is shrunk by a factor of
$e^{-s^\star}$.
We can flip the
surface of $e^{-s^\star L_R^\star}$ around the axis
$\Re(z)=\half $ to obtain the result shown in Figure~\ref{figemsL}(c).
We have thus found that $e^{-s^\star L_R^\star}$ attaches
the left boundary of the slanted wedge
$[\half (1-e^{-s^\star});e^{-s^\star}]$ to the right
boundary of $\ket{F}$.  On general slanted wedges
\begin{equation}\label{eslstarr}
\begin{split}
    e^{-s^\star L^\star_R}
     \,[a;b]&=[a;b]\ast[\half (1-e^{-s^\star});e^{-s^\star}]\,.
\end{split}
\end{equation}
Similarly, we obtain
\begin{equation}\label{eslstarl}
       e^{-s^\star L^\star_L}\, [a;b]=
[\half (e^{s^\star}-1);e^{s^\star}]\ast[a;b]\,.
\end{equation}
We notice from~(\ref{eslr}), (\ref{esll}), (\ref{eslstarr}), and (\ref{eslstarl}) that the slanted wedges associated 
with $e^{-s^\star L^\star_L}$ and $e^{-s^\star L^\star_R}$ can be obtained from those 
associated with $e^{-s L_R}$ and $e^{-s L_L}$, respectively, by 
letting $s\to s^\star$.  
This is not a peculiarity of Schnabl gauge;  
 it follows because 
  the surface $\RR^\star(s^\star)$ generated
$e^{-s^\star L[v]^\star}$ can be obtained from the surface $\RR(s)$ generated by $e^{-s L[v]}$ from a reflection in 
the $w$ frame~\cite{Kiermaier:2007jg}. 

The action of $e^{-s^\star L^\star}$ on a given slanted wedge can be
calculated from
\begin{equation}
\label{eroivkjbjk}
e^{-s^\star L^\star} [a; b] \equiv  
e^{-s^\star L_L^\star}e^{-s^\star L_R^\star} [a;b]= [\half (e^{s^\star}-1);e^{s^\star}]\ast[a;b] \ast
[\half (1-e^{-s^\star});e^{-s^\star}]
\ee
and gives
\begin{equation}
\label{eroivkjbjk67}
e^{-s^\star L^\star} [a; b]
= [\half (1+b) (e^{s^\star}-1) + a e^{s^\star} \,; b \,]\,.
\end{equation}
For the case of surface states the above reduces to the identity
\begin{equation}
\boxed{\phantom{\biggl(}
     e^{-s^\star L^\star}\, [a;1]=[(1+a)e^{s^\star}-1;1] \,.~}
\end{equation}
that is readily confirmed
by familiar methods (see eqn.~(A.28) of~\cite{0708.2591}).

\medskip
{}From~(\ref{eslr}),~(\ref{esll}),~(\ref{eslstarr}), and~(\ref{eslstarl})
we find that the left and right parts of the
classical propagator act on a slanted wedge $[a;b]$ as
\begin{equation}\label{propLR}
\boxed{\begin{split}
     ~~\phantom{\biggl(} e^{-sL_R}e^{-s^\star L^\star_R}\,[a;b]
    &=[a;b]\ast [\half (1+e^{s-s^\star})-e^{-s^\star}\,;\,e^{s-s^\star}]\,,
    ~~\\%[0.5ex]
      e^{-sL_L}e^{-s^\star L^\star_L}\,[a;b]
    &= [\half (1+e^{s^\star-s})-e^{-s}\,;\,e^{s^\star-s}]\ast [a;b] \phantom{\biggl)}.
\end{split}}
\end{equation}
It now follows
that the action  of the classical propagator~(\ref{classprop1})
on a slanted wedge $[a;b]$ is given by:
\begin{equation}
\label{classprop}
  e^{-sL}e^{-s^\star
L^\star}[a;b]
   =[\, \half (1+b)(1+e^{s^\star-s}-2e^{-s})+a\, e^{s^\star-s}\,;\,b\,]\,.
\end{equation}
On
 wedge
states $[a;1]$, this simplifies to
\begin{equation}\label{proponwedge}
 \boxed{\phantom{\biggl(}   e^{-sL}e^{-s^\star L^\star}[a;1]
   =\bigl[1-2e^{-s}+(1+a)e^{s^\star-s}\,;\,1\bigr]\,.~}
\end{equation}

\medskip
As we have emphasized,  there are no  states
associated with  slanted wedges
 $[a;b]$
  for
 $b\neq 1$. Such surfaces are incomplete,
they have a \emph{hidden} vertical boundary segment at $i\infty$.
Since eventually no hidden boundary can remain,
 a surface $[a;b]$
with $b\neq 1$,  will have its hidden boundary glued to
that of a compensating surface $[\hat a; 1/b]$ with inverse slant
factor. This will be especially relevant
when we build general one-loop diagrams in Section~\ref{secloop}.
 There we construct compensating slanted
wedges
 $[a;e^{s_{\rm{eff}}}]$ and $[\hat a;e^{-s_{\rm{eff}}}]$
 which can then be mapped to the annulus.
For tree diagrams the situation is simpler: the
total surface representing the diagram is always
of the form $[a;1]$, with horizontal identifications
applied to the vertical edges.

\subsection{Keeping track of insertions on slanted wedges}
The open string moduli are encoded in
the positions of punctures on the corresponding Riemann surfaces.
As we will use slanted wedges to describe these surfaces,
we need to keep track of operator insertions on slanted wedges.
We denote by
\begin{equation}
[\,a ; b \,| x_1, x_2, \ldots ,  x_k ]  \,, \quad\hbox{with}
\quad  \half \leq  x_1 \leq x_2 \leq \ldots
\leq x_k \leq \half + a\,,
\end{equation}
a slanted wedge with marked points at real coordinates $x_i$. The
wedge is presented, as usual, with its left boundary above $z = \half$.
When star multiplying two slanted wedges, the position $x$ of a marked point on the first slanted wedge
is unaffected:
\begin{equation}
    [\,a ; b \,| x] \ast [\,a' ; b']  =  [a + b a'; bb' \, |\, x \,  ] \,.
\end{equation}
A puncture at $x'$ on the second wedge, on the other hand, is displaced  and experiences scaling:
\begin{equation}
\label{iwltsktnpplsfaj}
[\,a ; b \,] \ast [\,a' ; b' \,| x']  =  [a + b a'; bb' \, |\, \half + a + b (x'- \half) \,  ] \,.
\end{equation}
 From this
one can readily verify that
\begin{equation}
\label{iwltksscl}
\begin{split}
e^{-sL_L} [\,a ; b \,| x]  &=  [~ \ldots \,; \,\ldots \, \, | ~1- e^{-s} + e^{-s}x \,] \,, \\
e^{-sL_R} [\,a ; b \,| x]  &= [ ~ \ldots \,; \,\ldots \, \, | ~x \,]  \,, \\
e^{-s^\star L_L^\star} [\,a ; b \,| x]  &=  [~ \ldots  \,; \, \ldots \, \, | ~e^{s^\star}x \,] \,, \\
e^{-s^\star L_R^\star} [\,a ; b \,| x]  &= [~ \ldots  \,; \, \ldots \, \, | ~x \,]  \,.
\end{split}
\end{equation}
We put dots $\ldots$ on the width and slant parameters of the resulting wedges
 because these values are unaffected by the punctures and were given earlier.
Note that the exponentials of right operators do not affect the position
since they add a slanted wedge from the right.  It follows from the above
relations that
\begin{equation}
\label{iwltptmnstvmss}
e^{-s L} \, e^{-s^\star L^\star}   [\,a ; b \,| x]  =  [~ \ldots  \,; \, \ldots \, \, | ~
1- e^{-s} + e^{s^\star-s}x \,]
 \,.
\end{equation}
This formula generalizes easily to the case of additional punctures:
all $x_i \to 1- e^{-s} + e^{s^\star-s}x_i$.

\subsection{Representation of the $L$, $L^\star$ algebra on slanted wedges}

It is known that one can view  $L, L^\star,  L^+ = L+L^\star$ as well
as $L^+_L$ and $L^+_R$
as differential operators acting on the familiar wedge states. As we have
learned, the left and right parts of $L$ and $L^\star$ only act naturally
on slanted wedges (acting on
 an ordinary  wedge they will give a slanted wedge).
In this section we represent $L_L, L_R, L_L^\star$ and $L^\star_R$, as  differential operators on slanted wedges.  This provides some further insight into slanted wedges,
a check of this formalism and, as a by product, a tool to derive
(or rederive) some identities.

Let us focus on the right part of the $L$ and $L^\star$
operators. From~(\ref{eslr}) and~(\ref{eslstarr}) we have
\begin{equation}\label{Lrep}
\begin{split}
   L_R[a;b]&= -\frac{d}{\,ds\,}\Bigr|_{s=0~}e^{-sL_R~}[a;b]
    =(-\half b\del_a-b\del_b)[a;b]\,,\\
   L^\star_R[a;b]&= -\frac{d}{ds^\star}\Bigr|_{s^\star=0}e^{-s^\star
L^\star_R}[a;b]
   =(-\half b\del_a+b\del_b)[a;b]\,,
\end{split}
\end{equation}
and we identify the representation
\begin{equation}
   L_R=  -\half b\del_a-b\del_b\,,
   \qquad L^\star_R= -\half b\del_a+b\del_b\,.
\end{equation}
For the left counterparts a similar calculation gives
\begin{equation}
   L_L=  (a-\half)\del_a +b\del_b\,,
   \qquad L^\star_L= -(a+\half) \del_a-b\del_b\,.
\end{equation}
One can readily confirm that acting on slanted wedges the operators
$L_R$ and $L_L$ commute, and so do $L_R^\star$ and $L_L^\star$.
One can now recover the more familiar
differential operators
\begin{equation}
L = -(\half(1+b)-a ) \partial_a\,, ~~
L^\star = -(\half(1+b)+a ) \partial_a\,,~~ L^+ = L+ L^\star= -(1+b) \partial_a\,.
\end{equation}
It is now possible to calculate the commutator $[L, L^\star]$ by imagining
it acting on a wedge state,
\begin{equation}
[L, L^\star] = [ -(\half(1+b)+a ) \partial_a\,, -(\half(1+b)-a ) \partial_a]
= -(1+b) \partial_a =  L^+\,,
\end{equation}
which is the expected result.  In fact, even the right parts of $L,L^\star$,
and $L^+$ obey the
same equation, as one would expect,
\begin{equation}\label{LLstaralg}
   [L_R,L_R^\star] =
\bigl[-\half b\del_a+b\del_b\,,-\half b\del_a-b\del_b\bigr]=- b\del_a =
L_R^+ \,.
\end{equation}

\medskip
Let us illustrate how one derives identities using the above
representation of operators.  Note first  that
\begin{equation}
   L_R^+\equiv L_R+L^\star_R=
   -b\del_a\,.
\end{equation}
It then follows that
\begin{equation}
\label{iwltfksscl}
   e^{-s^+ L_R^+}[a;b] =[a+b s^+;b]  = [a;b]\ast[s^+;1]\,,
\end{equation}
showing that $e^{-s^+ L_R^+}$ acts by right multiplying a wedge
of length $s^+$.
Additionally, for wedge states
$[a;1]$
of width $a$
we  have
\begin{equation}
     -L_R^+[a;1]=\del_a[a;1]\,, \qquad
[a;1]=e^{-a L^+_R}\,[0;1]\,,
\end{equation}
where the zero-length wedge
$[0;1]$ is the identity state $\ket{\mathcal{I}}$.
Another example uses
(\ref{propLR})
and (\ref{iwltfksscl}):
\begin{equation}
\label{iwltfkvm}
      e^{-tL_R}e^{-t L^\star_R}\,[a;b]
      =[a;b]\ast [1-e^{-t}\,;\,1]
      =e^{-(1-e^{-t})L_R^+}\, [a;b]\,,
\end{equation}
leading us to conclude that
$     e^{-tL_R}e^{-t L^\star_R}= e^{-(1-e^{-t})L_R^+}\,.$

\section{Riemann surfaces for tree-level diagrams}\label{sectree}
\setcounter{equation}{0}
In this section we will use the technology of slanted wedges developed
in the previous section to construct the punctured disks associated with tree diagrams.  We start with the particularly simple
case of a diagram with  five external lines. We then sketch the construction for
arbitrary tree-level diagrams.

\subsection{The five-point diagram}

\begin{figure}
\begin{center}
\epsfig{figure=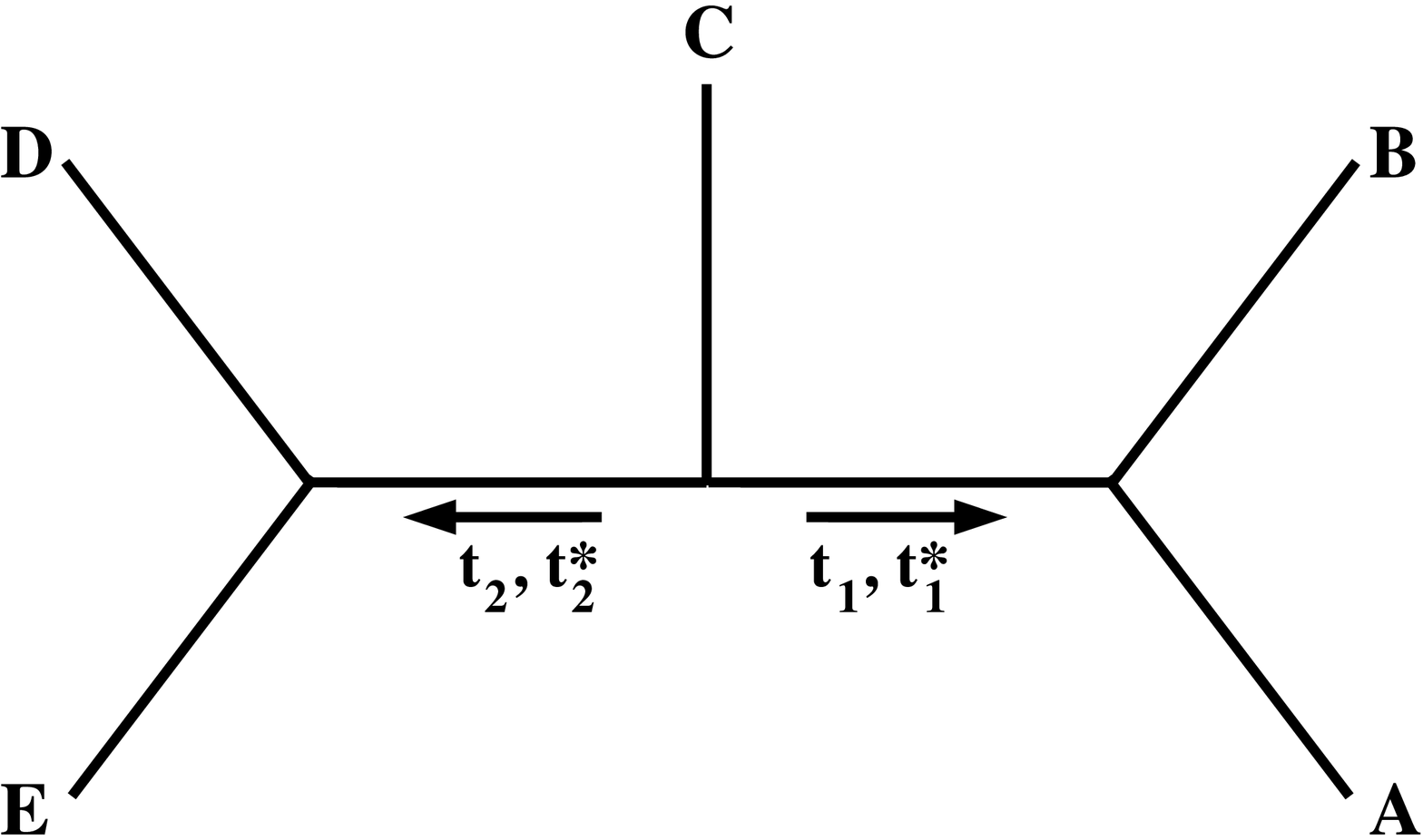, height=5cm}
\hskip 1.5cm
\epsfig{figure=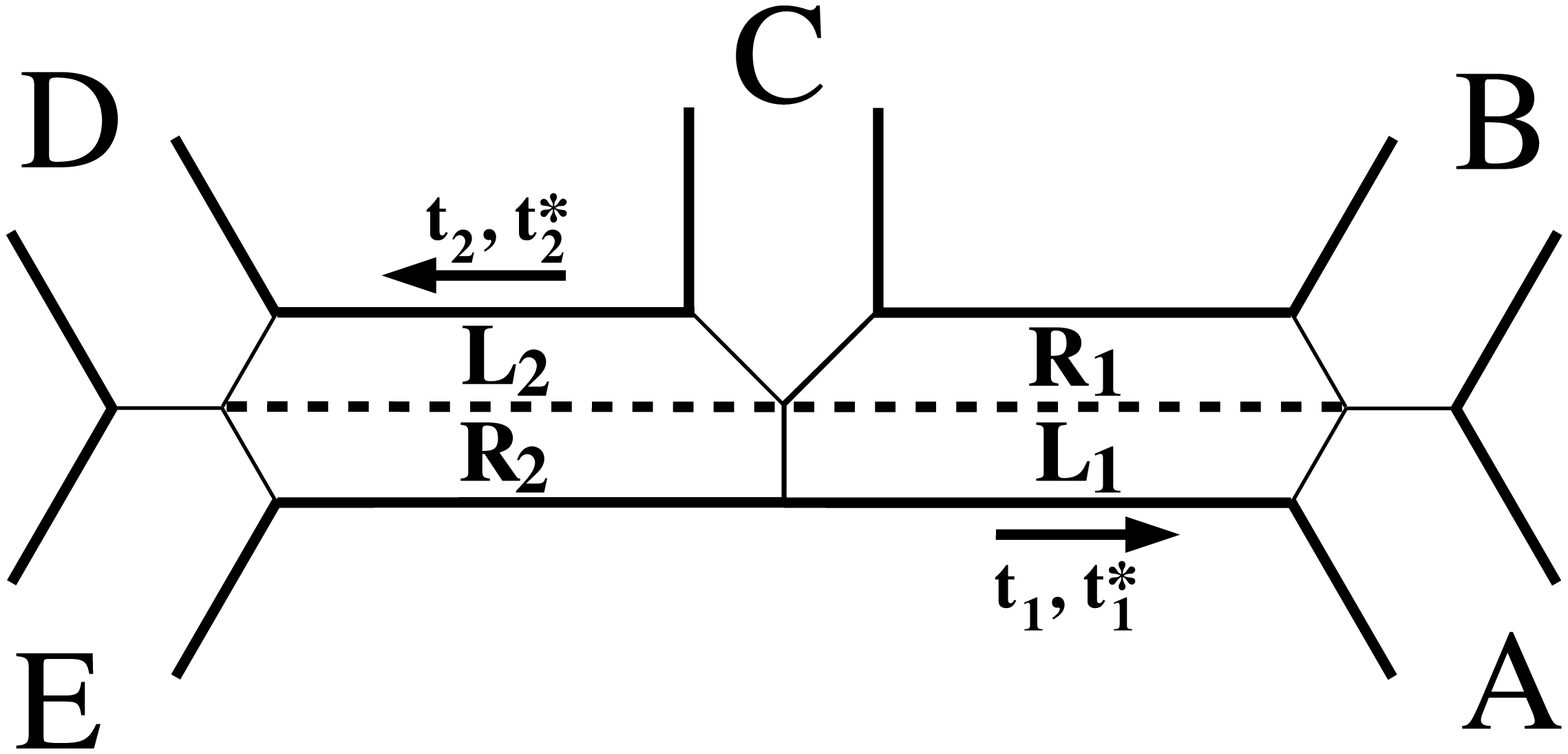, height=4cm}
\end{center}
\centerline{\hskip1cm(a)\hskip 8cm(b)}
\caption{
(a) A Feynman diagram for a five-point amplitude. (b) The topology of the corresponding Riemann surface.
}
\label{fig5point}
\end{figure}

Let us use the formalism of Section~\ref{secsw} to construct the surface corresponding to the
tree-level five-point diagram shown in
Figure~\ref{fig5point}. Our goal is to determine the relative angles of the operator insertions on the unit disk.  These are the open string
moduli.  There are, of course, no closed string moduli.
All other
diagrams contributing to the five point
function are permutations of the external states in the  diagram of
Figure~\ref{fig5point}.

\medskip
The diagram contains two internal propagators, parameterized by the
Schwinger parameters $t_i, t^\star_i$ with $i=1,2$.
    Here and in the
following we use the letter $t$ for Schwinger parameters of propagators in tree diagrams (or subtrees of
loop diagrams).
As this is a tree-level diagram, the classical propagator~(\ref{classprop1}) must be used on both lines.
We use arrows
to assign a direction to each propagator line in the diagram.
At fixed Schwinger parameters
the classical propagator on the $i$-th line
acts as the
operator $e^{-t_iL}BQB^\star e^{-t_i^\star L^\star}$ in the indicated
direction, {\em i.e.} on the state representing the surface in the
direction of
the arrow.  Equivalently, it acts as the BPZ conjugate operator $e^{-t_i^
\star L}BQB^\star e^{-t_i L^\star}$ in the direction opposite to the
arrow. Since the full
 propagator is BPZ invariant, amplitudes do not depend on
this assignment after
integration over Schwinger parameters.
The selection of specific arrows
 is simply a convention that fixes which Schwinger
parameter we call $t_i$ and which one we call~$t_i^\star$.

Let us consider the  part of the  diagram consisting of the
first propagator ($t_1$,$t^\star_1$) and the Fock
space states $\ket{F_A}$, $\ket{F_B}$. Each Fock space state is
of the form $[1;1|1]$. Together, and
 acted by
the propagator, they
form the twice-punctured surface state
\begin{equation}
\label{iwltssprmdrptvmpss}
\begin{split}
  [a_1;1\,|~x_A, x_B]&
  \equiv e^{-t_1L}e^{-t_1^\star L^\star}\bigl([1; 1|1]_A\ast [1;
1\,|1]_B\bigr) \\
&=  e^{-t_1L}e^{-t_1^\star L^\star}\,[2; 1|1,2]\\
 &  =[1-2e^{-t_1}+3e^{t_1^\star-t_1};1\,|\,x_A,x_B]\,\,,\\[1.0ex]
 \hbox{with} \phantom{jjkkkj} x_A&=1-e^{-t_1}+ e^{t_1^\star-t_1}  \\[1.0ex]
   x_B&=1-e^{-t_1}+2 e^{t_1^\star-t_1} \,,
\end{split}
\end{equation}
where we used~(\ref{proponwedge}) to calculate the wedge parameters
and (\ref{iwltptmnstvmss}) to calculate the positions $x_A$ and $x_B$
of the punctures on the resulting wedge.
Similarly, we can analyze the part of
 the diagram with the second propagator
($t_2$,$t^\star_2$) and the
Fock space states $\ket{F_D}$, $\ket{F_E}$. They form the surface state
\begin{equation}
\label{iwltsmllychrlppssy}
\begin{split}
  [a_2;1\,|~ x_D , x_E]
  \equiv e^{-t_2L}e^{-t_2^\star L^\star}\bigl([1; 1|1]_D\ast [1; 1|1]_E\bigr)=
  [1-2e^{-t_2}+3e^{t_2^\star-t_2};1|x_D, x_E\,]\,,
\end{split}
\end{equation}
with the operator insertions corresponding to  $\ket{F_D}$ and $\ket{F_E}$ located at
\begin{equation}
\begin{split}
   x_D&=1-e^{-t_2}+ e^{t_2^\star-t_2}\,,\\
   x_E&=1-e^{-t_2}+2 e^{t_2^\star-t_2} \,.
\end{split}
\end{equation}

To assemble the Riemann surface corresponding to the five-point diagram we
 glue the surfaces $[a_1;1|x_A, x_B]$, $[1; 1|1]_C$ (corresponding to
 $\ket{F_C}$) and $[a_2;1|x_D, x_E]$. We obtain the surface $\Sigma$ given by
\begin{equation}
\label{ylchrlpans}
\begin{split}
   \Sigma&\equiv
   [a_1;1|x_A, x_B]\ast [1; 1|1]_C \ast [a_2;1|\,x_D, x_E]\\[1.0ex]
   &=[a_1+a_2+1;1|\, x_A,\, x_B,\, \,a_1+1,\,\, a_1 + 1 + x_D, \,
   a_1 + 1 + x_E]\,.
   \end{split}
\end{equation}
 In particular, we notice that the wedge $\Sigma$ is not slanted.
The two vertical boundaries of $\Sigma$
 are thus identified horizontally
and the resulting surface is mapped to
the  unit disk $\eta$ via
\begin{equation}\label{diskframe}
    \eta=\exp\left({\frac{2\pi iz}{a}}\right)\,, \quad  a = a_1 + a_2 + 1\,.
\end{equation}
A horizontal distance $\Delta x$ along the boundary of $\Sigma$ translates into an angular separation $\Delta \phi$ on the unit disk given by
\begin{equation}
    \frac{\Delta \phi}{2\pi}=\frac{\Delta x}{a}\,.
\end{equation}
For the relative angles of the operator insertions on the boundary of the unit
 disk
we thus obtain
\begin{equation}\label{angles5point}
\begin{split}
   \frac{\phi_B-\phi_A}{2\pi}  =\qquad\frac{x_B-x_A}{a}\qquad
   &=
\frac{e^{t_1^\star-t_1}}{3-2e^{-t_1}-2e^{-t_2}+3e^{t_1^\star-t_1}+3e^{t_2^\star-t_2}
}\,,\\
   \frac{\phi_C-\phi_A}{2\pi}  =\quad\frac{a_1+1-x_A}{a}~\quad
   &=
\frac{1-e^{-t_1}+2e^{t_1^\star-t_1}}{3-2e^{-t_1}-2e^{-t_2}+3e^{t_1^\star-t_1}+3e^{t_2^\star-t_2}
}\,,\\
   \frac{\phi_D-\phi_A}{2\pi}  =\frac{a_1+1+x_D-x_A}{a}
   &=
\frac{2-e^{-t_1}+2e^{t_1^\star-t_1}-e^{-t_2}+e^{t_2^\star-t_2}}{3-2e^{-t_1}-2e^{-t_2}+3e^{t_1^\star-t_1}+3e^{t_2^\star-t_2}
}\,,\\
   \frac{\phi_E-\phi_A}{2\pi}  =\frac{a_1+1+x_E-x_A}{a}
   &=
\frac{2-e^{-t_1}+2e^{t_1^\star-t_1}-e^{-t_2}+2e^{t_2^\star-t_2}}{3-2e^{-t_1}-2e^{-t_2}+3e^{t_1^\star-t_1}+3e^{t_2^\star-t_2}
}\,.
\end{split}
\end{equation}
This concludes the computation of angles for the string diagram
in Figure~\ref{fig5point}.  Of course, the positions of three of the punctures
can be fixed arbitrarily, so there are just two open string moduli.
As usual for amplitudes in non BPZ-invariant gauges, we have twice as many
Schwinger parameters as moduli of the corresponding
Riemann surface.
Indeed,  we have four Schwinger parameters ($t_1$,
$t_1^\star$, $t_2$, $t_2^\star$).
This is not  a problem because each of the two propagators is
accompanied by a BRST operator $Q$.
In~\cite{0708.2591,Fuji:2006me}, the classical propagator~(\ref{classprop1}) was rewritten~as
\begin{equation}
   {\cal P} = \frac{B}{L} +\textrm{ other terms }\,,
\end{equation}
and it turned out that the $B/L$ term by itself covered the moduli space of
on-shell amplitudes for the four-point function.
All other terms only contributed off-shell.

It is therefore interesting to ask if there is  an assignment of $B/L$ and
$B^\star/L^\star$ to the
 propagator lines
in the five-point diagram
which produces all the requisite open string degenerations:
as a Schwinger parameter becomes large the associated line
produces the degeneration represented by a long strip.
That degeneration, moreover,  must occur independent of the
values of the other Schwinger parameters, even if they also go
to infinity.
Not every assignment works. If we assign $B^\star/L^\star$ to {\em both} propagators (namely, $t_1=t_2=0$)
making one Schwinger parameter large is
not sufficient to guarantee an open string degeneration.
 In fact, for $t_1^\star=t^\star_2\to\infty$
  the angles of the five insertions on the unit circle
 spread out over the circle,
a configuration that is clearly not degenerate.
This is not too surprising. If we had regarded the left propagator as
acting to the right, we would have encountered the operator combination
$e^{-t_2^\star L_L}e^{-t_1^\star L_L^\star}$ acting on the Fock space state
$\ket{F_A}$. This product
of operators has been noticed to produce
interesting subtleties in~\cite{0708.2591}. Although both Schwinger parameters
in the operator diverge, the resulting Riemann surface
is perfectly regular.
 Indeed, acting on any slanted wedge we have
  from~(\ref{propLR}):
\begin{equation}
\begin{split}
  \lim_{t_1^\star=t^\star_2\to\infty}e^{-t_2^\star L_L}e^{-t_1^\star L_L^\star}[a;b]
& = \lim_{t_1^\star=t^\star_2\to\infty}
 [\half  + \half
e^{t_1^\star - t_2^\star}- e^{-t_2^\star} \,; e^{t_1^\star - t_2^\star} ] \ast [a;b] = [1;1]\ast [a;b]\, \\
&= [a+1;b] \,.
\end{split}
\end{equation}
In this limit the operator  simply inserts the unit wedge $[1;1]$.
This is a surface of finite width and finite rescaling and  cannot
induce an open string degeneration in any diagram. For all other choices of assignments of $B/L$ and $B^\star/L^\star$ to the two
propagator lines, open
string degenerations are always produced when we make any Schwinger parameter large. Details of this analysis are given in
appendix~\ref{app5pt}.

\subsection{General tree diagrams}
The construction of the surface for the five-string diagram was particularly simple. For general
tree-level diagrams we need to be more systematic. As we did for the
five-string diagram we assign an arrow to each
propagator, indicating the direction in which it acts. This assignment
is arbitrary and will not affect the total set of surfaces created as the
Schwinger parameters vary over their full range because the propagator is BPZ-invariant.

We now rewrite the five-string diagram in a way that makes the case for
the general rules to be stated below.
Let us revisit
the surface considered in (\ref{iwltssprmdrptvmpss}):
\begin{equation}
 e^{-t_1L}e^{-t_1^\star L^\star}\bigl([1; 1|1]_A\ast [1;
1\,|1]_B\bigr) =  e^{-t_1L_L}
 e^{-t_1^\star L^\star_L}e^{-t_1L_R}
 e^{-t_1^\star L^\star_R}\bigl([1; 1|1]_A\ast [1;
1\,|1]_B\bigr)\,.
\end{equation}
Recalling~(\ref{propLR}), we then find
\begin{equation}
\begin{split}
 &e^{-t_1L}e^{-t_1^\star L^\star}\bigl([1; 1|1]_A\ast [1;1\,|1]_B\bigr)\\
    \quad &= [\half (1+e^{t_1^\star-t_1})-e^{-t_1}\,;\,e^{t_1^\star-t_1}]
    \ast[1; 1|1]_A\ast [1;1\,|1]_B\ast
    [\half (1+e^{t_1-t_1^\star})-e^{-t_1^\star}\,;\,e^{t_1-t_1^\star}]\\[1.0ex]
    \quad &= L_1 \ast  [1; 1|1]_A\ast [1;1\,|1]_B\ast  R_1\,,
\end{split}
\end{equation}
where we have defined
 the slanted wedges $L_i$ and $R_i$ associated with the left and right part of the $i$-th
 propagator:
\begin{equation}\label{LiRitree}
  L_i \equiv [\half (1+e^{t_i^\star-t_i})-e^{-t_i}\,;\,e^{t_i^\star-t_i}]\,,\qquad
  R_i \equiv [\half (1+e^{t_i-t_i^\star})-e^{-t_i^\star}\,;\,e^{t_i-t_i^\star}]\,.
\end{equation}
It is now clear that  (\ref{iwltsmllychrlppssy}) becomes:
\begin{equation}
e^{-t_2L}e^{-t_2^\star L^\star}\bigl([1; 1|1]_D\ast [1; 1|1]_E\bigr)=
L_2 \ast  [1; 1|1]_D\ast [1;
1\,|1]_E\ast  R_2\,.
\end{equation}
Assembling now the full surface $\Sigma$ as in (\ref{ylchrlpans}) we have
\begin{equation}
\Sigma=L_1 \ast  [1; 1|1]_A\ast [1;
1\,|1]_B\ast  R_1\ast  [1;1|1]_C\ast L_2 \ast  [1; 1|1]_D\ast [1;
1\,|1]_E\ast  R_2\,.
\end{equation}
Note that $\Sigma$ is a wedge of unit slant factor (an ordinary
wedge) because the slant factors of $L_i$ and $R_i$
are multiplicative inverses of each other.
Since the right and left edges of $\Sigma$ are to be identified, we can
slide part of the wedge state from left to right, cyclically.  We write, for
convenience,
\begin{equation}
\label{ylchrlpsttsrgrt}
\Sigma=[1; 1|1]_A\ast [1;1\,|1]_B\ast  R_1\ast  [1;1|1]_C\ast L_2 \ast  [1; 1|1]_D\ast [1;1\,|1]_E\ast  R_2\ast L_1\,.
\end{equation}
The rule for building
 the surface $\Sigma$ for general tree diagrams
is now clear:  {\em Begin at some external
state and trace around the diagram in the counterclockwise
direction.
 For each external state add the factor $[1;1|1]$.
For each line $i$ traversed against the propagator arrow
add a factor $R_i$. For each line $i$ traversed along the propagator arrow add a factor $L_i$.
All factors are added from the right.}
The formula in (\ref{ylchrlpsttsrgrt}) results from the application
of this rule to the diagram in Figure~\ref{fig5point},
 starting at the external state $\ket{F_A}$.
Note that the
rules build the surface using half strings.
If we are tracing in the direction of the arrow on the $i$ line, we multiply
by the surface $L_i$
because the propagator acts from the left on the surface to be built. If we are tracing against the arrow of the propagator, we
multiply by the surface
 $R_i$  because, in this case, the propagator acts from the right on the surface we have already built.

For the more complicated diagram in Figure~\ref{figtracetree} the rules are still simple to follow. Although it is redundant information, the labels
 $L_i$ and $R_i$
in Figure~\ref{figtracetree} represent the factors that must be included, as a result of the chosen
assignment of arrows, when
we follow the grey dotted curve along the diagram.
We have,
\begin{figure}
\centerline{
\epsfig{figure=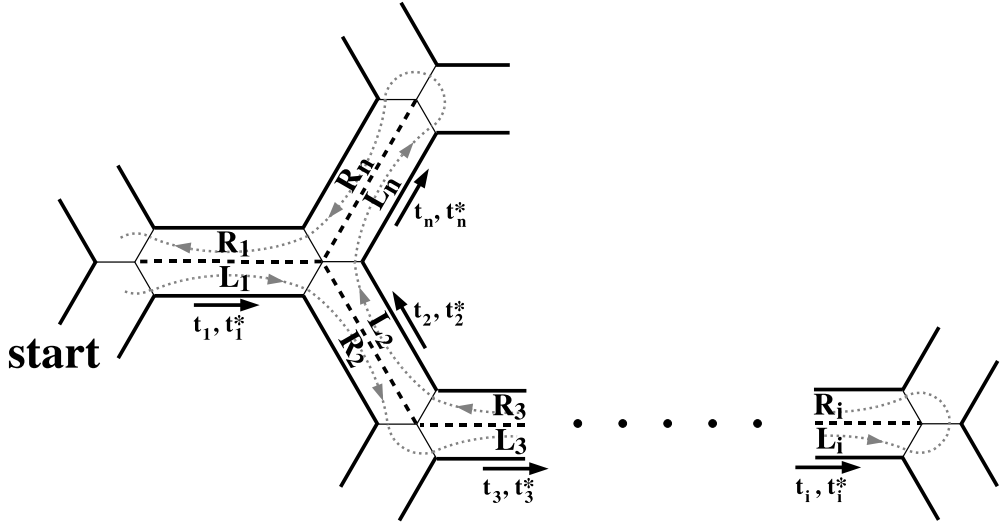, height=7cm}}
\caption{The surface $\Sigma$ for general tree diagrams is built by tracing the grey dotted line counterclockwise around the
diagram, starting at an arbitrary external state. When tracing along
 the $i$-th
propagator one picks up either the slanted wedge corresponding
to the
 left part ($L_i$) or right part ($R_i$)
of the
propagator. This depends on the direction of the propagator (black arrows). }
\label{figtracetree}
\end{figure}
\begin{equation}
\begin{split}
\label{iwltstckmnsnvmpssy}
\Sigma& =  [1;1|1]\ast L_1 \ast R_2 \ast[1;1|1] \ast L_3 \ast \ldots \ast L_i \ast [2;1|1,2]\ast R_i \ast \ldots \\[1.0ex]
&~~~\ldots \ast R_3\ast L_2 \ast L_n \ast [2;1|1,2] \ast R_n \ast R_1 \ast [1;1|1]\,.
\end{split}
\end{equation}
The resulting surface is of unit slant (for each $L_i$ there is an
$R_i$) and takes
the form
\begin{equation}
   \Sigma = [a;1\,|\, x_1, x_2, \ldots x_k\,]\,,
\end{equation}
for some calculable width $a$ and some calculable positions $x_i$.
The left and right boundaries of $\Sigma$ are identified
through translation by $a$.  We can therefore map the glued $\Sigma$
surface to a unit disk using
$\eta=\exp(2\pi iz/a)$.
To determine the moduli of the surface, we only need to know the angular separation  between operator
insertions on the unit circle. If insertions are separated by $\Delta x$ on $\Sigma$, their relative angle $\Delta\phi$ on the unit circle
is simply given by
\begin{equation}
  \frac{ \Delta\phi}{2\pi}  =\frac{\Delta x}{a} \,.
\end{equation}
This concludes our discussion of Riemann surfaces for general tree-level diagrams.

\section{Riemann surfaces for general one-loop diagrams}\label{secloop}
\setcounter{equation}{0}

In Section~\ref{sectp} we built the surface corresponding to the tadpole
diagram in Schnabl gauge using the simplified propagator $B/L$ (i.e. $s^\star=0$).
Using the $z$ frame we  built separately the parts of the surface that
contain the inner and outer boundary components of the annulus,
 as displayed in Figure~\ref{sl04fig}(b).
Let us now review this construction using the algebra of slanted wedges.

On the outer boundary
 (the right strip in the $z$ frame)
there is a Fock space state $[1;1|1]$. It is acted
by the right part $e^{-sL_R}$ of the propagator
so we get a slanted wedge $\Sigma$ given by
\begin{equation}
    \Sigma = e^{-sL_R} [1;1|1] = [\half (1+e^{s});e^{s}|1\,]\,,
\end{equation}
where we made use of (\ref{eslr}) and (\ref{iwltksscl}).
On the inner boundary there is only the remaining left part $e^{-sL_L}$ of the propagator so the resulting slanted wedge $\widehat \Sigma$ is just
\begin{equation}
     \widehat\Sigma\equiv[\half (1-e^{-s});e^{-s}]\,,
\end{equation}
making use of (\ref{esll}).
The slanted wedges  $\Sigma$ and $\widehat\Sigma$
are glued to each other at their
hidden boundaries at $i\infty$, as discussed
 in Section~\ref{sectp} using $\lambda$-regularization.

We need to place the surfaces $\Sigma$ and $\widehat \Sigma$ in the $z$-plane in
such a way that:  (i) their hidden boundaries at $i\infty$ glue correctly,
and
(ii)  the slanted identifications
become translations in
  the $w$ frame ($z=-\half e^{-w}$). We refer to the result
as the natural $w$-picture.
For the identifications on $\Sigma$ to be simple
translations in the $w$ frame, we
shift $\Sigma$ horizontally, so that
the position of its right
 boundary is a rescaling by $e^{s}$ of its left boundary. The translation is uniquely
 determined and $\Sigma$ lands between the vertical lines
 given
 in (\ref{iwltfkchlw}).
 Similarly, we need to shift the position of $\widehat\Sigma$ in such a way
  that the position of its right
 boundary is a rescaling by $e^{-s}$ of its left boundary. As $e^{-s}<1$, this
implies that we need to position $\widehat\Sigma$ in the region $\Re (z) <0$.
In fact, we readily see that  $\widehat\Sigma$ must be placed as the region between $\Re(z)=-\half $ and
$\Re(z)=-\half e^{-s}$.

 But we are not done yet. The boundaries at $i\infty$ now do \emph{not} glue
correctly.
 By the definition of slanted wedges, the parameterizations of
the left boundaries of
 $\Sigma$ and $\widehat\Sigma$ match -- indeed, they both
 have unit scaling factor.
 But for the hidden boundaries at $i\infty$ to glue nicely in the $w$ frame,
the parameterizations of the left boundary of $\Sigma$ needs to match the
parametrization of the \emph{right} boundary
 of $\widehat\Sigma$.\footnote{This requirement will lead to the established
 result for this diagram, but will be justified in more generality using $\lambda$-regularization in Section~\ref{sechidden}.} We can achieve this simply by rescaling
 the shifted
$\widehat\Sigma$
by $e^s$. Then $\widehat\Sigma$ is positioned
 between $\Re(z)=-\half e^{s}$ and $\Re(z)=-\half $. This is precisely the
configuration of surfaces that we ended up with and fully justified
in Section~\ref{sectp}.

The above steps can easily be
generalized to one-loop diagrams
 of arbitrary complexity. We will now show
how this is done.
 A detailed justification of the procedure is given in Section~\ref{aregvieononelooam}, where
 we discuss the $\lambda$-regulation explicitly.

\subsection{The natural $w$-picture}\label{naturalw}

For a one-loop diagram,
we construct two complementary  surfaces
\begin{equation}
\label{iwltfkvkmt}
\Sigma\equiv[a;e^{s_{\rm{eff}}}|
 \,\vec{x}\,]
\quad\hbox{and}\quad
\widehat{\Sigma}\equiv[\hat a; e^{-s_{\rm{eff}}}\,|
 \,\vec{\hat x}\,
]\,,
\end{equation}
 where $\vec{x}$ and $\vec{\hat x}$ collectively represent the positions of all punctures
 on $\Sigma$ and $\widehat \Sigma$, respectively.
These surfaces are said to be complementary because their scaling factors
multiply to one.
 On each surface, the left and
the right boundaries are identified,
and the two surfaces are glued to each other at their hidden boundaries 
at $i\infty$.

The \emph{natural $w$ picture} is one in which
the hidden boundaries of $\Sigma$ and $\widehat{\Sigma}$ glue
nicely and the identifications on $\Sigma$ and $\widehat \Sigma$ are
horizontal translations by $s_{\rm{eff}}$.
To obtain this picture
we need to place the surfaces $\Sigma$
and $\widehat\Sigma$ correctly in
 the $z$ frame.
First, we shift the surfaces $\Sigma$ and
$\widehat \Sigma$
by real constants $a_0$ and $\hat a_0$, respectively, so that the position of
their right boundaries is
a rescaling of the left boundaries by $e^{s_{\rm{eff}}}$ and $e^{-s_{\rm{eff}}}$,
respectively. Recall that by definition all
slanted wedges start out with their left boundary at $\Re(z)=\half $. The
required shifts are thus determined by the relations
\begin{equation}
    e^{s_{\rm{eff}}}=\frac{a_0+a+\half }{a_0+\half }\,,\qquad e^{-s_{\rm{eff}}}=\frac{\hat
a_0+\hat a+\half }{\hat a_0+\half }\,.
\end{equation}
Thus
\begin{equation}\label{defa0}
   a_0
   =\frac{a}{e^{s_{\rm{eff}}}-1}-\half\,, \qquad
   \hat a_0
=-\frac{\hat a }{1-e^{-s_{\rm{eff}}}}-\half \,.
\end{equation}
 This shift places the surface $\Sigma$ at
 \begin{equation}\label{sigmaregion}
  \hbox{Final}~ \Sigma ~
  \hbox{region:}~~~ a_0 + \half \leq \Re (z) \leq  e^{s_{\rm{eff}}}(a_0 + \half) \,.
 \end{equation}
As we discussed for the tadpole example above,
we then rescale  $\widehat{\Sigma}$ by a factor of
$e^{s_{\rm{eff}}}$ so that it has the
 canonical parametrization on its \emph{right} boundary.
 With this scaling $\widehat\Sigma$ ends up 
 in the location 
 \begin{equation}\label{hatsigmaregion}
\hbox{Final}~ \widehat\Sigma ~
 \hbox{region:}~~~ e^{s_{\rm{eff}}} (\hat{a}_0 + \half) \leq \Re (z) \leq 
  \hat{a}_0 + \half  \,.
 \end{equation}
 After positioning $\Sigma$ and $\widehat \Sigma$ in this way we
 map to the
 $w$ frame via (\ref{wzrel})
and to the annulus frame $\zeta$ via
\begin{equation}\label{zetadef}
   \zeta=e^{-\frac{2\pi i}{s_{\rm{eff}}} (w-i\pi)} \,  .
\end{equation}
The modulus $M$ of the annulus was defined in~(\ref{defM}).
We can read it off from~(\ref{zetadef}) as
\begin{equation}\label{annulusmodulus}
    M=\frac{\pi}{|s_{\rm{eff}}|}\,.
\end{equation}
 A point $x$ on $\Sigma$ with $\half \leq x\leq \half +a$ is located
 at $z =  x+ a_0$ in the shifted $\Sigma$ region~(\ref{sigmaregion}). 
 Using (\ref{iwltktcntaj99}), we see that it
ends on  a boundary of the annulus
 at an  angle
\begin{equation}
\label{389eiu}
   \phi=\frac{2\pi}{s_{\rm{eff}}}\ln (2|a_0+x|)=\frac{2\pi}{s_{\rm{eff}}}\ln
\Bigl( 2\Bigl|x-\half+\frac{a}{e^{s_{\rm{eff}}}-1}\Bigr|\Bigr)\,.
\end{equation}
Similarly, any
point $\hat x$ on $\widehat{\Sigma}$ is located  
at $z = e^{s_{\rm{eff}}}
( \hat x + \hat{a}_0)$ 
 in the final $\widehat\Sigma$ region~(\ref{hatsigmaregion})
and lands at
 an angle
\begin{equation}
\label{3948ufd}
     \hat\phi= \frac{2\pi}{s_{\rm{eff}}}\ln(2e^{s_{\rm{eff}}}|\hat x+\hat
a_0|)=\frac{2\pi}{s_{\rm{eff}}}\ln \Bigl( 2\Bigl|\half-\hat x+\frac{\hat a
}{1-e^{-s_{\rm{eff}}}}\Bigr|\Bigr)\,.
\end{equation}
All points in $\vec{x}$ land on the same boundary component and all points
in $\vec{\hat x}$ land on the other boundary component.  With the maps
written here, if $s_{\rm{eff}}>0$
the points in $\vec{x}$ lie on the outer component and if  $s_{\rm{eff}}<0$
they lie on the inner component. Of course, there is no invariant
distinction between these components as they can be exchanged by
a conformal map.

\subsection{General one-loop diagrams}
Let us now build the complementary
surfaces $\Sigma$ and $\widehat\Sigma$ in (\ref{iwltfkvkmt})
for a general
one-loop diagram.
Let
$n$ be the number of propagators running in the loop.
It follows that there are also $n$ vertices
within the loop. These  vertices will be labeled $1$ to $n$
as we move counterclockwise in the loop (see Figure~\ref{figloopn}(a)).
Two lines of each cubic vertex in the loop connect to loop propagators.
The remaining line can lead to a single external state
or to a subtree diagram with a set of external states.
In either case, the additional external state(s) at this vertex are all on one
specific boundary of the annulus.
We let $\Sigma$  represent the part of the  diagram which is
drawn on the outer side of the loop and
 $\widehat\Sigma$  represent the part of the diagram on the inside
of the loop.
Eventually, we will glue the surfaces $\Sigma$ and $\widehat\Sigma$ along their
hidden boundaries, shown
as dashed lines in Figure~\ref{figloopn}(b).
In Section~\ref{naturalw} we have  
learned how to perform this gluing.
\begin{figure}
\centerline{
\epsfig{figure=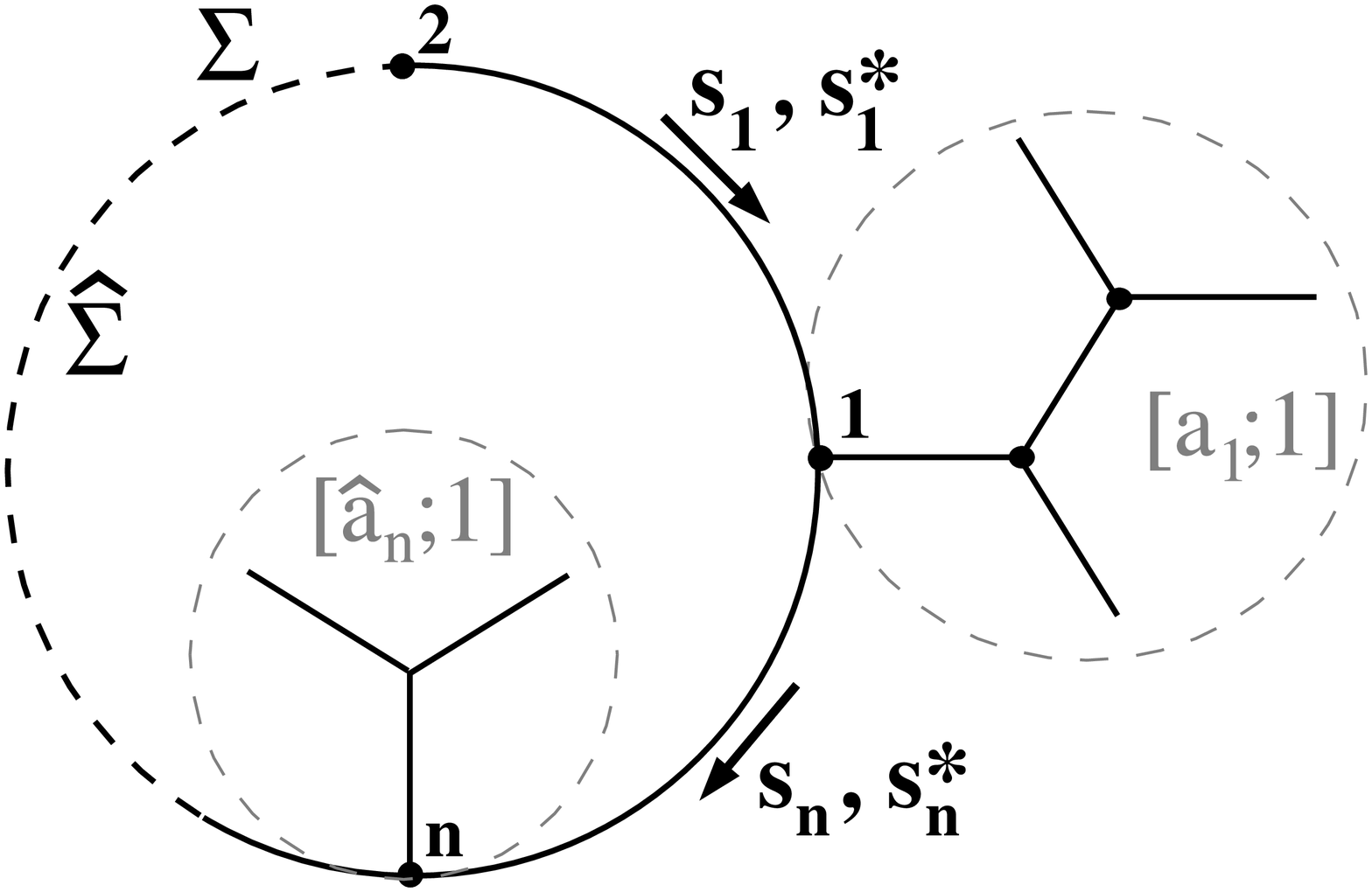, height=6.5cm}\hskip 3.5cm \epsfig{figure=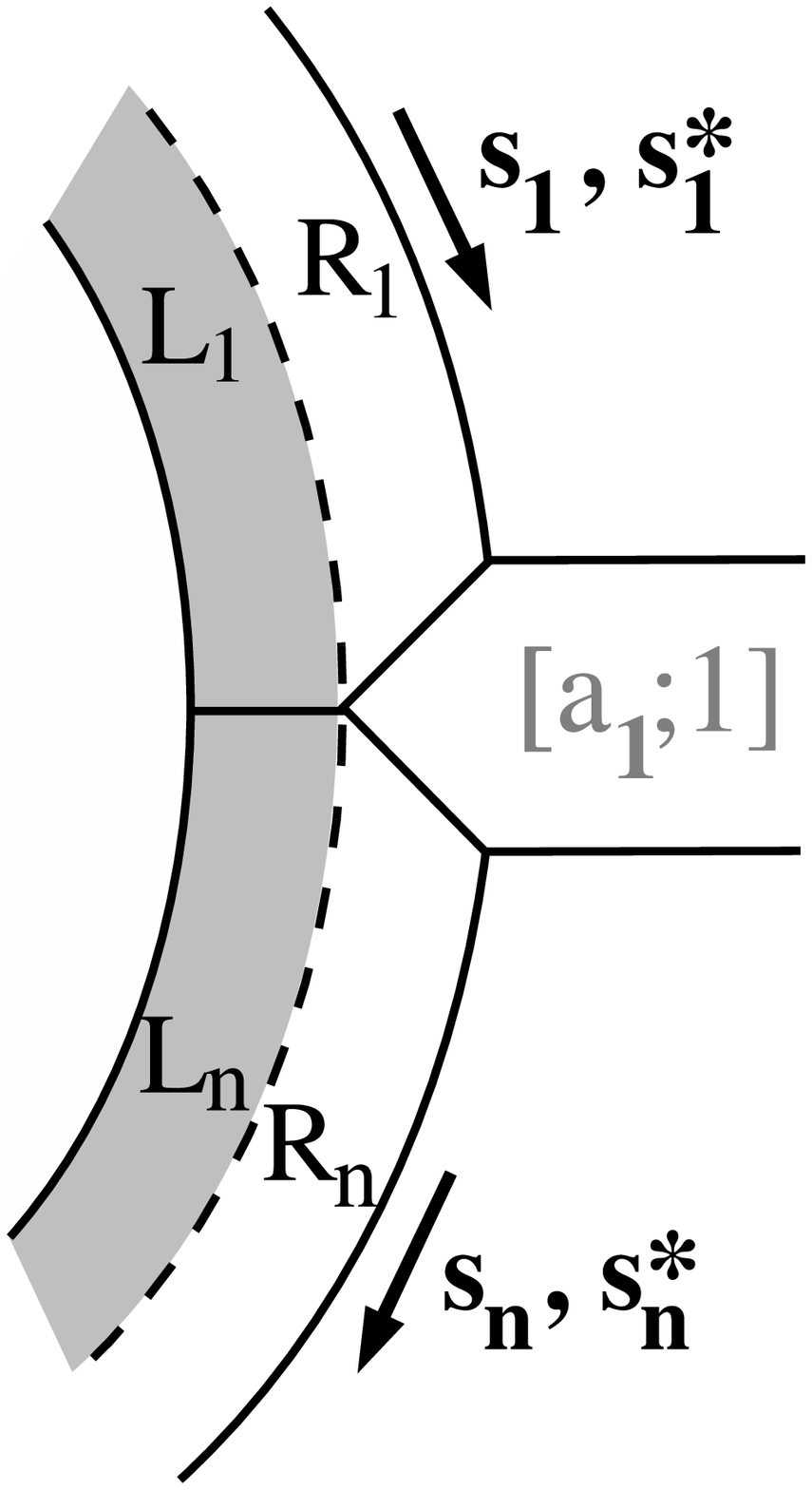,
height=6.5cm}
}
\centerline{\hskip3cm(a)\hskip9cm (b)}
\caption{(a) A general one-loop  diagram with $n$ vertices and $n$
propagators in the loop.
At each vertex the external states can either contribute to
$\Sigma$ or $\widehat{\Sigma}$ and
thus end up on either boundary of the annulus. (b) The topology of the
surfaces at vertex one.  The right part $R_i$ of
 the $i$-th propagator
contributes to $\Sigma$. The left part
  $L_i$ of the $i$-th propagator
(shaded) contributes to $\widehat\Sigma$. $\Sigma$
and $\widehat \Sigma$ are glued along the hidden boundaries
 represented by the dashed lines.}
\label{figloopn}
\end{figure}

If  the external states at the $i$-th vertex are on the $\Sigma$-side of the annulus, they add  to $\Sigma$ a surface
\begin{equation}
\label{iwltkthftfvm}
   [a_i;1\,|\,
   \vec{y_i}\,]  =  [a_i;1\,|\, y_i^1,\ldots\,, y_i^{m_i}\,] ~~\text{ with } a_i>0\,,
\end{equation}
where the $y_i^\alpha$ are the positions of the punctures and $\alpha= 1 ,
\ldots m_i$ is an
index that enumerates them.
If only 
a Fock space state  is connected to
vertex $i$, the surface in (\ref{iwltkthftfvm}) is $[1;1|1]$.
As shown in Figure~\ref{figloopn}(a), the surface in (\ref{iwltkthftfvm}) can
 in general
represent a complicated subtree diagram.
The slant factor is one because the subdiagram is a tree.
As the external states of the $i$-th vertex are on the $\Sigma$-side of the
diagram, they do not affect the $\widehat \Sigma$-side.
In order to treat
both $\Sigma$ and $\widehat\Sigma$ symmetrically, we also insert a surface
on $\widehat\Sigma$,
 the trivial ``identity surface"
\begin{equation}
    [\hat a_i;1]= [0;1]\,.
\end{equation}
If, on the other hand, the external states are on the boundary corresponding to
$\widehat\Sigma$, we have a surface insertion
\begin{equation}
\label{iwltkthfetfvm}
[\hat a_i;1\,|\,
   \vec{\hat {y}}_i\,]= [\hat a_i;1\,|\, \hat y_i^1,\ldots\,, \hat y_i^{\hat m_i}\,]
   ~~\text{ with } \hat a_i>0
   \end{equation}
    on $\widehat\Sigma$ accompanied by a trivial
   insertion $[a_i;1]=[0;1]$ on the $\Sigma$ side.
Thus the $n$ vertices in the loop are described by $2n$ wedges,
only $n$ of which are non-trivial.

\medskip
The propagators in the loop are in general  complicated, because their
geometric action depends
on the ghost number of the state that they act on~\cite{Kiermaier:2007jg}.
Since states of all ghost numbers circulate in the loop
we cannot use the classical propagator (\ref{classprop1}).
We choose the alternating
gauge introduced in~\cite{Kiermaier:2007jg} for the FP gauge fixing procedure
of Schnabl gauge.
This yields the propagator
\begin{equation}
    {\cal P}={\cal P}_{+}\,\Pi_++{\cal P}_{-}\,\Pi_-\,,
\end{equation}
where $\Pi_+$ ($\Pi_-$) is the projector on even (odd) ghost number,
and ${\cal P}_{\pm}$, for the $i$-th propagator,
 are
defined by
\begin{equation}~\label{altprop}
    {\cal P}_{+}
    =\int ds_i ds^\star_i\, e^{-s_iL}\,B\,Q\,B^\star e^{-s^\star_i L^\star}\,,\qquad
    {\cal P}_{-}
    =\int ds_i ds^\star_i\, e^{-s^\star_iL^\star}\,B^\star\,Q\,B e^{-s_i L}\,.
\end{equation}
The classical propagator is $\mathcal{P}_+$ since it acts on ghost number two sources to give ghost number one classical states.
With external physical states of ghost number one, all
 non-trivial
surface insertions at the loop insert states of ghost number one.  Since the three string vertex
couples states whose ghost numbers add up to three, the states on the
two loop-propagators that attach to the vertex must both have either even or
odd ghost number.  Consequently the states
running over {\em all}  
 the propagators in the loop are either of even ghost number
or odd ghost number.
It follows that in alternating gauge we only need to consider two types of Riemann surfaces for every diagram at one
loop level.
 The first surface is constructed by including a projector onto states of even ghost number anywhere in the loop and using
 ${\cal P}_{+}$  for all lines in the loop. The second surface is constructed
with a projector onto
odd ghost numbers in the loop and using  ${\cal P}_{-}$  for all
lines in the loop.

As mentioned before, the final set of surfaces is independent of the chosen
direction on the propagator on each line.  For simplicity, however,
we 
orient all propagators in the loop clockwise.
Tracing the outer loop counterclockwise, the  
right part of the propagator 
contributes to the surface $\Sigma$ on each line.
The inner loop must be traced clockwise so the left  part of the propagator
contributes to the surface $\widehat\Sigma$ on each line. 
This has been illustrated in
Figure~\ref{figloopn}(b).

At fixed Schwinger parameters,  the right part of 
${\cal P}_{+}$
adds to $\Sigma$ the slanted wedge corresponding to the operator
$e^{-s_iL_R}e^{-s^\star_i L_R^\star}$, which is calculated in
(\ref{propLR}).
The right part of 
${\cal P}_{-}$, on the other hand,  
 adds the slanted wedge   
corresponding to $e^{-s^\star_i L_R^\star}e^{-s_iL_R}$ to $\Sigma$,
which is calculated using (\ref{eslr}) and (\ref{eslstarr}).
We conclude
that the  $i$-th propagator contributes to
$\Sigma$ the slanted wedge  $R_i$ 
given by
\begin{equation}
\label{iwltkssajj}
 R_i\equiv
[r_i;e^{s_i-s^\star_i}]
\quad\text{with}\quad
\left\{
\begin{array}{l}
r_i=\half (1+e^{s_i-s^\star_i})-e^{-s^\star_i}
\quad\,\text{ for even ghost number }({\cal P}_+)\\\\
 r_i=e^{s_i}-\half(1+e^{s_i-s^\star_i})
\qquad\text{ for \,odd\, ghost number  }({\cal P}_-)\,\,.
\end{array}
\right.
\end{equation}
Similarly, the left part of the propagator contributes to $\widehat \Sigma$
the slanted wedge $L_i$ given by  $e^{-s_iL_L}e^{-s^\star_i L_L^\star}$
for $\mathcal{P}_+$ and $e^{-s^\star_i L_L^\star}e^{-s_iL_L}$
for $\mathcal{P}_-$.  We readily find   
\begin{equation}
\label{iwltkssajjj}
 L_i\equiv
[ l_i;e^{s^\star_i-s_i}]
\quad\text{with}\quad
\left\{
\begin{array}{l}
 l_i=\half (1+e^{s^\star_i-s_i})-e^{-s_i}
\quad~~\text{ for even ghost number }({\cal P}_+)\\\\
 l_i=e^{s^\star_i}-\,\half(1+e^{s^\star_i-s_i})
\qquad\text{ for \,odd\, ghost number  }({\cal P}_-)\,.
\end{array}
\right.
\end{equation} 
The definitions~(\ref{iwltkssajj}) and~(\ref{iwltkssajjj}) generalize~(\ref{LiRitree}) 
from the classical propagators of tree diagrams to loop propagators in alternating gauge.

\medskip
We   now assemble the complete surfaces $\Sigma$ and $\widehat \Sigma$. We
 construct
$\Sigma$ by gluing the surfaces of propagators and external states
counterclockwise, starting at vertex $1$. We obtain
\begin{equation}~\label{Sigmagen}
\begin{split}
\Sigma\equiv[a;e^{s_{\rm{eff}}}|\vec x\,]&=[a_1;1| \vec{y}_1]\ast
 R_1
\ast\cdots\ast[a_n;1| \vec{y}_n]\ast
 R_n
\,,\\
\end{split}
\end{equation}
Similarly, we construct $\widehat \Sigma$ by gluing the surfaces of propagators
and external states.
We trace clockwise starting right below vertex $1$ and get
\begin{equation}\label{hatSigmagen}
    \widehat{\Sigma}\equiv [\hat a;e^{-s_{\rm{eff}}}|\, \vec{\hat{x}}\,]=
 L_n
\ast [\hat a_n;1|\,\vec{\hat y}_n\,]\ast\cdots \ast
 L_1
\ast[\hat a_1;1|\,\vec{\hat y}_1\,]\,.
\end{equation}
It is clear from ~(\ref{Sigmagen}) and~(\ref{hatSigmagen}) that
\begin{equation}\label{seffdefined}
    s_{\rm{eff}}=\sum_{i=1}^n (s_i-s^\star_i)\,.
\end{equation}
It follows that the modulus $M$ of the annulus is given by
\begin{equation}\label{oneloopM}
        \boxed{\phantom{\Bigl(} M=\frac{\pi}{\left|s_{\rm{eff}}\right|}\,.~~}
\end{equation}
To calculate the positions of the punctures we first determine the total lengths $a$ and $\hat a$ of $\Sigma$ and $\widehat \Sigma$.  Looking
at (\ref{Sigmagen})  we see that
\be 
a = a_1 + r_1 + e^{s_1 - s_1^\star} \bigl(a_2 + r_2 + e^{s_2 - s_2^\star} (a_3 + r_3 + \ldots  \,\,\,.
\ee
This is readily seen to give
\begin{equation}\label{valueofa99}
    a=\sum_{i=1}^n e^{\sum_{j=1}^{i-1} s_j-s^\star_j}(a_i+r_i)
    = \sum_{i=1}^n b_i\, (a_i+r_i)\,.
\end{equation}
where we defined
\begin{equation}\label{bindexed}
   b_i \equiv e^{\sum_{j=1}^{i-1}s_j-s_j^\star}
   \,. 
\end{equation}
We can view $b_i$ as a \emph{local scaling factor}.  
It is the product of 
the slant factors of the surfaces $R_1$, $R_2, \ldots$, up to
$R_{i-1}$.  It is the scaling factor that must apply to $R_i$ when
it is glued in to form $\Sigma$ in (\ref{Sigmagen}).

The value of $\hat a$ is computed similarly.   
Looking at (\ref{hatSigmagen}) we write
\be 
\hat a =  e^{s_n^\star-s_n} \hat{a}_n + l_n + e^{s_n^\star-s_n}
\bigl( e^{s_{n-1}^\star-s^{\phantom{\star}}_{n-1}}\,\hat{a}_{n-1} + l_{n-1} + \ldots  \,\,\,,
\ee
which gives 
\begin{equation}\label{valueofhata}
    \hat a= \sum_{i=1}^n e^{\sum_{j=i+1}^{n}
s^\star_j-s_j}(e^{s^\star_i-s_i}\hat a_i+ l_i)\,.
\end{equation}
Noticing that $e^{s_i-s^\star_i} l_i=r_i$  (see (\ref{iwltkssajj}) and (\ref{iwltkssajjj}))
we can also rewrite $\hat a$ as
\begin{equation}
        \hat a= e^{-s_{\rm{eff}}}\sum_{i=1}^n
        e^{\sum_{j=1}^{i-1} s_j-s^\star_j}(\hat a_i+r_i) =
        e^{-s_{\rm{eff}}}\sum_{i=1}^n
        b_i\,(\hat a_i+r_i)\,.
\end{equation}
For insertions at positions $y_k^\alpha$ and $\hat{y}_k^\alpha$ we denote
by $x_k^\alpha$ and $\hat{x}_k^\alpha$
 their final coordinates  on $\Sigma$ and $\widehat \Sigma$.
 The collection of these insertions were represented by
$\vec{x}$ and $\vec{\hat x}$ in
(\ref{Sigmagen}) and (\ref{hatSigmagen}). Short calculations
using (\ref{iwltsktnpplsfaj}) show that
these positions are given by
\begin{equation}\label{pospunctonsigma}
\begin{split}
     x^\alpha_k-\half  &= \,b_k   \, (y^\alpha_k-\half ) +  \sum_{i=1}^{k-1}
     b_i\, (a_i+r_i)
     \,, 
    \\
     \hat x^\alpha_k-\half  &=
      e^{-s_{\rm{eff}}}\Bigl(\,b_k\,(\hat y_k^\alpha-\half +r_k)
      + \sum_{i=k+1}^n
      b_i\,( \hat a_i+r_i)
      \Bigr)\,.
\end{split}
\end{equation}
It follows immediately from (\ref{389eiu}) and (\ref{3948ufd})
that in the annulus frame $\zeta$ these positions translate into the angles
\begin{equation}\label{allangles}
 \boxed{\phantom{\Biggl(}
   \phi^\alpha_k=\frac{2\pi}{s_{\rm{eff}}}\ln \Bigl( 2
\Bigl|x^\alpha_k-\half+\frac{a}{e^{s_{\rm{eff}}}-1}\Bigr|\Bigr)
   \,,  \qquad
   \hat\phi^\alpha_k= \frac{2\pi}{s_{\rm{eff}}}\ln \Bigl( 2\Bigl|\half-\hat
x^\alpha_k+\frac{\hat a}{1-e^{-s_{\rm{eff}}}}\Bigr|\Bigr)\,.
~~}
\end{equation}
 Up to a trivial overall rotation of the annulus, the angles~(\ref{allangles}) represent
the open string moduli of the one-loop diagram. This concludes our construction of moduli for
general one-loop diagrams in Schnabl gauge.

\section{The one-loop two-point diagram}\label{secloop2pt}
\setcounter{equation}{0}
We now apply the general construction of Section~\ref{secloop} to the one-loop
two-point diagram. In the following analysis we will restrict ourselves to the Riemann surfaces
generated by even ghost-number propagators running in the loop, i.e. we use the propagator
${\cal P}_+$ in the loop. The other Riemann surface, which is generated by putting ${\cal P}_-$ on all
loop propagator lines, can be calculated
analogously.

\begin{figure}
\centerline{
\epsfig{figure=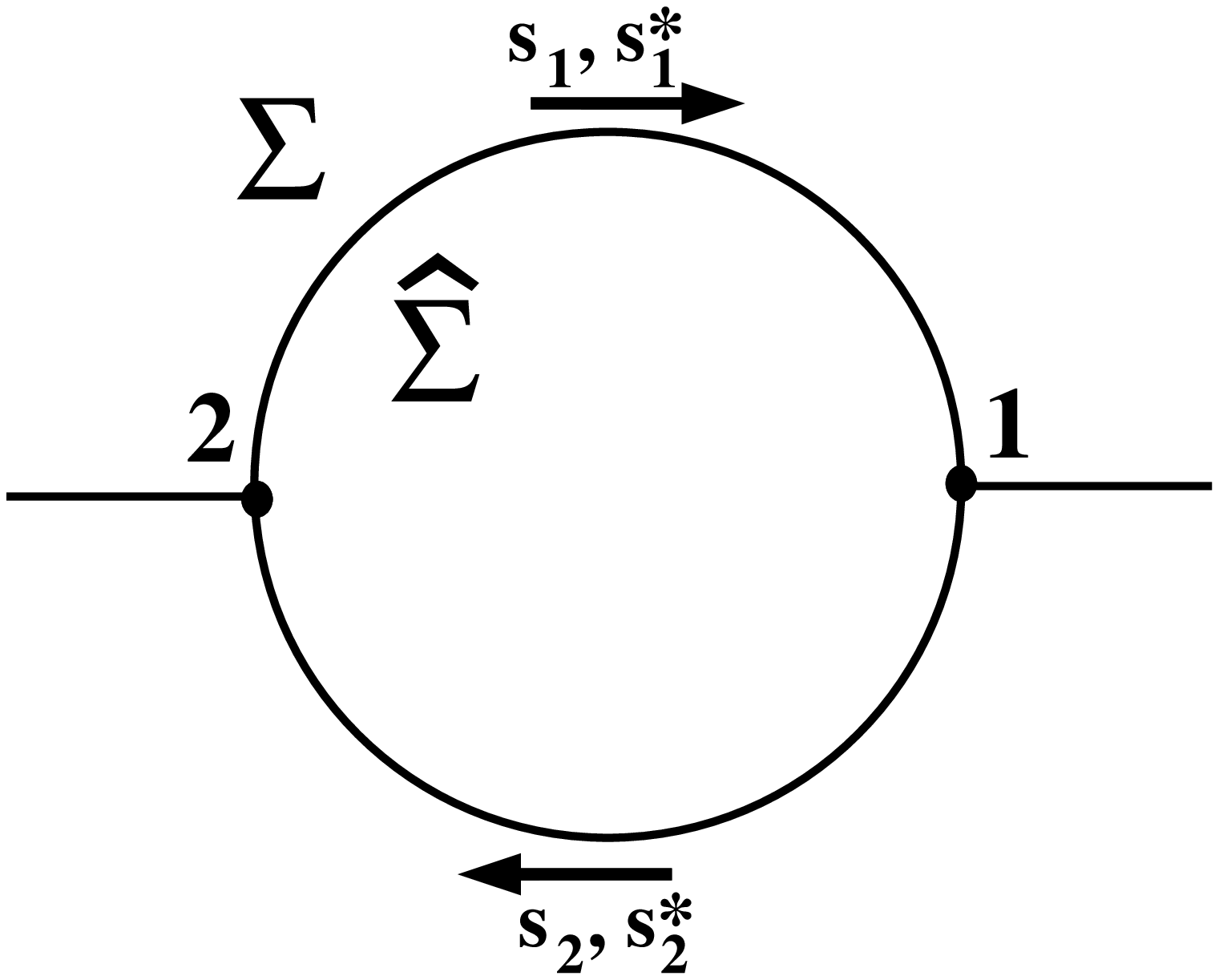, height=5.5cm}
\hskip 2.0cm
\epsfig{figure=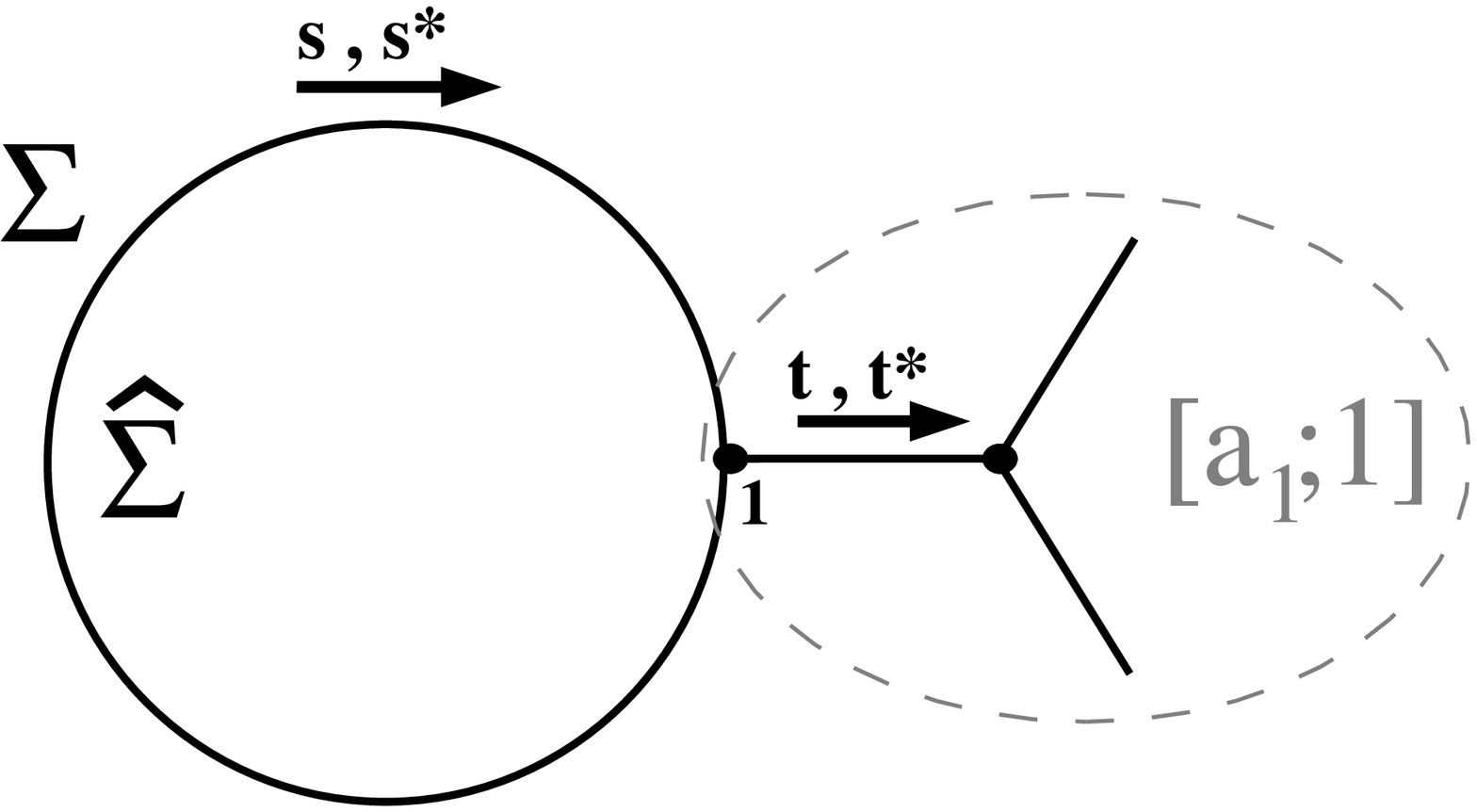, height=5.5cm}}
 \centerline{(a)\hskip 9.3cm (b) \hskip .9cm}
\caption{The two diagrams contributing to the two-point function with both
insertions on the same
boundary.}
\label{figloop2ptA}
\end{figure}

\subsection{Riemann surfaces with both insertions on the same
boundary}

Let us consider the one-loop two-point function with both insertions on the same boundary component
of the annulus -- the so-called planar contributions.
 There are two diagrams that contribute,
as shown in~Figure~\ref{figloop2ptA}.

\subsubsection{First diagram}
 In the first diagram (Figure~\ref{figloop2ptA}(a))
  we have two propagators in the loop $(n=2)$ and two Fock space
surfaces $[1;1|1]$ connected directly
to vertices in the loop, on the side that we choose to call
the surface $\Sigma$. The Fock space surfaces do not
contribute to $\widehat \Sigma$.
 In the notation of (\ref{iwltkthftfvm})
\begin{equation}
    a_1=a_2=1\,, \qquad y_1^1=y_2^1=1\,,\qquad \hat a_1=\hat a_2=0\,.
\end{equation}
We label the Schwinger parameters of the two propagators by $s_1$, $s^\star_1$,
$s_2$, $s^\star_2$ and (\ref{seffdefined}) gives
\begin{equation}
s_{\rm{eff}}=s_1-s^\star_1+s_2-s^\star_2\,.
\end{equation}
The length $a$ of the $\Sigma$ surface follows from (\ref{valueofa99})
and (\ref{iwltkssajj}).  We find
\begin{equation}
\label{alengthfor2pt}
a =  {\textstyle{3\over 2} } + 2 e^{s_1 - s_1^\star} - e^{-s_1^\star}
+ \half \,e^{s_{\rm{eff}}} - e^{s_1 -s_1^\star-s_2^\star}\,.
\end{equation}
The position of the punctures on $\Sigma$ are found using
(\ref{pospunctonsigma}) and (\ref{iwltkssajj}).  We obtain:
\begin{equation}
x_1^1 = 1 \,,\qquad
x_2^1 = 2  + e^{s_1 - s_i^\star} - e^{-s_1^\star} \,.
\end{equation}
The relevant open string modulus is the relative angle between the insertions.
Making use of (\ref{allangles}) a short calculation gives
\begin{equation}
\label{delhiv}
\begin{split}
   \Delta\phi=\phi^1_2-\phi^1_1
   &= \frac{2\pi}{s_{\rm{eff}}}\ln\frac{
     2-e^{-s^\star_1}+e^{s_1-s^\star_1}
    + e^{s^\star_2-s_2}-e^{-s_2}}
    {1-e^{-s^\star_1}+2e^{s_1-s^\star_1}-
e^{s_1-s^\star_1-s^\star_2}+e^{s_{\rm{eff}}}}\,.
 \end{split}
\end{equation}

\medskip
Just like for the
five-point diagram, let us consider
the Riemann surfaces generated by the simplified propagator $B/L$.
We thus set $s^\star_1=s^\star_2=0$ and (\ref{delhiv}) becomes
\begin{equation}
\begin{split}
   \Delta \phi
   = \frac{2\pi}{s_1+s_2} \ln\Bigl(\frac{ 1+e^{s_1}}{e^{s_1}+e^{s_1+s_2}}\Bigr)
   \,=\,\pi+\frac{2\pi}{s_1+s_2} \ln \Bigl(\frac{
\cosh{\textstyle {{s_1}\over 2}}}{\cosh {\textstyle {{s_2}\over 2}}}\Bigr)\,.
\end{split}
\end{equation}
It is convenient to study this angle for fixed modulus of the
annulus:
$s_{\rm{eff}}=s_1+s_2=\text{const}$. We then have
\begin{equation}
\begin{split}
   \Delta \phi
    &=\pi+\frac{2\pi}{s_{\rm{eff}}} \ln \Bigl(\frac{
\cosh{\textstyle {{s_1}\over 2}}}{\cosh{\textstyle {{s_{\rm{eff}}-s_1}\over 2}}}\Bigr)\,.
\end{split}
\end{equation}
For $s_1=\half s_{\rm{eff}}$ the two
punctures are maximally separated: $\Delta \phi=\pi$.
As we vary $s_1$ the separation angle varies within an interval
centered at $\pi$. The maximal (minimal) angle $\Delta\phi_+$ ($\Delta\phi_-$)  is obtained for
$s_1=s_{\rm{eff}}$ ($s_1=0$), and it is given by
\begin{equation}
\label{excurmodpos}
   \Delta \phi_{\pm}
    =\pi\pm\frac{2\pi}{s_{\rm{eff}}} \ln
    \cosh{\textstyle {{s_{\rm{eff}}}\over 2}}\,.
\end{equation}
Close to closed string degeneration, i.e. for $s_{\rm{eff}}\ll1$, we obtain the
simplified expression
\begin{equation}
   \Delta \phi_{\pm}
    =\pi\pm\frac{\pi}{4}s_{\rm{eff}} \qquad\text{ for }\quad s_{\rm{eff}}\ll1\,.
\end{equation}
Thus near closed string degeneration the diagram just generates a
region of moduli in which the punctures are nearly opposite.   Close to
open string degeneration ($s_{\rm{eff}} \to \infty$) equation~(\ref{excurmodpos}) shows that almost
 the entire range of the position modulus
is covered.
In general, not all of the range of the position modulus is obtained.
In Figure~\ref{figloop2ptmod} we show the region of  the
complete two-dimensional moduli
space $(s_{\rm{eff}}, \Delta \phi)$ that the present, first diagram covers.
\begin{figure}
\centerline{
\epsfig{figure=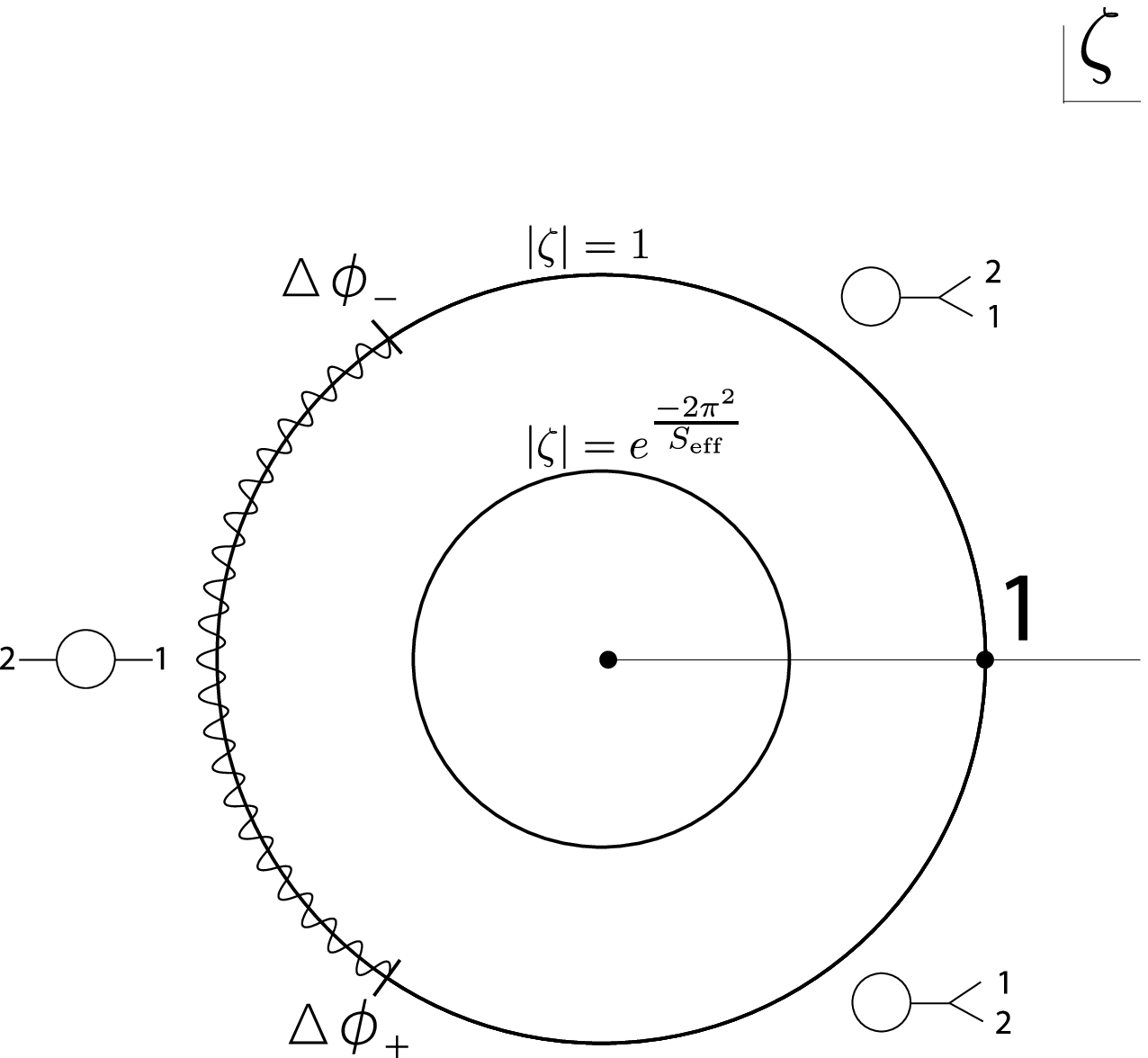, height=5.5cm}
\hskip 1.5cm
\epsfig{figure=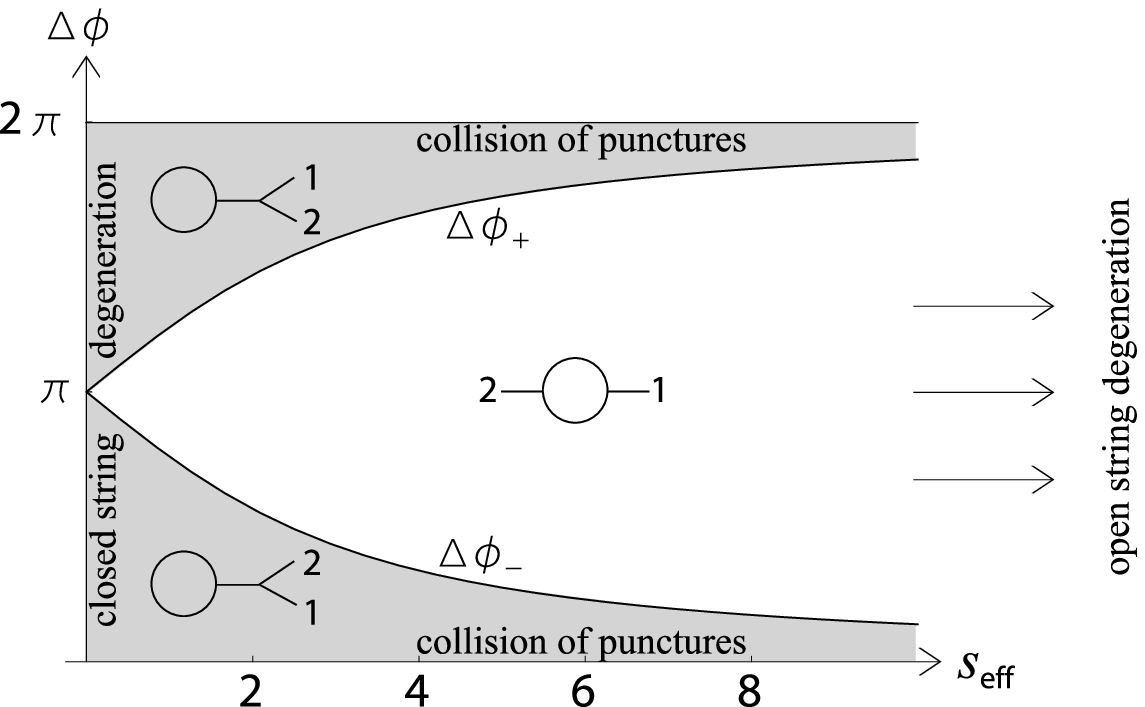, height=5.5cm}}
\centerline{
\hskip.5 cm (a)\hskip 8cm (b)\hskip 2.5cm}
\caption{
(a)
The $\zeta$-frame annulus for the planar one-loop two-point function.
Insertion $1$ is fixed at angle $\phi=0$ and the position modulus
is the angle $\Delta \phi$ for insertion 2.
 Both string diagrams in Figure~\ref{figloop2ptA}
are needed to generate
 the full position range
  $0\leq\Delta\phi \leq 2\pi$.
(b) The space $(s_{\rm{eff}}, \Delta \phi)$
of closed and open moduli is  covered fully by the indicated string diagrams.}
\label{figloop2ptmod}
\end{figure}
The remaining region, as we will see now, is generated by
second diagram contributing to the amplitude.

\subsubsection{Second diagram}

In the second diagram (Figure~\ref{figloop2ptA}(b)) there is only one propagator in the loop ($s^\star$, $s$)
and both external states are connected to the loop through
another internal propagator ($t$, $t^\star$).
We then have
\begin{equation}
   [a_1;1|\, y_1^1, y_1^2]
   = e^{-t L}e^{-t^\star L^\star}[2;1|1,2]
    =[1-2e^{-t}+3e^{t^\star-t};1| y_1^1, y_1^2]\,,
\end{equation}
where we use  
 (\ref{classprop}) and (\ref{iwltptmnstvmss}) to obtain
\begin{equation}\label{insposa1}
   a_1= 1-2e^{-t}+3e^{t^\star-t}\,, \quad
   y^1_1= 1-e^{-t}+e^{t^\star-t}\,, \quad
   y^2_1= 1-e^{-t}+2e^{t^\star-t}\,,\quad
   \hat a_1=0\,.
\end{equation}
To build  $\Sigma$ 
we use (\ref{Sigmagen}) with $n=1$ and find
\begin{equation}
   \Sigma\equiv[a;e^{s-s^\star}| x_1^1 ,x_1^2] =
    [a_1;1|\, y_1^1, y_1^2] \ast [r_1;e^{s-s^\star}]\,, \qquad 
r_1=\half (1+e^{s-s^\star})-e^{-s^\star}\,,
\end{equation}
and thus
\begin{equation}
   a=a_1+r_1  
   ={\textstyle {3\over 2}}-2e^{-t}+3e^{t^\star-t}-e^{-s^\star}+\half e^{s-s^\star}\,.
\end{equation}
As $ [a_1;1|\, y_1^1, y_1^2]$ is the left-most surface in $\Sigma$, the positions of insertions on $\Sigma$ coincide with the positions of
insertions on $[a_1;1]$:  we have $ x^1_1=y^1_1$ and $ x^2_1=y^2_1$.
For the relative angle $\Delta\phi$ between the two
insertions we use (\ref{allangles}) to obtain
\begin{equation}
\begin{split}
   \Delta \phi
   &  = \frac{2\pi}{s_{\rm{eff}}}\ln\biggl(\frac
{1-e^{-s^\star}+e^{s-s^\star}-e^{-t}+e^{t^\star-t}-e^{s-s^\star-t}+2
e^{s-s^\star+ t^\star-t}}
{1-e^{-s^\star}+e^{s-s^\star}-e^{-t}+2e^{t^\star-t}-e^{s-s^\star-t}+e^{s-s^\star+
t^\star-t}}\biggr)\,,
\end{split}
\end{equation}
with $s_{\rm{eff}}=s-s^\star$.

\medskip
Let us again focus on the surfaces generated by the simplified propagators
$B^\star/L^\star$ or $B/L$. We have two options.
\begin{itemize}
  \item
    $s^\star=t=0$ (The case $s=t=0$ gives similar results)\\
    In this case 
     $s_{\rm{eff}}=s$  
    and $\Delta \phi$ reduces to  
    \begin{equation}  
      \Delta \phi  = \frac{2\pi}{s_{\rm{eff}}}\ln\biggl(\frac
      {e^{t^\star}+2 e^{s_{\rm{eff}}+ t^\star}-1}
      {2e^{t^\star}+e^{s_{\rm{eff}}+ t^\star}-1}\biggr)\,.
    \end{equation}
    For $t^\star\to0$, this gives
    \begin{equation}  
      \Delta \phi  = \frac{2\pi}{s_{\rm{eff}}}\ln\biggl(\frac
      {2 e^{s_{\rm{eff}}}}
      {1+e^{s_{\rm{eff}}}}\biggr)=\pi-\frac{2\pi}{s_{\rm{eff}}}\ln\cosh{\textstyle {s_{\rm{eff}}\over 2}}  \quad \textrm{ for }
t^\star\to0\,,
    \end{equation}
and matches smoothly to the first diagram's  $\Delta \phi_-$  
as given in (\ref{excurmodpos}).

    For $t^\star\to\infty$, however,
    there is no collision between
    the insertions. Instead, we obtain
    \begin{equation}  
      \Delta \phi  = \frac{2\pi}{s_{\rm{eff}}}\ln\biggl(\frac
      {1+2 e^{s_{\rm{eff}}}}
      {2+e^{s_{\rm{eff}}}}\biggr) \quad \textrm{ for }~ t^\star\to\infty\,.
    \end{equation}
    Thus diagram two with propagator $B^\star/L^\star$ in the subtree does
\emph{not} cover moduli space together with diagram one.
This is not surprising because tracing  
along the $\Sigma$ boundary of the Feynman diagram we encounter the operator combination
$e^{-t^\star L_L}e^{-s_{\rm{eff}} L^\star_L}e^{-t^\star L^\star_L}$. At fixed annulus modulus, i.e. for $s_{\rm{eff}}=const$,  
 this operator does not produce open string degeneration for
 $t^\star\to\infty$. In fact,
\begin{equation}
    \lim_{t^\star\to\infty}e^{-t^\star L_L}e^{-s_{\rm{eff}} L^\star_L}e^{-t^\star L^\star_L}[a;b]
    =[a+\half(e^{s_{\rm{eff}}}+1);e^{s_{\rm{eff}}}b]\,, 
    \end{equation}
which is a perfectly regular surface.

  \item
    $s^\star=t^\star=0$ (The case $s=t^\star=0$ gives similar results)\\
 Again, 
  $s_{\rm{eff}}=s$  
 and this choice  
 corresponds to $B/L$ as the propagator in the tree.  This
 time we obtain
  \begin{equation}  
    \Delta \phi  = \frac{2\pi}{s_{\rm{eff}}}\ln\biggl(\frac
    {e^{s_{\rm{eff}}}+e^{s_{\rm{eff}}-t}}
    {e^{s_{\rm{eff}}}+e^{-t}}\biggr)
    = \pi-\frac{2\pi}{s_{\rm{eff}}}\ln\biggl(\frac
    {\cosh {\textstyle {{t+s_{\rm{eff}}}\over 2}}}{\cosh {\textstyle {t\over 2}}}\biggr)\,.
  \end{equation}
  For $t=0$ we again match to $\Delta \phi_-$ in
  the first diagram. 
  This time, all
angles $0<\Delta\phi<\Delta\phi_-$ are covered.
  Indeed,
  \begin{equation}
    \Delta \phi  \to 0 \quad \textrm{ for } \quad t\to\infty\,.
  \end{equation}
  This is sufficient to cover moduli space together with diagram one,
  as shown in Figure~\ref{figloop2ptmod}.  The diagram gives the
  shaded region $0<\Delta\phi<\Delta\phi_-$.
Of course, since external states are distinguishable,
the region \
$\Delta\phi_+<\Delta\phi<2\pi$ is generated by the string diagram
in which the order of the Fock space state insertions is reversed.
\end{itemize}

\subsection{Riemann surfaces with insertions on both boundaries}
Let us consider a one-loop amplitude with one Fock space state insertion on the
outer boundary and one Fock space state insertion on the inner boundary.
This nonplanar string diagram in shown in
Figure~\ref{figloop2ptB}.
\begin{figure}
\centerline{
\epsfig{figure=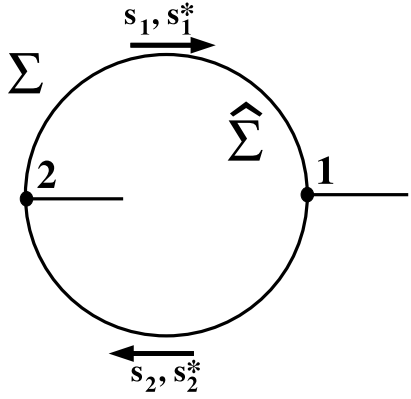, height=5.5cm}}
\caption{The diagram of the two-point function with insertions on both
boundaries.}
\label{figloop2ptB}
\end{figure}
As $[a_1;1|1]$ and $[\hat a_2;1|1]$ are the Fock space surfaces, we have
\begin{equation}
    a_1=\hat a_2=1\,,\qquad
    a_2=\hat a_1=0\,,\qquad
    y^1_1=1\,,\qquad
    \hat y^1_2=1\,.
\end{equation}
We have two propagators running in the loop. The relevant surfaces,
using (\ref{Sigmagen}) and (\ref{hatSigmagen}) are
\begin{equation}
\begin{split}
\Sigma\equiv[a;1|x_1^1] &=  [1;1|1] * R_1 * R_2 \\
\widehat \Sigma\equiv [\hat a;1|\hat{x}_2^1] &= L_2  * [1;1|1] * L_1\,.
\end{split}
\end{equation}
The relevant parameters above are readily calculated:
\begin{equation}
\begin{split}
    a
&={\textstyle {3\over 2}}-e^{-s^\star_1}+e^{s_1-s^\star_1}(1-e^{-s^\star_2}+\half e^{s_2-s^\star_2})\\
    \hat a
&=e^{-s_{\rm{eff}}}\bigl(\half -e^{-s^\star_1}+e^{s_1-s^\star_1}(2-e^{-s^\star_2}+\half e^{s_2-s^\star_2})\bigr)\,,\\  s_{\rm{eff}} &= s_1-s^\star_1+s_2-s^\star_2\,,  \\
    x^1_1&=1\,,\qquad
    \hat x^1_2=e^{s^\star_2-s_2}-e^{s_2}+1\,.
\end{split}
\end{equation}
 A calculation using the above results and (\ref{allangles})
 gives us the difference in insertion angles
\begin{equation}
    \Delta\phi=\hat\phi^1_2-\phi^1_1
    =\frac{2\pi}{s_{\rm{eff}}}\ln\frac{
     1-e^{-s^\star_1}+e^{s_1-s^\star_1}-e^{-s_2}+e^{s^\star_2-s_2} }
{1-e^{-s^\star_1}+e^{s_1-s^\star_1}-e^{s_1-s^\star_1-s^\star_2}+e^{s_{\rm{eff}}}}\,.
\end{equation}

\medskip
Let us consider two cases of simplified propagators. If both propagators are
$B/L$, then we can set
$s^\star_1=s^\star_2=0$ (similarly for $B^\star/L^\star$ and $s_1=s_2=0$) and
obtain
\begin{equation}
    \Delta\phi
    =\frac{2\pi}{s_1+s_2}\ln\frac{e^{s_1}}{e^{s_{\rm{eff}}}}
    =2\pi\,\frac{s_1}{s_1+s_2}\,.
\end{equation}
Moduli space is covered: for fixed $s_{\rm{eff}}=s_1+s_2$, $\Delta \phi$
takes on all values between $0$ and $2\pi$.

To examine the case of mixed propagators $B/L$ and
$B^\star/L^\star$ we set
$s^\star_1=s_2=0$. Then,
\begin{equation}
    \Delta\phi
    =\frac{2\pi}{s_1-s^\star_2}\ln(e^{s_1-s^\star_2}-e^{-s^\star_2}+1) \,.
\end{equation}
Since $s_1 = s_{\rm{eff}} + s_2^\star$ and $s_1, s_2^* \geq 0$, for
 fixed
 $s_{\rm{eff}}>0$ we have a constraint in the range of $s_1$.  Moduli
 space is not fully covered.
 In fact, for $s_{\rm{eff}}\to\infty$ we have $\Delta \phi\to 2\pi$ for the entire range of permissible
 $s_1$, $s_2^\star$. In this limit the open string modulus is stuck at the collision of punctures.

\bigskip

\section{A regularized view on one-loop diagrams}\label{aregvieononelooam}
\setcounter{equation}{0}
In Section~\ref{secloop}
we presented a prescription
to map the Riemann surface
of a general one-loop diagram to the annulus
while keeping track of the operator
insertions of external states. This allowed us to calculate the closed and open string
moduli of the surface as  simple functions of the Schwinger parameters.
The treatment was entirely in Schnabl gauge and used
the formalism of slanted wedges.
To justify our prescription, however,
we need to revisit the construction by regularizing
Schnabl gauge. 
 This analysis extends the proof for the one-loop 
 tadpole given in Section~\ref{dofeoruhk} to general one-loop diagrams.
Again, we use the $\lambda$-regularization introduced in~\cite{Kiermaier:2007jg}.
To confirm our prescription we need  
 to examine the three types of operations that 
 are used.  
These operations are the multiplication of slanted wedges, the gluing between
left and right boundaries on both $\Sigma$ and $\widehat\Sigma$, and the
gluing of $\Sigma$ and $\widehat \Sigma$ to each other at their hidden boundaries. 
Before we check these operations, let us analyze the relevant gluing boundaries in more detail.

\subsection{The boundaries of regularized slanted wedges}
To examine the gluing curves, it is
convenient to represent
 the coordinate curve $f^\lambda(e^{i\theta})$ in the $z$ frame
in terms of the parameterized curves
$\gamma^\lambda_R$ and $\gamma^\lambda_L$
 defined in (\ref{gammalLR}) and shown in Figure~\ref{sl01fig}(c).
 Regarded as the regulated surface $[1;1|1]$, a
Fock space state is  then
bounded by $\half+\gamma^\lambda_L$ and $\frac{3}{2}+\gamma^\lambda_R$.
The boundaries touch at the midpoint $\theta=\pi/2$.

Similarly, the regularized slanted wedge corresponding to $e^{-sL^\lambda_R}$
is bounded by
the curves $\half +\gamma^\lambda_R$ and $e^s(\half+\gamma^\lambda_R)$. Its left boundary glues nicely to
a Fock space state. The right boundary is a simple rescaling of the left boundary by $e^s$. This was illustrated in
the context of the tadpole graph in Figure~\ref{sl03fig}(b). 
The two boundaries of $e^{-sL^\lambda_R}$  do not touch for
$\theta=\frac{\pi}{2}$.
In fact, $e^{-sL^\lambda_R}$ has a
vertical boundary on the imaginary axis from $i\Lambda$ to $ie^s\Lambda$,
as discussed in Section~\ref{sectp}. This vertical line segment connects the endpoints of the left and right boundary of $e^{-sL^\lambda_R}$.

The regularized slanted wedge corresponding to
$e^{-s^\star (L^\lambda_R)^\star}$ is more delicate. Recall that in Figure~\ref{figemsL}
we \emph{flipped} the surface around its right vertical
boundary to be able to
interpret $e^{-s^\star L_R^\star}$ as the slanted wedge
$[\half(1-e^{-s^\star});e^{-s^\star}]$. We conclude that the regularized boundaries of
$e^{-s^\star (L^\lambda_R)^\star}$
are given by $\half+\gamma^\lambda_L$ and $\half+\half(1-e^{-s^\star})+e^{-s^\star}\gamma^\lambda_L$. The surface of
$e^{-s^\star (L^\lambda_R)^\star}$
also has a hidden vertical boundary. It
is located between $1+ie^{-s^\star}\Lambda$ and $1+i\Lambda$. These
facts are illustrated in Figure~\ref{figregesstar}, where we also show
the surface for $e^{-s^\star (L_L^{\lambda})^\star}$, which needs
further displacement and rescaling to be presented as a
 regularization of a conventional slanted wedge.

\begin{figure}
\centerline{\epsfig{figure=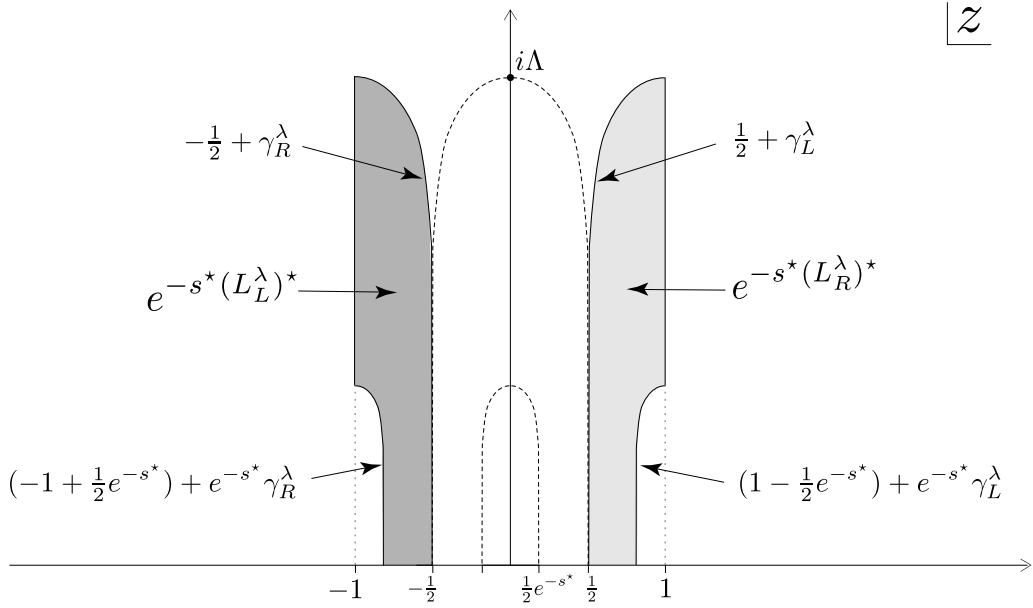, height=8cm}}
\caption{The regularized version of the
slanted wedges corresponding to $e^{-s^\star (L_L^{\lambda})^\star}$ and
$e^{-s^\star (L_R^\lambda)^\star}$.
}
\label{figregesstar}
\end{figure}

\medskip
 In our construction, we build surfaces from slanted wedges
 associated with propagators and Fock space states.
From this 
it is clear, that 
the slanted wedges $[a;b]$ relevant to our construction 
will, after regularization, be bounded on the left 
by either $\half+\gamma^\lambda_L$ or $\half+\gamma^\lambda_R$ and will  be bounded on the right by either $\half+a+b\gamma^\lambda_L$ or
by $\half+a+b\gamma^\lambda_R$. Furthermore, 
the slanted wedges  associated with the left and right parts of $e^{-sL}$ and $e^{-s^\star L^\star}$ carry a hidden boundary
that needs to be glued to the hidden boundary of a complementary surface.

For $\lambda\to 0$, the curves $\gamma_{L/R}^\lambda(\theta)$ both approach the canonical vertical sliver parametrization
$\gamma(\theta)$ defined in~(\ref{gamma0}).
One may therefore wonder why it
is  not trivial that regularized slanted wedges glue nicely for $\lambda\to0$. The problem is that the convergence
of $\gamma_{L/R}^\lambda (\theta)$ to the curve $\gamma(\theta)$ in the limit $\lambda\to0$ is not \emph{uniform} on the full interval
 $0\leq\theta\leq\frac{\pi}{2}$. In fact, for any $\lambda>0$, the curves $\gamma^\lambda_{L/R}$ start deviating significantly
from $\gamma$ in the region where $\Im(z)$ is of order $\Lambda$,
 as discussed in Section~\ref{sec21}.
This effect can be neglected for tree-level amplitudes. In fact, the relevant frame for the
calculation of moduli and correlators in tree amplitudes is the disk frame $\eta$, which is related to the $z$ frame through
\begin{equation}\label{}
     \eta=\exp\left(\frac{2\pi iz}{a}\right)\,.
\end{equation}
Here, $a>0$ is a function of the Schwinger parameters and independent of $\lambda$.
All points with large imaginary values in the $z$ frame converge to the point $\eta=0$.
In the $\eta$-frame, the deviations of $\gamma^\lambda_{L/R}$ from $\gamma$ are thus
suppressed by a factor of $e^{-\frac{2\pi \Lambda}{a}}$, which vanishes for $\lambda\to0$.

For one-loop diagrams the natural frame to consider 
for the gluing is either the $\zeta$-frame of the annulus
or the $w$ frame, in which the annulus is ``unwrapped''. 
The $w$ coordinate is given~by
\begin{equation}\label{}
   w=-\ln(2z)+i\pi\,.
\end{equation}
Clearly, not all points $z$ with large imaginary values converge to a single point in the $w$ frame. Therefore
the analysis for one-loop diagrams is more subtle than for trees.
The
 mapping of the
boundary curves $\gamma^\lambda_{L/R}$
 to the $w$ frame
have been analyzed
in the context of the
simplified tadpole diagram.
Indeed, in Section~\ref{sectp} we have proven that
 when mapped to the $w$ frame,
the left boundary
of the Fock space state is a translation by $s$ of the right boundary
of $e^{-sL_R}$.
Concretely, we showed  in claim (2) of Section~\ref{dofeoruhk} that
\begin{equation}\label{limit1}
   \lim_{\lambda\to 0} w\Bigl(
   a_0 + \half +\gamma^\lambda_L\Bigr)-w\Bigl( e^{s}
   \bigl(  a_0 + \half+\gamma^\lambda_R\bigr)\Bigr)  =s\,.
\end{equation}
The limit holds uniformly on $0\leq\theta\leq\frac{\pi}{2}$.
This result can be easily generalized to the uniform convergence
\begin{equation}\label{limit} 
\boxed{\begin{split}
   \phantom{\biggl(}&\lim_{\lambda\to 0}~ w\bigl(e^{s'}(d+\gamma^\lambda_{L/R})\bigr)-w\bigl( e^{s''}(d+\gamma^\lambda_{R/L})\bigr)  
   =s''-s'\,~~ \\ &\quad \forall \textrm{
    $d,s',s''\in\mathbb{R}$
   independent of  $\lambda$\,;\,  $d\neq 0$}\,.\phantom{\Bigl(}
   \end{split}}
\end{equation}
Note that only this case of mixed curves $\gamma_L$ and $\gamma_R$ is non-trivial. 
If both curves are either of the type $\gamma_L$ or of the type $\gamma_R$,
the identity analogous to~(\ref{limit}) is exact for all $\lambda$:
\begin{equation}\label{limittrivial}
  w\bigl(e^{s'}(d+\gamma^\lambda_{L/R})\bigr)-w\bigl( e^{s''}(d+\gamma^\lambda_{L/R})\bigr)  =s''-s'
 \quad\text{ for all} \quad \lambda\geq0 \,.
\end{equation}
This follows from the definition of the map~(\ref{wzrel}) to 
the $w$ frame.

\subsection{Gluings and identifications on $\Sigma$}
For one-loop diagrams, we know from~(\ref{Sigmagen}) that the surface $\Sigma$ is constructed by the multiplication of $2n$ slanted wedges. Let as analyze the validity of the gluing between any pair of neighboring slanted wedges in this product. To this end,
we split the surface $\Sigma$ into two slanted wedges. One of them comprises all the surfaces to the left of the gluing we are interested in, the other one comprises all the surfaces to the right of this gluing. We thus have
\begin{equation}\label{}
   \Sigma =[a;e^{s_{\rm{eff}}}]=[a_1;b_1]\ast[a_2;b_2]\,.
\end{equation}
When multiplying the slanted wedges $[a_1;b_1]$ and $[a_2;b_2]$, we need to glue the right boundary of $[a_1;b_1]$ to the
left boundary $[a_2;b_2]$. To see that our usual multiplication prescription is valid, we analyze this gluing
of the two boundaries using $\lambda$-regulated slanted wedges. If these boundaries are either both of the type $\gamma^\lambda_L$ or both of the type $\gamma^\lambda_R$, the gluing is
natural for all $\lambda$ and no limit needs to be taken. If the boundaries are of mixed type, the gluing curves do not match
for $\Im(z)$ of order $\Lambda$. The surfaces thus either start overlapping or separating in this region.
But using~(\ref{limit}) in the form
\begin{equation}
     \lim_{\lambda\to 0}~ w\bigl(\half+a_0+a_1+b_1\gamma^\lambda_L\bigr)-w\bigl(\half+a_0+a_1+b_1\gamma^\lambda_R\bigr)  =0 \,,
\end{equation}
 we see that the boundaries match in the  $w$ frame for $\lambda\to0$. The shift $a_0$, defined in~(\ref{defa0}), is
 independent of $\lambda$ so that uniform convergence is
guaranteed.
We have thus shown that all the slanted wedges which are multiplied
 in the construction of $\Sigma$ glue nicely to each other  in the 
 $w$ frame when $\lambda\to0$.

Eventually, we also have to glue the left and right boundaries of $\Sigma$ to each other. This is done by the map from the 
$w$ frame to the annulus $\zeta$ through~(\ref{zetadef}). This map has the periodicity
$w\sim w-s_{\rm{eff}}$.
It is thus sufficient to show
that in the limit $\lambda\to0$, the left and right boundaries of $\Sigma$ are related through a translation by $s_{\rm{eff}}$ in the $w$ frame. 
 As shown in~(\ref{limittrivial}), if 
both boundaries of $\Sigma$ are 
of  type $\gamma_L^\lambda$ or both are of  type $\gamma_R^\lambda$, this
relation in the 
$w$ frame is exact by construction, even for finite $\lambda$.
If one 
boundary is of the type $\gamma^\lambda_L$ and the other  
boundary is of the type $\gamma_R^\lambda$, we can
use~(\ref{limit}) in the form
\begin{equation}
   \lim_{\lambda\to 0}~ w\bigl(\half+a_0+\gamma^\lambda_{L/R}\bigr)-w\bigl( e^{s_{\rm{eff}}}(\half+a_0+\gamma^\lambda_{R/L})\bigr)=s_{\rm{eff}}\,
\end{equation}
to see that the two boundaries of $\Sigma$ are related through a translation by $s_{\rm{eff}}$ in the $w$ frame. 

\medskip
In summary, we have shown that the $w$-frame image of 
$\Sigma$, as $\lambda \to 0$,  
represents a smooth surface which is foliated by horizontal 
lines of length $|s_{\rm{eff}}|$. Clearly, the same arguments as above also apply to the surface $\widehat\Sigma$. To complete the proof
of our prescription, we still need to show that the surfaces $\Sigma$ and $\widehat \Sigma$ also glue smoothly to each other at
their hidden boundaries.

\subsection{Gluing the hidden boundaries}\label{sechidden}
In constructing one-loop amplitudes for Schnabl gauge, we cut the surfaces associated with $e^{-s_iL}$ and $e^{-s^\star_iL^\star}$ into two
pieces
associated with their left and right parts.
As we saw 
 using 
$\lambda$-regularization, these surfaces really have a hidden boundary at
$i\infty$ at which they were cut, and we need to ensure that these hidden boundaries glue nicely when we form the annulus.

In the $\lambda$-regularized construction
 the hidden boundaries 
are of the general form
\begin{equation}\label{genhiddbnd}
     \boxed{\phantom{\biggl(}
     z=d+i x\Lambda  \qquad \textrm{ with } \quad e^{s'}\leq x\leq e^{s''}\,, \qquad
     d, s', s''\in\mathbb{R}~\textrm{ independent of }\lambda\,.}
\end{equation}
The parameters $d$, $s'$, and $s''$ are thus suitable to characterize general hidden boundaries of slanted wedges. They emerge
as actual vertical boundary segments once the slanted wedge is regularized,
but we will still call them hidden boundaries, to avoid confusion with
other types of boundaries. 
The hidden boundary of 
 $e^{-s^\star {(L^\lambda)}^\star_R}$, for example,
  stretches from $1+ie^{-s^\star}\Lambda$ to $1+i\Lambda$,
  as we can see in Figure~\ref{figregesstar}. We thus have $d=1$, $s'=-s^\star$, and $s''=0$ as the parameters of the hidden boundary of $e^{-s^\star L^\star_R}$.  
   Note also that the parameters 
   $s'$ and $s''$  
   are just
  the logarithms of the scaling factors that apply to the left or right boundaries
  of the surface
  associated with $e^{-s^\star L^\star_R}$. 
   The parameter $s''$ that defines the top endpoint 
    of the  hidden boundary arises from the left boundary which has a scale 
  factor of one, thus $s''=0$.  
  The parameter $s'$ that defines the bottom endpoint 
  of the hidden boundary arises from the right boundary, which
  has a scale factor of $e^{-s^\star}$, thus $s' = -s^\star$.
  
  Let us summarize the parameters of hidden boundaries for 
 slanted
wedges associated with propagators:
\begin{align}\label{listhidden1} 
  d&=0 & s'&=0 & s''&=s 
  &&\textrm{for the hidden boundary of }e^{-sL_R}\,,\\
\label{listhidden2}  d&=1 & s'&=-s & s''&=0 
  &&\textrm{for the hidden boundary of }e^{-sL_L}\,,\\
\label{listhidden3}  d&=1 & s'&=-s^\star & s''&=0 
  &&\textrm{for the hidden boundary of }e^{-s^\star L^\star_R}\,,\\
\label{listhidden4}  d&=0 & s'&=0 & s''&=s^\star 
  &&\textrm{for the hidden boundary of }e^{-s^\star L^\star_L}\,.
\end{align}
When we multiply regulated 
slanted wedges to form the surfaces $\Sigma$ and $\widehat\Sigma$, 
hidden boundaries 
get shifted and rescaled. Of course they are then still of the general form~(\ref{genhiddbnd}).

In proving  claim (1) of Section~\ref{dofeoruhk} 
--that the hidden boundaries of $e^{-sL_L}$ and $e^{-sL_R}$ glue nicely in the 
tadpole graph-- we showed that 
\begin{equation}\label{showedgoodglue} 
   \lim_{\lambda\to0}~ w(i x\Lambda)-w(a_0+1+i x\Lambda) =0 \qquad \textrm{ for all } \quad 1\leq x\leq e^{s}\,.
\end{equation}
More generally, 
consider two hidden boundaries of the form~(\ref{genhiddbnd}) with identical ranges of $x$  so that they are related by
just a horizontal translation.  
If the horizontal distance $\Delta d\in\mathbb{R}$ that separates
these hidden boundaries is independent of $\lambda$,  
 they glue nicely in the 
$w$ frame in the limit $\lambda\to 0$.
 A straightforward generalization of the proof in Section~\ref{dofeoruhk} indeed shows that 
the following limit holds uniformly
\begin{equation}\label{goodglue} 
   \boxed{\phantom{\biggl(}
 \lim_{\lambda\to0}~ w(d+i x\Lambda)-w(d+\Delta d+i x\Lambda) =0  
 \quad \forall ~ 
e^{s'}\leq x\leq e^{s''} ~ \text{ with } ~
d,\Delta d,s',s''\in\mathbb{R} 
 \,.
 ~}
\end{equation}

We now show that for one-loop diagrams all gluings of hidden boundaries work nicely in the  $w$ frame.
Each propagator in the loop has two hidden boundaries, one from cutting $e^{-sL}$ and one from cutting $e^{-s^\star L^\star}$.
For definiteness, we analyze the $k$-th propagator in the loop and assume it is of type ${\cal P}_+$.  This propagator is responsible for the insertion
of the slanted wedge 
associated with
\be
\label{insert1} 
e^{-s_kL_R}\, \underline{e^{-s^\star_kL^\star_R}}\,\qquad \hbox{into} ~~\Sigma\,. 
\ee
The same propagator will be responsible for the insertion 
of the slanted wedge associated with
\be 
\label{insert2}
e^{-s_kL_L}\, \underline{e^{-s^\star_k L^\star_L}}\,\qquad \hbox{into} ~~ 
\widehat\Sigma\,.
\ee
We will focus on the underlined operators in the two expressions
above.  The first produces a hidden boundary in $\Sigma$ and the
second a  hidden boundary in $\widehat \Sigma$.   We aim to show that these hidden
boundaries
appear at the same height and have the same 
 vertical range 
so that
(\ref{goodglue}) implies that they glue correctly as the regulator is
removed. More concretely, we want to show that these two hidden boundaries
are characterized by (\ref{genhiddbnd}) with identical parameters
$s'$ and $s''$, both independent of $\lambda$.  
The 
value of $d$ for each boundary must also be $\lambda$-independent.

Let us begin 
with the hidden  boundary generated by
$e^{-s^\star_kL^\star_R}$ in $\Sigma$.
Just before $\Sigma$ is mapped to
 the $w$ frame, the 
associated  slanted wedge has its 
left boundary at a position $x^k_L$ that is independent of
$\lambda$ (as is familiar from our calculations of positions in
Section~6, positions just depend on Schwinger parameters).
According to~(\ref{listhidden3}),
the hidden boundary of  $e^{-s^\star_kL^\star_R}$, as a canonically presented slanted wedge 
is positioned a distance $d-\half =\half$
 to the right of its left boundary.
We conclude that on $\Sigma$, it is positioned
 a distance $\half b_k$ from its left boundary, i.e. at
 $d=x_L^k+\half b_k$.  The factor of $b_k$ is necessary because
 it represents the \emph{local scale factor}:  it is the product of the  
 scale factors of all the slanted wedges 
 that precede the insertion of 
 $e^{-s^\star_kL^\star_R}$ in $\Sigma$ (see (\ref{Sigmagen})).
 Note that the first operator in $(\ref{insert1})$ does not 
 contribute to the local scale factor 
 because its slanted wedge 
 ends up to the right of the one we are looking at. 
 As both $x_L^k$ and $b_k$ are manifestly $\lambda$ independent, 
 so is  the location $d$ of the hidden boundary.
 This is all that matters,  its specific value is not needed.

The parameters $s'$ and $s''$ for  $e^{-s^\star_kL^\star_R}$ listed in (\ref{listhidden3}) get a contribution from the logarithm of the
local scale factor $b_k$ at the insertion.  We thus have: 
\be
\label{8okdn}
s' = -s_k^\star  + \ln  b_k \,, \quad s'' = 0 + \ln b_k = \ln b_k \,,
 \qquad \text{for }e^{-s^\star_kL^\star_R}\text{ on }\Sigma\,.
\ee

Let us now consider the insertion of 
$e^{-s^\star_k L^\star_L}$ on $\widehat\Sigma$, just before 
$\widehat\Sigma$ is mapped to
the $w$ frame. The position 
$\hat{x}_L^k$ 
of the  associated slanted wedge, defined as the real value of its left boundary, is 
independent of $\lambda$.  What we need is the local scale factor
$b_{\rm{loc}}$
at this  position.  
For this we 
recall that $\widehat \Sigma$ is 
 rescaled by $e^{s_{\rm{eff}}}$ in such a way that its left boundary
has scaling factor  $e^{s_{\rm{eff}}}$ and the right boundary has 
unit scaling factor. It then follows from (\ref{hatSigmagen}) that, in addition
to  $e^{s_{\rm{eff}}}$,  
we get the multiplicative contribution from the slant parameters
 of $L_n, L_{n-1}, \ldots,
L_{k+1}$ {\em and} the slant parameter of the first 
operator in (\ref{insert2}).
This gives
\be
b_{\rm{loc}} = e^{s_{\rm{eff}}}  \cdot  e^{\sum_{j=k+1}^n (s_j^\star - s_j)}\cdot
e^{-s_k} =     e^{\sum_{j=1}^k (s_j - s_j^\star)}\cdot
e^{-s_k} = b_k e^{-s_k^\star} \,.
\ee
This time the $s'$ and $s''$ parameters in (\ref{listhidden4}) are modified
by the addition of the logarithm of~$b_{\rm{loc}}$.  We thus get
\be
\label{eohckug}
s' = 0 + \ln b_{\rm{loc}} = -s_k^\star + \ln b_k \,, \quad
s'' = s_k^\star + \ln b_{\rm{loc}} = \ln b_k \,, 
 \qquad \text{for }e^{-s^\star_kL^\star_L}\text{ on }\widehat\Sigma\,.
\ee
Comparing~(\ref{8okdn}) with~(\ref{eohckug}), we see that the 
parameters $s'$ and $s''$ match precisely. 
Therefore we conclude from~(\ref{goodglue})
that the hidden boundaries of $e^{-s^\star_k L^\star_R}$ and $e^{-s^\star_k L^\star_L}$ glue nicely in the  $w$ frame.

\bigskip
\noindent
A few remarks are in order.
\begin{itemize}
  \item
    The hidden boundaries of $e^{-s_k L_R}$ and $e^{-s_k L_L}$ also
    glue seamlessly in the  $w$ frame. The proof is completely
    analogous to the one presented above.
  \item
    The propagators in  subtrees also have hidden boundaries. The hidden boundaries associated with a subtree propagator are either
    both on $\Sigma$ or both on $\widehat \Sigma$.  These boundaries cannot be simply ignored
    because unlike in tree-amplitudes, the subtree is mapped to the annulus and \emph{not} to the disk.
    Still, it is easy to  see by a similar analysis as for loop propagators that these hidden boundaries glue nicely in the
   $w$ frame.
  \item
  One might wonder if the question 
  of gluing hidden boundaries could  have been ignored.
  After all,   these hidden  
    boundaries are pushed off to $i\infty$ in the $z$ frame and to $\frac{\pi}{2}-i\infty$ in the $w$ frame and these seem to be well defined points. 
    This naive argument, however, 
     leads to wrong conclusions.  It would allow
\emph{independent}  $z$-frame rescalings of   $\Sigma$ and $\widehat\Sigma$, which is equivalent to shifting their relative horizontal
    position in the $w$ frame. On the annulus, this corresponds to rotating the inner and outer boundaries of the annulus with respect to
    each other. But if we have insertions on both the inner and the outer boundary, the configuration after such a relative rotation
    is \emph{not} conformally equivalent to the original one. 
    We conclude that the naive
    expectation that the gluing in Schnabl gauge works out correctly automatically, leaves an ambiguity in the open string moduli.
     This ambiguity is fixed when we regulate and demand the gluing to work out nicely in the limit $\lambda\to0$.
  \item
    While we did our analysis of the gluing of hidden boundaries using the $\lambda$-regulated gauges, any other family of
    regular linear $b$-gauges associated with zero modes in the frames $z=\tilde f^\lambda(\xi)$ could have been used, as long
    as this family approaches Schnabl gauge in the limit $\lambda\to 0$. By construction, for such
    a family the frames satisfy
    \begin{equation}\label{}
       \lim_{\lambda\to0} \tilde\Lambda=\infty \quad \textrm{ with } \quad 
       i\tilde\Lambda\equiv \tilde f^\lambda(i)\,.
    \end{equation}
    The proof of consistent gluing of hidden boundaries goes through with 
     $\Lambda\to\tilde \Lambda$.
\end{itemize}

\medskip
\noindent
This concludes the proof of our prescription for the construction of general one-loop Riemann surfaces in Schnabl gauge.

\section{Concluding remarks}\label{secconcl}
\setcounter{equation}{0}

The open
string midpoint has played a very subtle and important role
in covariant open string field theory.
The midpoint makes it non-trivial to formulate open string
field theory as a theory of half-strings (see~\cite{Gross:2001yk}).
Spacetime diffeomorphisms are not quite open-string gauge
symmetries because of the special status of the
midpoint in the star product~\cite{Witten:1986gi}.
  Nevertheless, closed string poles appear
in open string loop diagrams, again because of the special
role of the midpoint.  Naively, the star algebra was
expected to have no projectors.  But again, open string surface
states with singular behavior at the midpoint give rise to
projectors that seem to be completely consistent.

It is perhaps no surprise then that the tachyon vacuum solution
uses a gauge, Schnabl gauge, that is described by the
 conformal frame of a projector.  So does the rolling tachyon solution that
describes the decay of a D-brane. Since observables associated with
 these solutions
probe closed string 
physics~\cite{0804.1131,Hashimoto:2001sm,Gaiotto:2001ji}
 it is natural to ask if
 the use of
Schnabl gauge
allows the correct incorporation of
 closed string physics.
 As a first step,
we ask if Schnabl gauge, just like Siegel gauge, leads to correct
loop amplitudes.  Indeed, naive
arguments suggested that the singular midpoint behavior in Schnabl
gauge could ruin the validity of the gauge at
loop level, precisely
where closed string physics is revealed.  In a nutshell, the string diagrams
for one loop appeared to give a surface that is disconnected into two pieces,
each of which contains one of the boundary components of the annulus.

The analysis presented in this paper gives reason for optimism
and teaches us a few facts:

\begin{itemize}

\item   The left and
right parts of  the operator $L$
(the Virasoro zero-mode in sliver frame) fail to commute.  This non-commutation is required by consistency:
it introduces a finite hidden boundary to each of the two disconnected surfaces that form the annulus.  The gluing across the hidden boundaries
restores the closed string moduli.

\item  Schnabl gauge string diagrams at one loop cover the 
(one-dimensional) closed string moduli space. 
This is no proof of complete consistency, but
dispels the fear of inconsistency due to subtle midpoint effects.

\item All moduli, open and closed, of one-loop amplitudes with arbitrary numbers of open string states are calculable in closed form.
Schnabl gauge off-shell amplitudes may  ultimately be recognized as simpler than those
in the familiar Siegel or light-cone gauges.

\item  Wedge surfaces have a natural generalization in the form of
{\em slanted} wedges.  Only on slanted wedges we have a natural
action of the left and right parts of the operators $L$ and $L^\star$.  
The use of these surfaces allows us to give (for the first time) an explicit algorithm
to construct arbitrarily complicated tree and one-loop diagrams.

\end{itemize}

The focus in this paper has been narrow.
We have studied the moduli of the diagrams generated in Schnabl gauge.
We have not calculated any loop amplitude in detail.  For this one must,
of course, deal with the antighost and BRST insertions.  Even
 regarding
moduli we have not answered everything.
 Though the specific examples we have analyzed in this paper are encouraging,
 it is not yet clear whether open string moduli are covered in general.
 This problem is in fact still unsolved at tree-level.
We are lacking proof that even
tree amplitudes are correctly reproduced in Schnabl gauge.
The open string propagator has moduli associated with the operators 
$B/L$ and $B^\star/L^\star$,
but also contains the BRST operator $Q$, which acts as a total derivative on moduli space.
Our analysis of the tree-level five-point function and the one-loop two-point function
suggests that there might be an assignment of  simplified propagators $B/L$ and $B^\star/L^\star$
to the propagator lines so that the string diagram has all the requisite degenerations.
Finding such an assignment
 could be the next 
step in a proof of consistency of Schnabl gauge.

 The $\lambda$-regularized gauges are fully consistent and Schnabl gauge amplitudes can in principle
 be defined by
 the limit  $\lambda\to 0$ of
   $\lambda$-regulated amplitudes.
 Calculating regularized amplitudes is problematic, because even at small (but fixed) $\lambda$,
 the geometry differs significantly from the Schnabl geometry when any Schwinger parameter becomes large, i.e.
 of order
  $\ord{\log\log\lambda^{-1}}$.
 When one imposes cutoffs on the integration region of Schwinger parameters, the limit
 of removing these cutoffs and the limit $\lambda\to0$ do not, in general, commute. It would be interesting to determine
 a cutoff prescription for which these limits  commute and thus define consistent amplitudes for Schnabl gauge.
 A possible candidate for such a prescription is a generalization of the symmetric limit defined for the four-point amplitude in~\cite{0708.2591}.
 Note that, in this paper, we took the limit $\lambda\to0$ at fixed Schwinger parameters and any amplitude calculated using
 these surfaces is thus a true Schnabl-gauge amplitude and needs to be supplemented with a suitable prescription
 on the  integration over Schwinger parameters.

The conformal field theory boundary state of the rolling tachyon has been studied
 to extract the time-dependent pressure profile of tachyon condensation  (see~\cite{Sen:2004nf} and references therein). The result
suggests that the pressure goes to zero at late times, consistent with the 
expectation that the D-brane decays into heavy non-relativistic closed strings. The conformal field theory analysis of the closed
string production in the background of the rolling tachyon encounters
UV divergences~\cite{Lambert:2003zr}.
As the corresponding analytic solution of string field theory has been found, this problem can now also be studied within open string field theory.\footnote{For an interesting recent analysis of observables associated with on-shell closed string states, see~\cite{0804.1131}.}
It would be interesting to extract a boundary state from the one-loop 
open-string vacuum amplitude in the background of the
rolling tachyon solution. This string field theory
boundary state may confirm the
expected late time behavior of the pressure and could help us 
understand the role of observables in open string field theory.

All in all, our work shows that Schnabl gauge is not only
a convenient gauge for analytic solutions in string field theory but
also simplifies string perturbation theory considerably. While the
ultimate consistency proof is still pending, we hope that the tools
developed here will  help construct this proof and lead to
new insights into the role of closed strings in open string field theory.

\vspace{0.6cm}

{\bf \large Acknowledgments:}
We would like to thank Ashoke Sen for helpful discussions and
for collaboration in the initial stages of this project.
 We also thank Yuji Okawa and Martin Schnabl for valuable comments on a draft
 version of this paper.
The work of M.K. and
B.Z. is supported in part by the U.S.
DOE grant DE-FC02-94ER40818.

\appendix
\section{Covering of Moduli space in the five-point diagram}\label{app5pt}
\setcounter{equation}{0}
In this appendix we will analyze which assignment of $B/L$ and $B^\star/L^\star$
to the propagators in the five-point amplitude
 always produces
open string degenerations when a Schwinger parameter becomes
large. To do so, we will set one of the Schwinger parameters of each propagator
to zero in our result for the angles of operator insertions on the unit
circle~(\ref{angles5point}).
Notice that the only degenerations we expect from the diagram in
Figure~\ref{fig5point}
 are
the collision of the insertions of  $\ket{F_A}$ and $\ket{F_B}$, and the collision
of the insertions of  $\ket{F_D}$ and $\ket{F_E}$.
There are three distinct cases of $B/L$ and $B^\star/L^\star$ assignments.
\begin{itemize}
  \item
    {\bf case 1: $t_1=t_2=0$ (propagator 1: $B^\star/L^\star$; propagator 2:
$B^\star/L^\star$)}\\
    In this case we obtain for the angles of operator insertions:
in~(\ref{angles5point})\\
    \begin{center}\small
    \begin{tabular}{|c|c|c|c|c|}
    \hline
    & finite $t_1^\star$, $t_2^\star\phantom{\Bigl{(}}$ & $t_1^\star\to\infty$
    & $t_2^\star\to\infty$ & $t_1^\star=t_2^\star\to\infty$\\
    \hline
    $\phantom{\biggl{(}}\frac{\phi_B-\phi_A}{2\pi}$ &
$\frac{e^{t_1^\star}}{3e^{t_1^\star}+3e^{t_2^\star}-1}$
    & $\frac{1}{3}$ & $0$ & $\frac{1}{6}$\\
    \hline
    $\phantom{\biggl{(}}\frac{\phi_C-\phi_A}{2\pi}$ &
$\frac{2e^{t_1^\star}}{3e^{t_1^\star}+3e^{t_2^\star}-1}$
    & $\frac{2}{3}$ & $0$ & $\frac{1}{3}$\\
    \hline
    $\phantom{\biggl{(}}\frac{\phi_D-\phi_A}{2\pi}$ &
$\frac{2e^{t_1^\star}+e^{t_2^\star}}{3e^{t_1^\star}+3e^{t_2^\star}-1}$
    & $\frac{2}{3}$ & $\frac{1}{3}$ & $\frac{1}{2}$\\
    \hline
    $\phantom{\biggl{(}}\frac{\phi_E-\phi_A}{2\pi}$ &
$\frac{2e^{t_1^\star}+2e^{t_2^\star}}{3e^{t_1^\star}+3e^{t_2^\star}-1}$
    & $\frac{2}{3}$ & $\frac{2}{3}$ & $\frac{2}{3}$\\
    \hline
    \end{tabular}
    \end{center}
    The angles $\phi_C$, $\phi_D$, and $\phi_E$
    approach each other for
$t_1^\star\to\infty$, if $t_2^\star$ stays finite. This is conformally
equivalent to $\phi_A$ and $\phi_B$ coming close together. But this
    cannot be a stable degeneration, because if $t_2^\star$ also becomes large,
the angles $\phi_A$ and $\phi_B$ are no longer
    degenerate. In fact, for $t_1^\star=t_2^\star\to\infty$
 all insertions are separated by finite angles from each other!
Thus this is not a
consistent assignment of $B/L$ and $B^\star/L^\star$. 

  \item
    {\bf case 2: $t_1^\star=t_2^\star=0$ (propagator 1: $B/L$; propagator 2:
$B/L$)}\\
    In this case we obtain for the angles of operator insertions
in~(\ref{angles5point}):\\
    \begin{center}\small
    \begin{tabular}{|c|c|c|c|c|}
    \hline
    & finite $t_1$, $t_2\phantom{\Bigl{(}}$ & $t_1\to\infty$ & $t_2\to\infty$
     & $t_1=t_2\to\infty$\\
    \hline
     $\phantom{\biggl{(}}\frac{\phi_B-\phi_A}{2\pi}$   &
$\quad\frac{e^{-t_1}}{3+e^{-t_1}+e^{-t_2}}\quad$
    & $0$
    & $\quad\frac{e^{-t_1}}{3+e^{-t_1}}\quad$
    & $0$\\
    \hline
     $\phantom{\biggl{(}}\frac{\phi_C-\phi_A}{2\pi}$   &
$\quad\frac{1+e^{-t_1}}{3+e^{-t_1}+e^{-t_2}}\quad$
    & $\frac{1}{3+e^{-t_2}}$
    &$\quad\frac{1+e^{-t_1}}{3+e^{-t_1}}\quad$
    & $\frac{1}{3}$\\
    \hline
     $\phantom{\biggl{(}}\frac{\phi_D-\phi_A}{2\pi}$   &
$\quad\frac{2+e^{-t_1}}{3+e^{-t_1}+e^{-t_2}}\quad$
    & $\frac{2}{3+e^{-t_2}}$
    &$\quad\frac{2+e^{-t_1}}{3+e^{-t_1}}\quad$
    & $\frac{2}{3}$\\
    \hline
     $\phantom{\biggl{(}}\frac{\phi_E-\phi_A}{2\pi}$   &
$\quad\frac{2+e^{-t_1}+e^{-t_2}}{3+e^{-t_1}+e^{-t_2}}\quad$
    & $\frac{2+e^{-t_2}}{3+e^{-t_2}}$
    &$\quad\frac{2+e^{-t_1}}{3+e^{-t_1}}\quad$
    & $\frac{2}{3}$\\
    \hline
    \end{tabular}
    \end{center}
    The angles $\phi_A$ and $\phi_B$ come close together for $t_1\to\infty$.
Making $t_2$ also large
    cannot prevent the degeneration.
    Similarly, the degeneration of  $\phi_D$ and $\phi_E$ in the limit
$t_2\to\infty$ cannot be undone
    by making $t_1$ comparably large.
    Thus, this is a good assignment of propagators.

  \item
    {\bf case 3: $t_1^\star=t_2=0$ (propagator 1: $B/L$ , propagator 2:
$B^\star/L^\star$)}\\
    In this case we obtain for the angles of operator insertions
in~(\ref{angles5point}):\\
    \begin{center}\small
    \begin{tabular}{|c|c|c|c|c|}
    \hline
    & finite $t_1$, $t_2^\star\phantom{\Bigl{(}}$ & $t_1\to\infty$ &
$t_2^\star\to\infty$
     & $t_1=t^\star_2\to\infty$\\
    \hline
     $\phantom{\biggl{(}}\frac{\phi_B-\phi_A}{2\pi}$   &
$\quad\frac{e^{-t_1}}{1+e^{-t_1}+3e^{t_2^\star}}\quad$
    &  $0$ & $0$ & $0$\\
    \hline
     $\phantom{\biggl{(}}\frac{\phi_C-\phi_A}{2\pi}$   &
$\quad\frac{1+e^{-t_1}}{1+e^{-t_1}+3e^{t_2^\star}}\quad$
    &  $\frac{1}{1+3e^{t_2^\star}}$ & $0$ & $0$\\
    \hline
     $\phantom{\biggl{(}}\frac{\phi_D-\phi_A}{2\pi}$   &
$\quad\frac{1+e^{-t_1}+e^{t_2^\star}}{1+e^{-t_1}+3e^{t_2^\star}}\quad$
    &  $\frac{1+e^{t_2^\star}}{1+3e^{t_2^\star}}$ & $\frac{1}{3}$ &
$\frac{1}{3}$\\
    \hline
     $\phantom{\biggl{(}}\frac{\phi_E-\phi_A}{2\pi}$   &
$\quad\frac{1+e^{-t_1}+2e^{t_2^\star}}{1+e^{-t_1}+3e^{t_2^\star}}\quad$
    &  $\frac{1+2e^{t_2^\star}}{1+3e^{t_2^\star}}$ & $\frac{2}{3}$ &
$\frac{2}{3}$\\
    \hline
    \end{tabular}
    \end{center}
    The angles $\phi_A$ and $\phi_B$ come close together for $t_1\to\infty$.
    Making $t^\star_2$ also large, cannot prevent the degeneration.
    Similarly, for $t_2^\star$ very large, $\phi_A$,  $\phi_B$ and $\phi_C$
approach each other.
This is conformally equivalent to
    $\phi_D$ and $\phi_E$ coming close together. Again, this cannot be undone by
making $t_1$ comparably large.
    Thus, this is also a good assignment of propagators.

\end{itemize}

\end{document}